\documentstyle[12pt]{report}

\newcommand{\be}{\begin{equation}}
\newcommand{\ee}{\end{equation}}
\newcommand{\bea}{\begin{eqnarray}}
\newcommand{\eea}{\end{eqnarray}}
\newcommand{\ptp}{P\"oschl-Teller }

\newcommand{\vrp}{\varepsilon}
\newcommand{\th}{\theta}
\newcommand{\s}{{\cal S}}

\newcommand{\half}{{1 \over 2}}
\newcommand{\g}{\gamma}
\newcommand{\p}{\prime}
\newcommand{\al}{\alpha}
\newcommand{\ba}{\beta}
\newcommand{\de}{\delta}

\newcommand{\ov}{\over}
\newcommand{\ve}{\varepsilon}

\newread\epsffilein    
\newif\ifepsffileok    
\newif\ifepsfbbfound   
\newif\ifepsfverbose   
\newdimen\epsfxsize    
\newdimen\epsfysize    
\newdimen\epsftsize    
\newdimen\epsfrsize    
\newdimen\epsftmp      
\newdimen\pspoints     
\def\epsfbox#1{\global\def\epsfllx{72}\global\def\epsflly{72}%
   \global\def\epsfurx{540}\global\def\epsfury{720}%
   \def\lbracket{[}\def\testit{#1}\ifx\testit\lbracket
   \let\next=\epsfgetlitbb\else\let\next=\epsfnormal\fi\next{#1}}%
\def\epsfgetlitbb#1#2 #3 #4 #5]#6{\epsfgrab #2 #3 #4 #5 .\\%
   \epsfsetgraph{#6}}%
\def\epsfnormal#1{\epsfgetbb{#1}\epsfsetgraph{#1}}%
\def\epsfgetbb#1{%
\openin\epsffilein=#1
\ifeof\epsffilein\errmessage{I couldn't open #1, will ignore it}\else
   {\epsffileoktrue \chardef\other=12
    \def\do##1{\catcode`##1=\other}\dospecials \catcode`\ =10
    \loop
       \read\epsffilein to \epsffileline
       \ifeof\epsffilein\epsffileokfalse\else
          \expandafter\epsfaux\epsffileline:. \\%
       \fi
   \ifepsffileok\repeat
   \ifepsfbbfound\else
    \ifepsfverbose\message{No bounding box comment in #1; using defaults}\fi\fi
   }\closein\epsffilein\fi}%
\def\epsfclipstring{}
\def\epsfsetgraph#1{%
   \epsfrsize=\epsfury\pspoints
   \advance\epsfrsize by-\epsflly\pspoints
   \epsftsize=\epsfurx\pspoints
   \advance\epsftsize by-\epsfllx\pspoints
   \epsfxsize\epsfsize\epsftsize\epsfrsize
   \ifnum\epsfxsize=0 \ifnum\epsfysize=0
      \epsfxsize=\epsftsize \epsfysize=\epsfrsize
      \epsfrsize=0pt
     \else\epsftmp=\epsftsize \divide\epsftmp\epsfrsize
       \epsfxsize=\epsfysize \multiply\epsfxsize\epsftmp
       \multiply\epsftmp\epsfrsize \advance\epsftsize-\epsftmp
       \epsftmp=\epsfysize
       \loop \advance\epsftsize\epsftsize \divide\epsftmp 2
       \ifnum\epsftmp>0
          \ifnum\epsftsize<\epsfrsize\else
             \advance\epsftsize-\epsfrsize \advance\epsfxsize\epsftmp \fi
       \repeat
       \epsfrsize=0pt
     \fi
   \else \ifnum\epsfysize=0
     \epsftmp=\epsfrsize \divide\epsftmp\epsftsize
     \epsfysize=\epsfxsize \multiply\epsfysize\epsftmp   
     \multiply\epsftmp\epsftsize \advance\epsfrsize-\epsftmp
     \epsftmp=\epsfxsize
     \loop \advance\epsfrsize\epsfrsize \divide\epsftmp 2
     \ifnum\epsftmp>0
        \ifnum\epsfrsize<\epsftsize\else
           \advance\epsfrsize-\epsftsize \advance\epsfysize\epsftmp \fi
     \repeat
     \epsfrsize=0pt
    \else
     \epsfrsize=\epsfysize
    \fi
   \fi
   \ifepsfverbose\message{#1: width=\the\epsfxsize, height=\the\epsfysize}\fi
   \epsftmp=10\epsfxsize \divide\epsftmp\pspoints
   \vbox to\epsfysize{\vfil\hbox to\epsfxsize{%
      \ifnum\epsfrsize=0\relax
        \includegraphics{#1}%
      \else
        \epsfrsize=10\epsfysize \divide\epsfrsize\pspoints
        \includegraphics{#1}%
      \fi
      \hfil}}%
\global\epsfxsize=0pt\global\epsfysize=0pt}%
{\catcode`\%=12 \global\let\epsfpercent=
\long\def\epsfaux#1#2:#3\\{\ifx#1\epsfpercent
   \def\testit{#2}\ifx\testit\epsfbblit
      \epsfgrab #3 . . . \\%
      \epsffileokfalse
      \global\epsfbbfoundtrue
   \fi\else\ifx#1\par\else\epsffileokfalse\fi\fi}%
\def\epsfempty{}%
\def\epsfgrab #1 #2 #3 #4 #5\\{%
\global\def\epsfllx{#1}\ifx\epsfllx\epsfempty
      \epsfgrab #2 #3 #4 #5 .\\\else
   \global\def\epsflly{#2}%
   \global\def\epsfurx{#3}\global\def\epsfury{#4}\fi}%
\def\epsfsize#1#2{\epsfxsize}
\makeatletter
\typeout{%
Enhancements to Picture Environment. Version 1.2 - Released June 1, 1986}
\def\lop#1\to#2{\expandafter\lopoff#1\lopoff#1#2}
\long\def\lopoff,#1,#2\lopoff#3#4{\def#4{#1}\def#3{,#2}}
\def\@@mlistempty{,}
\newif\iflistnonempty
\def\multiputlist(#1,#2)(#3,#4){\@ifnextchar
[{\@imultiputlist(#1,#2)(#3,#4)}{\@imultiputlist(#1,#2)(#3,#4)[]}}

\long\def\@imultiputlist(#1,#2)(#3,#4)[#5]#6{{%
\@xdim=#1\unitlength \@ydim=#2\unitlength
\listnonemptytrue \def\@@mlist{,#6,} 
\loop
\lop\@@mlist\to\@@firstoflist
\@killglue\raise\@ydim\hbox to\z@{\hskip
\@xdim\@imakepicbox(0,0)[#5]{\@@firstoflist}\hss}
\advance\@xdim #3\unitlength\advance\@ydim #4\unitlength
\ifx\@@mlist\@@mlistempty \listnonemptyfalse\fi
\iflistnonempty
\repeat\relax
\ignorespaces}}
\newcount\@@multicnt
\def\matrixput(#1,#2)(#3,#4)#5(#6,#7)#8#9{%
\ifnum#5>#8\@matrixput(#1,#2)(#3,#4){#5}(#6,#7){#8}{#9}%
\else\@matrixput(#1,#2)(#6,#7){#8}(#3,#4){#5}{#9}\fi}

\long\def\@matrixput(#1,#2)(#3,#4)#5(#6,#7)#8#9{{\@killglue%
\@multicnt=#5\relax\@@multicnt=#8\relax%
\@xdim=0pt%
\@ydim=0pt%
\setbox\@tempboxa\hbox{\@whilenum \@multicnt > 0\do {%
\raise\@ydim\hbox to \z@{\hskip\@xdim #9\hss}%
\advance\@multicnt \m@ne%
\advance\@xdim #3\unitlength\advance\@ydim #4\unitlength}}%
\@xdim=#1\unitlength%
\@ydim=#2\unitlength%
\@whilenum \@@multicnt > 0\do {%
\raise\@ydim\hbox to \z@{\hskip\@xdim \copy\@tempboxa\hss}%
\advance\@@multicnt \m@ne%
\advance\@xdim #6\unitlength\advance\@ydim #7\unitlength}%
\ignorespaces}}
\newcount\d@lta
\newdimen\@delta
\newdimen\@@delta
\newcount\@gridcnt
\def\grid(#1,#2)(#3,#4){\@ifnextchar [{\@igrid(#1,#2)(#3,#4)}%
{\@igrid(#1,#2)(#3,#4)[@,@]}}

\long\def\@igrid(#1,#2)(#3,#4)[#5,#6]{%
\makebox(#1,#2){%
\@delta=#1pt\@@delta=#3pt\divide\@delta \@@delta\d@lta=\@delta%
\advance\d@lta \@ne\relax\message{grid=\the\d@lta\space x}%
\multiput(0,0)(#3,0){\d@lta}{\hbox to\z@{\hskip -\@halfwidth \vrule
	 \@width \@wholewidth \@height #2\unitlength \@depth \z@\hss}}%
\ifx#5@\relax\else%
\global\@gridcnt=#5%
\multiput(0,0)(#3,0){\d@lta}{%
\makebox(0,-2)[t]{\number\@gridcnt\global\advance\@gridcnt by #3}}%
\global\@gridcnt=#5%
\multiput(0,#2)(#3,0){\d@lta}{\makebox(0,0)[b]{\number\@gridcnt\vspace{2mm}%
\global\advance\@gridcnt by #3}}%
\fi%
\@delta=#2pt\@@delta=#4pt\divide\@delta \@@delta\d@lta=\@delta%
\advance\d@lta \@ne\relax\message{\the\d@lta . }%
\multiput(0,0)(0,#4){\d@lta}{\vrule \@height \@halfwidth \@depth \@halfwidth
	 \@width #1\unitlength}%
\ifx#6@\relax\else
\global\@gridcnt=#6%
\multiput(0,0)(0,#4){\d@lta}{%
\makebox(0,0)[r]{\number\@gridcnt\ \global\advance\@gridcnt by #4}}%
\global\@gridcnt=#6%
\multiput(#1,0)(0,#4){\d@lta}{%
\makebox(0,0)[l]{\ \number\@gridcnt\global\advance\@gridcnt by #4}}%
\fi}}
\def\picsquare{\hskip -0.5\@wholewidth%
\vrule height \@halfwidth depth \@halfwidth width \@wholewidth}
\def\picsquare@bl{\vrule height \@wholewidth depth \z@  width \@wholewidth}
\newif\if@jointhem \global\@jointhemfalse
\newif\if@firstpoint \global\@firstpointtrue
\newcount\@joinkind
\def\dottedjoin{\global\@jointhemtrue \global\@joinkind=0\relax
  \bgroup\@ifnextchar[{\@idottedjoin}{\@idottedjoin[\picsquare@bl]}}
\def\@idottedjoin[#1]#2{\gdef\dotchar@join{#1}\gdef\dotgap@join{#2}}
\def\enddottedjoin{\global\@jointhemfalse \global\@firstpointtrue\egroup}
\def\dashjoin{\global\@jointhemtrue \global\@joinkind=1\relax
  \bgroup\@ifnextchar[{\@idashjoin}{\@idashjoin[\dashlinestretch]}}
\def\@idashjoin[#1]#2{\edef\dashlinestretch{#1}\gdef\dashlen@join{#2}%
\@ifnextchar[{\@iidashjoin}{\gdef\dotgap@join{}}}
\def\@iidashjoin[#1]{\gdef\dotgap@join{#1}}

\def\drawjoin{\global\@jointhemtrue \global\@joinkind=2\relax
  \bgroup\@ifnextchar[{\@idrawjoin}{}}
\def\@idrawjoin[#1]{\def\drawlinestretch{#1}}

\long\def\jput(#1,#2)#3{{\@killglue\raise#2\unitlength\hbox to \z@{\hskip
#1\unitlength #3\hss}\ignorespaces}
\if@jointhem
 \if@firstpoint \gdef\x@one{#1} \gdef\y@one{#2} \global\@firstpointfalse
 \else\ifcase\@joinkind
	\@dottedline[\dotchar@join]{\dotgap@join\unitlength}%
(\x@one\unitlength,\y@one\unitlength)(#1\unitlength,#2\unitlength)
	\or\@dashline[\dashlinestretch]{\dashlen@join}[\dotgap@join]%
(\x@one,\y@one)(#1,#2)
	\else\@drawline[\drawlinestretch](\x@one,\y@one)(#1,#2)\fi
    \gdef\x@one{#1} \gdef\y@one{#2}
 \fi
\fi}
\newdimen\@dotgap
\newdimen\@ddotgap
\newcount\@x@diff
\newcount\@y@diff
\newdimen\x@diff
\newdimen\y@diff
\newbox\@dotbox
\newcount\num@segments
\newcount\num@segmentsi
\newif\ifsqrt@done
\def\sqrtandstuff#1#2#3{
\ifdim #1 <0pt \@x@diff= -#1 \else\@x@diff=#1\fi
\ifdim #2 <0pt \@y@diff= -#2 \else\@y@diff=#2\fi
\@dotgap=#3 \divide\@dotgap \tw@
\advance\@x@diff \@dotgap \advance\@y@diff \@dotgap
\@dotgap=#3
\divide\@x@diff \@dotgap \divide\@y@diff \@dotgap
\sqrt@donefalse
\ifnum\@x@diff < 2
   \ifnum\@y@diff < 2 \num@segments=\@x@diff \advance\num@segments \@y@diff
		      \sqrt@donetrue
        \else\num@segments=\@y@diff \sqrt@donetrue\fi
   \else\ifnum\@y@diff < 2 \num@segments=\@x@diff \sqrt@donetrue\fi
\fi
\ifsqrt@done \ifnum\num@segments=\z@ \num@segments=\@ne\fi\relax
 \else \ifnum\@y@diff >\@x@diff
		 \@tempcnta=\@x@diff \@x@diff=\@y@diff \@y@diff=\@tempcnta
       \fi    		
  \num@segments=\@y@diff
  \multiply\num@segments \num@segments
  \multiply\num@segments by 457
  \divide\num@segments \@x@diff
  \advance\num@segments by 750 
  \divide\num@segments \@m
  \advance\num@segments \@x@diff
\fi}
\def\dottedline{\@ifnextchar [{\@idottedline}{\@idottedline[\picsquare@bl]}}
\def\@idottedline[#1]#2(#3,#4){\@ifnextchar (%
{\@iidottedline[#1]{#2}(#3,#4)}{\relax}}
\def\@iidottedline[#1]#2(#3,#4)(#5,#6){\@dottedline[#1]{#2\unitlength}%
(#3\unitlength,#4\unitlength)(#5\unitlength,#6\unitlength)%
\@idottedline[#1]{#2}(#5,#6)}
\long\def\@dottedline[#1]#2(#3,#4)(#5,#6){{%
\x@diff=#5\relax\advance\x@diff by -#3\relax
\y@diff=#6\relax\advance\y@diff by -#4\relax
\sqrtandstuff{\x@diff}{\y@diff}{#2}
\divide\x@diff \num@segments
\divide\y@diff \num@segments
\advance\num@segments \@ne     
\setbox\@dotbox\hbox{#1}
\@xdim=#3 \@ydim=#4
\ifdim\ht\@dotbox >\z@
  \advance\@xdim -0.5\wd\@dotbox
  \advance\@ydim -0.5\ht\@dotbox
  \advance\@ydim .5\dp\@dotbox\fi
\@killglue
\loop \ifnum\num@segments > 0
\unskip\raise\@ydim\hbox to\z@{\hskip\@xdim #1\hss}%
\advance\num@segments \m@ne\advance\@xdim\x@diff\advance\@ydim\y@diff%
\repeat
\ignorespaces}}
\def\dashlinestretch{0} 
\def\dashline{\@ifnextchar [{\@idashline}{\@idashline[\dashlinestretch]}}
\def\@idashline[#1]#2{\@ifnextchar [{\@iidashline[#1]{#2}}%
{\@iidashline[#1]{#2}[\@empty]}} 
\def\@iidashline[#1]#2[#3](#4,#5){\@ifnextchar (%
{\@iiidashline[#1]{#2}[#3](#4,#5)}{\relax}}
\def\@iiidashline[#1]#2[#3](#4,#5)(#6,#7){%
\@dashline[#1]{#2}[#3](#4,#5)(#6,#7)%
\@iidashline[#1]{#2}[#3](#6,#7)}
\long\def\@dashline[#1]#2[#3](#4,#5)(#6,#7){{%
\x@diff=#6\unitlength \advance\x@diff by -#4\unitlength
\y@diff=#7\unitlength \advance\y@diff by -#5\unitlength
\@tempdima=#2\unitlength \advance\@tempdima -\@wholewidth
\sqrtandstuff{\x@diff}{\y@diff}{\@tempdima}
\ifnum\num@segments <3 \num@segments=3\fi
\@tempdima=\x@diff \@tempdimb=\y@diff
\divide\@tempdimb by\num@segments
\divide\@tempdima by\num@segments
{\ifx#3\@empty \relax
    \ifdim\@tempdima < 0pt \x@diff=-\@tempdima\else\x@diff=\@tempdima\fi
    \ifdim\@tempdimb < 0pt \y@diff=-\@tempdimb\else\y@diff=\@tempdimb\fi
    \ifdim\x@diff < 0.3pt 
           \ifdim\@tempdimb > 0pt
	        \global\setbox\@dotbox\hbox{\hskip -\@halfwidth \vrule
		 \@width \@wholewidth \@height \@tempdimb}
	   \else\global\setbox\@dotbox\hbox{\hskip -\@halfwidth \vrule
		 \@width \@wholewidth \@height\z@ \@depth -\@tempdimb}\fi
       \else\ifdim\y@diff < 0.3pt 
               \ifdim\@tempdima >0pt
		  \global\setbox\@dotbox\hbox{\vrule \@height \@halfwidth
		 		\@depth \@halfwidth \@width \@tempdima}
		\else\global\setbox\@dotbox\hbox{\hskip \@tempdima
			 \vrule \@height \@halfwidth \@depth \@halfwidth
				 \@width -\@tempdima \hskip \@tempdima}\fi
	    \else\global\setbox\@dotbox\hbox{%
\@dottedline[\picsquare]{0.98\@wholewidth}(0pt,0pt)(\@tempdima,\@tempdimb)}
\fi\fi
\else\global\setbox\@dotbox\hbox{%
\@dottedline[\picsquare]{#3\unitlength}(0pt,0pt)(\@tempdima,\@tempdimb)}
\fi}
\advance\x@diff by -\@tempdima 
\advance\y@diff by -\@tempdimb
\@tempdima=\num@segments\@wholewidth \@tempdima=2\@tempdima 
\@tempcnta=\@tempdima \@tempdima=#2\unitlength \@tempdimb=0.5\@tempdima
\@tempcntb=\@tempdimb \advance\@tempcnta by \@tempcntb 
\divide\@tempcnta by\@tempdima \advance\num@segments by -\@tempcnta
\ifnum #1=0 \relax\else\ifnum #1 < -100
  \typeout{***dashline: reduction > -100 percent implies blankness!***}
\else\num@segmentsi=#1 \advance\num@segmentsi by 100
     \multiply\num@segments by\num@segmentsi \divide\num@segments by 100
\fi\fi
\divide\num@segments by 2 
\ifnum\num@segments >0 
 \divide\x@diff by\num@segments
 \divide\y@diff by\num@segments
 \advance\num@segments by\@ne 
 \else\num@segments=2 
\fi
\@xdim=#4\unitlength \@ydim=#5\unitlength
\@killglue
\loop \ifnum\num@segments > 0
\unskip\raise\@ydim\hbox to\z@{\hskip\@xdim \copy\@dotbox\hss}%
\advance\num@segments \m@ne\advance\@xdim\x@diff\advance\@ydim\y@diff%
\repeat
\ignorespaces}}
\newif\if@flippedargs
\def\lineslope(#1,#2){%
\ifdim #1 <0pt \@xdim= -#1 \else\@xdim=#1\fi
\ifdim #2 <0pt \@ydim= -#2 \else\@ydim=#2\fi
\ifdim\@xdim >\@ydim \@tempdima=\@xdim \@xdim=\@ydim \@ydim=\@tempdima
\@flippedargstrue\else\@flippedargsfalse\fi
\ifdim\@ydim >1pt \@tempcnta=\@ydim
            \divide\@tempcnta by 65536
            \divide\@xdim \@tempcnta\fi
\ifdim\@xdim <.083333pt \@xarg=1 \@yarg=0
 \else\ifdim\@xdim <.183333pt	\@xarg=6 \@yarg=1
 \else\ifdim\@xdim <.225pt 	\@xarg=5 \@yarg=1
 \else\ifdim\@xdim <.291666pt 	\@xarg=4 \@yarg=1
 \else\ifdim\@xdim <.366666pt 	\@xarg=3 \@yarg=1
 \else\ifdim\@xdim <.45pt 	\@xarg=5 \@yarg=2
 \else\ifdim\@xdim <.55pt 	\@xarg=2 \@yarg=1
 \else\ifdim\@xdim <.633333pt 	\@xarg=5 \@yarg=3
 \else\ifdim\@xdim <.708333pt 	\@xarg=3 \@yarg=2
 \else\ifdim\@xdim <.775pt 	\@xarg=4 \@yarg=3
 \else\ifdim\@xdim <.816666pt 	\@xarg=5 \@yarg=4
 \else\ifdim\@xdim <.916666pt 	\@xarg=6 \@yarg=5
       \else			\@xarg=1 \@yarg=1%
\fi\fi\fi\fi\fi\fi\fi\fi\fi\fi\fi\fi
\if@flippedargs\relax\else\@tempcnta=\@xarg \@xarg=\@yarg
			  \@yarg=\@tempcnta\fi
\ifdim #1 <0pt \@xarg= -\@xarg\fi
\ifdim #2 <0pt \@yarg= -\@yarg\fi
}
\newif\if@toosmall
\newif\if@drawit
\newif\if@horvline
\def\drawlinestretch{0} 
\def\drawline{\@ifnextchar [{\@idrawline}{\@idrawline[\drawlinestretch]}}
\def\@idrawline[#1](#2,#3){\@ifnextchar ({\@iidrawline[#1](#2,#3)}{\relax}}
\def\@iidrawline[#1](#2,#3)(#4,#5){\@drawline[#1](#2,#3)(#4,#5)
\@idrawline[#1](#4,#5)}
\def\@drawline[#1](#2,#3)(#4,#5){{%
\x@diff=#4\unitlength \advance\x@diff by -#2\unitlength
\y@diff=#5\unitlength \advance\y@diff by -#3\unitlength
\ifx\@linefnt\tenln \linethickness{0.5pt} \else \linethickness{0.9pt}\fi
\lineslope(\x@diff,\y@diff)
\@toosmalltrue
{\ifdim\x@diff <\z@ \x@diff=-\x@diff\fi
 \ifdim\y@diff <\z@ \y@diff=-\y@diff\fi
 \ifdim\x@diff >10pt \global\@toosmallfalse\fi
 \ifdim\y@diff >10pt \global\@toosmallfalse\fi}
\@drawitfalse\@horvlinefalse
\ifnum#1 <0 \relax\else\@horvlinetrue\fi
\if@toosmall\@horvlinetrue\fi
\if@horvline
 \ifdim\x@diff =0pt \put(#2,#3){\ifdim\y@diff >0pt \@linelen=\y@diff \@upline
 				\else\@linelen=-\y@diff \@downline\fi}%
 \else\ifdim\y@diff =0pt
          \ifdim\x@diff >0pt \put(#2,#3){\vrule \@height \@halfwidth \@depth
				\@halfwidth \@width \x@diff}
		\else \put(#4,#5){\vrule \@height \@halfwidth \@depth
				\@halfwidth \@width -\x@diff}\fi
       \else\@drawittrue\fi\fi 
\else\@drawittrue\fi
\if@drawit
\ifnum\@xarg< 0 \@negargtrue\else\@negargfalse\fi
\ifnum\@xarg =0 \setbox\@linechar%
\hbox{\hskip -\@halfwidth \vrule \@width \@wholewidth \@height 10.2pt
 \@depth \z@}
\else \ifnum\@yarg =0 \setbox\@linechar%
\hbox{\vrule \@height \@halfwidth \@depth \@halfwidth \@width 10.2pt}
\else \if@negarg \@xarg -\@xarg \@yyarg -\@yarg
        \else \@yyarg \@yarg\fi
\ifnum\@yyarg >0 \@tempcnta\@yyarg \else \@tempcnta -\@yyarg\fi
\setbox\@linechar\hbox{\@linefnt\@getlinechar(\@xarg,\@yyarg)}%
\fi\fi
\if@toosmall
  \@dottedline[\picsquare]{.98\@wholewidth}%
(#2\unitlength,#3\unitlength)(#4\unitlength,#5\unitlength)%
\else
\ifnum\@xarg=0\relax\else\ifdim\x@diff >\z@ \advance\x@diff -\wd\@linechar
  \else\advance\x@diff \wd\@linechar\fi\fi
\ifnum\@yarg=0\relax\else\ifdim\y@diff >\z@\advance\y@diff -\ht\@linechar
  \else\advance\y@diff \ht\@linechar\fi\fi
\ifdim\x@diff <\z@ \@x@diff=-\x@diff \else\@x@diff=\x@diff\fi
\ifdim\y@diff <\z@ \@y@diff=-\y@diff \else\@y@diff=\y@diff\fi
\num@segments=0 \num@segmentsi=0
\ifdim\wd\@linechar >1pt
 \num@segmentsi=\@x@diff \divide\num@segmentsi \wd\@linechar\fi
\ifdim\ht\@linechar >1pt
 \num@segments=\@y@diff \divide\num@segments \ht\@linechar\fi
\ifnum\num@segmentsi >\num@segments \num@segments=\num@segmentsi\fi
\advance\num@segments \@ne 
\ifnum #1=0 \relax \else\ifnum #1 < -99
  \typeout{***drawline: reduction <= -100 percent implies blankness!***}
\else\num@segmentsi=#1 \advance\num@segmentsi by 100
     \multiply\num@segments \num@segmentsi
     \divide\num@segments by 100
     \ifnum \num@segments=0 \num@segments=1 \fi
\fi\fi
\divide\x@diff \num@segments
\divide\y@diff \num@segments
\advance\num@segments \@ne 
\@xdim=#2\unitlength \@ydim=#3\unitlength
\if@negarg \advance\@xdim -\wd\@linechar\fi
\ifnum\@yarg <0 \advance\@ydim -\ht\@linechar\fi
\@killglue
\loop \ifnum\num@segments > 0
\unskip\raise\@ydim\hbox to\z@{\hskip\@xdim \copy\@linechar\hss}%
\advance\num@segments \m@ne\advance\@xdim\x@diff\advance\@ydim\y@diff%
\repeat
\ignorespaces
\fi
\fi}}
\long\def\splittwoargs#1 #2 {(#1,#2)}
\newif\if@stillmore
\newread\@datafile
\long\def\putfile#1#2{\openin\@datafile = #1
\@stillmoretrue
\loop
\ifeof\@datafile\relax\else\read\@datafile to\@dataline\fi
\ifeof\@datafile\@stillmorefalse
\else\ifx\@dataline\@empty \relax
     \else
\expandafter\expandafter\expandafter\put\expandafter\splittwoargs%
\@dataline{#2}
     \fi
\fi
\if@stillmore
\repeat
\closein\@datafile
}
\makeatother

\makeatletter
\typeout{%
Extension to Epic and LaTeX. Version 1.0 - Release August 14, 1988}
\newcount\@gphlinewidth
\newcount\@eepictcnt
\newdimen\@tempdimc
\@gphlinewidth\@wholewidth \divide\@gphlinewidth 4736

\def\thinlines{\let\@linefnt\tenln \let\@circlefnt\tencirc
    \@wholewidth\fontdimen8\tenln \@halfwidth .5\@wholewidth
    \@gphlinewidth\@wholewidth \divide\@gphlinewidth 4736\relax}
\def\thicklines{\let\@linefnt\tenlnw \let\@circlefnt\tencircw
    \@wholewidth\fontdimen8\tenlnw \@halfwidth .5\@wholewidth
    \@gphlinewidth\@wholewidth \divide\@gphlinewidth 4736
    \advance\@gphlinewidth\@ne   
    \relax}
\newif\if@nodotdef \global\@nodotdeftrue
\def\dottedjoin{\global\@jointhemtrue \global\@joinkind=0\relax
  \bgroup\@ifnextchar[{\global\@nodotdeffalse\@idottedjoin}%
                      {\global\@nodotdeftrue\@idottedjoin[\@empty]}}
\long\def\jput(#1,#2)#3{\@killglue\raise#2\unitlength\hbox to \z@{\hskip
#1\unitlength #3\hss}%
\if@jointhem \if@firstpoint \gdef\x@one{#1} \gdef\y@one{#2} 
 \global\@firstpointfalse\else\ifcase\@joinkind
    \if@nodotdef
        \@spdottedline{\dotgap@join\unitlength}%
(\x@one\unitlength ,\y@one\unitlength)(#1\unitlength,#2\unitlength)
    \else
	\@dottedline[\dotchar@join]{\dotgap@join\unitlength}%
(\x@one\unitlength,\y@one\unitlength)(#1\unitlength,#2\unitlength)
    \fi
	\or\@dashline[\dashlinestretch]{\dashlen@join\unitlength}[\dotgap@join]%
(\x@one,\y@one)(#1,#2)
	\else\@drawline[\drawlinestretch](\x@one,\y@one)(#1,#2)\fi
    \gdef\x@one{#1}\gdef\y@one{#2}%
 \fi
\fi\ignorespaces}
\def\dottedline{\@ifnextchar [{\@idottedline}{\@ispdottedline}}
\def\@ispdottedline#1(#2,#3){\@ifnextchar (%
{\@iispdottedline{#1}(#2,#3)}{\relax}}
\def\@iispdottedline#1(#2,#3)(#4,#5){\@spdottedline{#1\unitlength}%
(#2\unitlength,#3\unitlength)(#4\unitlength,#5\unitlength)%
\@ispdottedline{#1}(#4,#5)}
\def\@spdottedline#1(#2,#3)(#4,#5){%
    \@tempcnta \@gphlinewidth\relax
    \advance\@tempcnta by 2     
    \special{pn \the\@tempcnta}%
    \@tempdimc=#2\relax
    \@tempcnta \@tempdimc\relax \advance\@tempcnta 2368 \divide\@tempcnta 4736
    \@tempdimc=#3\relax
    \@tempcntb -\@tempdimc\relax\advance\@tempcntb -2368 \divide\@tempcntb 4736
    \@paspecial{\the\@tempcnta}{\the\@tempcntb}%
    \@tempdimc=#4\relax
    \@tempcnta \@tempdimc\relax \advance\@tempcnta 2368 \divide\@tempcnta 4736
    \@tempdimc=#5\relax
    \@tempcntb -\@tempdimc\relax\advance\@tempcntb -2368 \divide\@tempcntb 4736
    \@paspecial{\the\@tempcnta}{\the\@tempcntb}%
    \@tempdimc=#1\relax
    \@tempcnta \@tempdimc\relax \advance\@tempcnta 2368 \divide\@tempcnta 4736
    \@tempcntb \@tempcnta\relax \divide\@tempcntb 1000
    \multiply \@tempcntb 1000 \advance\@tempcnta -\@tempcntb
    \divide\@tempcntb 1000
    \ifnum\@tempcnta < 10
        \special{dt \the\@tempcntb.00\the\@tempcnta}%
    \else\ifnum\@tempcnta < 100
        \special{dt \the\@tempcntb.0\the\@tempcnta}%
    \else
        \special{dt \the\@tempcntb.\the\@tempcnta}%
    \fi\fi
    \ignorespaces
}
\def\@iiidashline[#1]#2[#3](#4,#5)(#6,#7){%
\@dashline[#1]{#2\unitlength}[#3](#4,#5)(#6,#7)%
\@iidashline[#1]{#2}[#3](#6,#7)}
\long\def\@dashline[#1]#2[#3](#4,#5)(#6,#7){{%
\x@diff=#6\unitlength \advance\x@diff by -#4\unitlength
\y@diff=#7\unitlength \advance\y@diff by -#5\unitlength
\@tempdima=#2\relax \advance\@tempdima -\@wholewidth
\sqrtandstuff{\x@diff}{\y@diff}{\@tempdima}%
\ifnum\num@segments <3 \num@segments=3\fi
\@tempdima=\x@diff \@tempdimb=\y@diff
\divide\@tempdimb by\num@segments
\divide\@tempdima by\num@segments
{\ifx#3\@empty \relax
    \ifdim\@tempdima < 0pt \x@diff=-\@tempdima\else\x@diff=\@tempdima\fi
    \ifdim\@tempdimb < 0pt \y@diff=-\@tempdimb\else\y@diff=\@tempdimb\fi
    \global\setbox\@dotbox\hbox{%
                \@absspdrawline(0pt,0pt)(\@tempdima,\@tempdimb)}%
    \else\global\setbox\@dotbox\hbox{%
        \@spdottedline{#3\unitlength}(0pt,0pt)(\@tempdima,\@tempdimb)}%
    \fi}%
\advance\x@diff by -\@tempdima 
\advance\y@diff by -\@tempdimb
\@tempdima=\num@segments\@wholewidth \@tempdima=2\@tempdima
\@tempcnta\@tempdima\relax \@tempdima=#2\relax \@tempdimb=0.5\@tempdima
\@tempcntb\@tempdimb\relax \advance\@tempcnta by \@tempcntb 
\divide\@tempcnta by\@tempdima \advance\num@segments by -\@tempcnta
\ifnum #1=0 \relax\else\ifnum #1 < -100
  \typeout{***dashline: reduction > -100 percent implies blankness!***}
\else\num@segmentsi=#1 \advance\num@segmentsi by 100
     \multiply\num@segments by\num@segmentsi \divide\num@segments by 100
\fi\fi
\divide\num@segments by 2 
\ifnum\num@segments >0 
 \divide\x@diff by\num@segments
 \divide\y@diff by\num@segments
 \advance\num@segments by\@ne 
 \else\num@segments=2 
\fi
\@xdim=#4\unitlength \@ydim=#5\unitlength
\@killglue
\loop \ifnum\num@segments > 0
\unskip\raise\@ydim\hbox to\z@{\hskip\@xdim \copy\@dotbox\hss}%
\advance\num@segments \m@ne\advance\@xdim\x@diff\advance\@ydim\y@diff%
\repeat}%
\ignorespaces}
\def\@drawline[#1](#2,#3)(#4,#5){{%
\@drawitfalse\@horvlinefalse
\ifnum#1 <0 \relax\else\@horvlinetrue\fi
\if@horvline
 \@spdrawline(#2,#3)(#4,#5)
\else\@drawittrue\fi
\if@drawit
\ifnum #1=0 \relax \else\ifnum #1 < -99
  \typeout{***drawline: reduction <= -100 percent implies blankness!***}%
\else\num@segmentsi=#1 \advance\num@segmentsi by 50
     \multiply\num@segmentsi 2
\fi\fi
\@dashline[\num@segmentsi]{10pt}[\@empty](#2,#3)(#4,#5)
\fi}\ignorespaces}
\def\@spdrawline(#1,#2)(#3,#4){%
   \@absspdrawline(#1\unitlength,#2\unitlength)(#3\unitlength,#4\unitlength)
   \ignorespaces
}
\def\@absspdrawline(#1,#2)(#3,#4){%
    \special{pn \the\@gphlinewidth}%
    \@tempdimc=#1\relax
    \@tempcnta \@tempdimc\relax \advance\@tempcnta 2368 \divide\@tempcnta 4736
    \@tempdimc=#2\relax
    \@tempcntb -\@tempdimc\relax\advance\@tempcntb-2368 \divide\@tempcntb 4736
    \@paspecial{\the\@tempcnta}{\the\@tempcntb}%
    \@tempdimc=#3\relax
    \@tempcnta\@tempdimc\relax \advance\@tempcnta 2368 \divide\@tempcnta 4736
    \@tempdimc=#4\relax
    \@tempcntb -\@tempdimc\relax\advance\@tempcntb-2368 \divide\@tempcntb 4736
    \@paspecial{\the\@tempcnta}{\the\@tempcntb}%
    \special{fp}%
    \ignorespaces
}
\def\@paspecial#1#2{%
    \special{pa #1 #2}%
}
\def\Thicklines{\let\@linefnt\tenlnw \let\@circlefnt\tencircw
    \@wholewidth\fontdimen8\tenlnw \@wholewidth 1.5\@wholewidth
    \@halfwidth .5\@wholewidth
    \@gphlinewidth\@wholewidth \divide\@gphlinewidth 4736\relax}
\def\@circlespecial#1#2#3#4{%
	      \special{pn \the\@gphlinewidth}%
	      \special{ar 0 0 #1 #2 #3 #4}
}
\def\@arc#1#2#3#4{%
	\@tempdima #1\unitlength
	\@tempdimb #2\unitlength
        \@tempcnta\@tempdima \advance\@tempcnta 4736 \divide\@tempcnta 9473
	\@tempcntb\@tempdimb \advance\@tempcntb 4736 \divide\@tempcntb 9473
	\setbox\@tempboxa\hbox{%
	    \@circlespecial{\the\@tempcnta}{\the\@tempcntb}{#3}{#4}}%
        \wd\@tempboxa\z@ \box\@tempboxa}
\def\circle{%
    \@ifstar{\special{bk}\@circle}{\@circle}}
\def\@circle#1{\@arc{#1}{#1}{0}{6.2832}}
\def\ellipse{%
    \@ifstar{\special{bk}\@ellipse}{\@ellipse}}
\def\@ellipse#1#2{{\@arc{#1}{#2}{0}{6.2832}}}
\def\arc#1#2#3{\@arc{#1}{#1}{#2}{#3}}
\def\@linespecial#1#2{%
	      \special{pn \the\@gphlinewidth}%
	      \special{pa 0 0}%
	      \special{pa #1 #2}%
	      \special{fp}%
}
\def\@sline{%
	\ifnum\@xarg< 0
	  \@negargtrue \@xarg -\@xarg \@tempdima -\@linelen
	\else
	  \@negargfalse \@tempdima\@linelen
	\fi
	\@tempcnta\@linelen \divide\@tempcnta 4736
        \@yyarg -\@yarg \multiply\@yyarg \@tempcnta \divide\@yyarg\@xarg
 	\if@negarg
	    \@tempcnta -\@tempcnta
	\fi
	\setbox\@linechar\hbox{\@linespecial{\the\@tempcnta}{\the\@yyarg}}%
	\wd\@linechar\@tempdima
	\@clnht\@linelen
        \multiply\@clnht\@yarg
        \divide\@clnht\@xarg
	\ifnum\@yarg< 0
	  \@clnht -\@clnht
	  \ht\@linechar\z@ \dp\@linechar\@clnht
	\else
	  \ht\@linechar\@clnht \dp\@linechar\z@
	\fi
	\box\@linechar
	\if@negarg
	  \@yyarg -\@yarg
	\else
	  \@yyarg \@yarg
	\fi
	\setbox\@linechar\hbox{\@linefnt\@getlinechar(\@xarg,\@yyarg)}%
	\ifnum\@yarg> 0
	  \let\@upordown\raise
	  \advance\@clnht -\ht\@linechar
	\else
	  \let\@upordown\lower
	\fi
	\if@negarg \kern\wd\@linechar \fi
}
\def\spline(#1,#2){%
    \special{pn \the\@gphlinewidth}%
    \@spline(#1,#2)%
}
\def\@spline(#1,#2){%
    \@tempdima #1\unitlength
    \@tempdimb #2\unitlength
    \@tempcnta \@tempdima \advance\@tempcnta 2368 \divide\@tempcnta 4736
    \@tempcntb -\@tempdimb \advance\@tempcntb -2368 \divide\@tempcntb 4736
    \@paspecial{\the\@tempcnta}{\the\@tempcntb}%
    \@ifnextchar ({\@spline}{\special{sp}}%
}
\def\path(#1,#2){%
    \special{pn \the\@gphlinewidth}%
    \@path(#1,#2)%
}
\def\@path(#1,#2){%
    \@tempdima #1\unitlength
    \@tempdimb #2\unitlength
    \@tempcnta \@tempdima \advance\@tempcnta 2368 \divide\@tempcnta 4736
    \@tempcntb -\@tempdimb \advance\@tempcntb -2368 \divide\@tempcntb 4736
    \@paspecial{\the\@tempcnta}{\the\@tempcntb}%
    \@ifnextchar ({\@path}{\special{fp}}%
}

\newdimen\maxovaldiam \maxovaldiam 40pt\relax

\def\@oval(#1,#2)[#3]{\begingroup\boxmaxdepth \maxdimen
  \@ovttrue \@ovbtrue \@ovltrue \@ovrtrue
  \@tfor\@tempa :=#3\do{\csname @ov\@tempa false\endcsname}\@ovxx
  #1\unitlength \@ovyy #2\unitlength
  \@tempdimb \ifdim \@ovyy >\@ovxx \@ovxx\else \@ovyy \fi
  \@ovro \ifdim \@tempdimb>\maxovaldiam \maxovaldiam\else\@tempdimb\fi\relax
  \divide \@ovro \tw@
  \@ovdx\@ovxx \divide\@ovdx \tw@
  \@ovdy\@ovyy \divide\@ovdy \tw@
  \setbox\@tempboxa \hbox{%
  \if@ovr \@ovverta\fi
  \if@ovl \kern \@ovxx \@ovvertb\kern -\@ovxx \fi
  \if@ovt \@ovhorz \kern -\@ovxx \fi
  \if@ovb \raise \@ovyy \@ovhorz \fi}
  \ht\@tempboxa\z@ \dp\@tempboxa\z@
  \@put{-\@ovdx}{-\@ovdy}{\box\@tempboxa}%
  \endgroup}

\def\@qcirc#1#2#3#4{%
    \special{pn \the\@gphlinewidth}%
    \@eepictcnt \@gphlinewidth \divide\@eepictcnt 2
    \@tempcnta #1 
      \advance\@tempcnta 2368 \divide\@tempcnta 4736
      \advance\@tempcnta\@eepictcnt
    \@tempcntb #2 \divide\@tempcntb 4736 \advance\@tempcntb 2
    \hbox{%
    \@qcircspecial{\the\@tempcnta}{-\the\@eepictcnt}{\the\@tempcntb}{#3}{#4}}%
}
\def\@qcircspecial#1#2#3#4#5{\special{ar #1 #2 #3 #3 #4 #5}}

\def\@ovverta{\vbox to \@ovyy{%
    \if@ovb
        \kern \@ovro
        \@qcirc{\@ovro}{\@ovro}{3.1416}{4.7124}\nointerlineskip
    \else
        \kern \@ovdy
    \fi
    \leaders\vrule width \@wholewidth\vfil \nointerlineskip
    \if@ovt
        \@qcirc{\@ovro}{\@ovro}{1.5708}{3.1416}\nointerlineskip
        \kern \@ovro
    \else
        \kern \@ovdy
    \fi
}\kern -\@wholewidth}

\def\@ovvertb{\vbox to \@ovyy{%
    \if@ovb
        \kern \@ovro
        \@qcirc{-\@ovro}{\@ovro}{4.6124}{6.2832}\nointerlineskip
    \else
        \kern \@ovdy
    \fi
    \leaders\vrule width \@wholewidth\vfil \nointerlineskip
    \if@ovt
        \@qcirc{-\@ovro}{\@ovro}{0}{1.6708}\nointerlineskip
        \kern \@ovro
    \else
        \kern \@ovdy
    \fi
}\kern -\@wholewidth}

\def\@ovhorz{\hbox to \@ovxx{%
    \if@ovr \kern \@ovro\else \kern \@ovdx \fi
    \leaders \hrule height \@wholewidth \hfil
    \if@ovl \kern \@ovro\else \kern \@ovdx \fi
    }}

\def\allinethickness#1{\let\@linefnt\tenlnw \let\@circlefnt\tencircw
    \@wholewidth #1 \@halfwidth .5\@wholewidth
    \@gphlinewidth\@wholewidth \divide\@gphlinewidth 4736\relax}
\makeatother

\begin{document}

\begin{titlepage}

 \begin{center}
  { \large  
TOPICS IN 2D INTEGRABLE  FIELD THEORIES WITH BOUNDARY
INTERACTIONS\footnote{Pedagogical review based on the PhD dissertation}}
\end{center}
 \vspace{2mm}

 \begin{center}Sergei Skorik \end{center}

 \begin{center}
{\it Physics Department,
University of Southern California, Los Angeles, CA 90089-0484}\end{center}

\vspace{20mm}

\begin{center} {\bf Abstract} \end{center}

We study different aspects of  integrable  boundary quantum field  
theories, focussing   mostly on the ``boundary  sine-Gordon model''
and its applications to condensed matter physics.

The first part of the review deals with formal problems. We analyze  
the classical limit and  perform semi-classical quantization. We show 
that 
the non-relativistic limit  corresponds to the Calogero-Moser
model with a boundary potential.  We construct a  lattice  
regularization of the problem via the XXZ chain. We  classify 
boundary bound states. We generalize the Destri de Vega method 
to compute the ground state energy of the theory  on a finite  
interval.

The second part deals with  some applications to 
condensed matter physics. 
We show how to compute analytically time and space dependent correlations in
one-dimensional quantum integrable systems with an impurity. Our approach is
based on a description of these systems in terms of massless scattering 
of quasiparticles. 
Correlators follow then from matrix elements of local
operators between multiparticle states -- the  massless form-factors.
Although, in general an infinite sum of these form-factors has to be considered,
we find that for the current, spin and energy operators only a few 
(two or three) are necessary to obtain an accuracy of more than 1\%. 
Our results hold for {\tt arbitrary impurity strength}, in contrary
to the perturbative expansions in the coupling constants. As an example,
we compute the frequency dependent condunctance, 
at zero temperature, in a Luttinger liquid
with an impurity, and also discuss the succeptibility in the Kondo model
and the time-dependent properties of the two-state problem with dissipation.

\vspace{2.7cm}

\noindent May 1996

\end{titlepage}

\begin{center} {\bf \Large Acknowledgements} \end{center}

\vspace{1cm}

   It is my pleasure to thank Hubert Saleur for his guidance, 
 understanding and 
the financial support. I am indebted to Hubert for much of my new skills
and acquisitions.
 I would like to thank also Nick Warner for sharing his
enthusiasm towards theoretical physics, Itzhak Bars for teaching 
an excellent class on Quantum Field Theory and K. Pilch for teaching 
a remarkable class on Quantum Mechanics at USC. 

I want to thank sincerely Anton Kapustin, Sergei Cherkis and Yuri Levin
 for an interesting collaboration and ``live discussion''
that sporadically
 took place in the kitchen around midnight. All three were graduate
students at Caltech.

Finally, I want to thank those few friends, whose warm company, good
jokes and optimism allowed me to survive four years and conduct this
research in the Wild West.
 These are Sergei Cherkis, Jonathan Cohen, Yuri Levin, Alexei
Shalopenok, Vyacheslav Solomatov, J\"{u}rgen Schulze, Naresh Talsania...




    \tableofcontents


\newpage

\topskip=3cm

\noindent{\Large \bf Chapter 0}

\vspace{1cm}

\noindent{\Large \bf Introduction}

\addcontentsline{toc}{chapter}{\protect\numberline{0}{Introduction}}

\vspace{2cm}

In one space dimension, a quantum field theory 
can be defined either on a circle,
or on an open interval with certain boundary conditions. In Hamiltonian
formulation, boundary conditions amount to the presence of additional
interaction term on the boundary \cite{GZ},
$$H=H_{bulk}+V(\{\phi_B\}),$$
where $\{\phi_B\}$ denotes the set of boundary degrees of freedom --
fluctuating fields at the boundary.
The reasons for studying the theories with boundary conditions
seem natural since in practice one has to deal often with a bounded
system having some interface with the external world, as in the problem
of polymer adsorption
\cite{polymer}. \footnote{E.g., for the Ising
model such an interface can be modeled by a boundary magnetic field 
\cite{McCoy}.            }
Also, the problems on the semi-infinite interval in 1D sometimes
appear as the reductions of 3D problems to the $s$-wave dependence as
e.g. in the case of the monopole-catalysed baryon decay 
\cite{monopol}.  
\topskip=0cm

One of the most appreciated and rich at the present
applications of the boundary 
field theory 
is to the impurity problems of condensed matter physics
\cite{AL,FLUDS,LSS,CMP}, provided that the scattering
on impurity can be mapped onto the scattering off the boundary with
some boundary potential. An incomplete list of examples includes 
the Kondo model \cite{AFL:Rev}, the
dissipative quantum mechanics \cite{dissip_qm} and  the quantum Hall
liquids with constriction \cite{moon}. All of the mentioned above
three models look alike from the point of view of the bulk part,
which is a free massless boson, but differ by the boundary interaction.
The role of integrability here is two-folded. First, it
allows to find the static properties (e.g. via the Bethe ansatz)
of the model from the bare Hamiltonian, including the mass spectrum, 
scattering matrices, or
the free energy. Second, it allows to take advantage of 
 working with the physical excitations in the formalism of the FS theory.
Namely, integrability suggests
a convenient basis of massless particle states
which are particular combinations of plane waves that scatter diagonally
off the boundary. \footnote{
Corresponding classical solutions are presented in
\cite{FSW}.} Working with these massless particles at first sight 
adds some complexity, but it is paid off by the final simple and manageable
results.
Let us stress that we deal only with such boundary interactions
that preserve the integrability of the bulk part. Fortunately, there
are quite a few of such models in the real world. \footnote{
The term ``integrable models'' in our context amounts to the model
having an infinite number of conserved charges in involution
(i.e. mutually commuting), both on quantum and classical levels. The integrability results to the factorized scattering property
by making use of the argument that S-matrix must commute with the infinite
number of charges \cite{ZZ}.}

Different integrable boundary theories have been 
introduced and solved throughout
the history of integrable models. A noticable contribution in this
field has been made by Cardy \cite{Cardy:BCFT,CardyLew}, who studied
the critical surface behavior withing the framework of conformal field theory.
Another substantial advancement  has been done by the work of Ghoshal and Zamolodchikov \cite{GZ}.
The authors have succeeded in formulating and (partially) solving the set 
of general constraints for the factorizable boundary field theory,
thus generating a powerful method of obtaining the integrable boundary
models from the integrable bulk field theories. The equations of
\cite{GZ} resolve the boundary integrability (boundary Yang-Baxter
equation) 
together with the physical constarints of 
unitarity and crossing symmetry and possess the restrictive power to
determine the scattering matrices up to ``CDD ambiguity''.
In a more simple words, given 
a FS theory in the bulk with its two-particle scattering matrices, 
we can derive all possible integrable boundary models compatible  
with this bulk theory. Of course, what we get is merely the FS description,
leaving the identification of the Hamiltonian structure to our intuition. 
But other methods, developed in \cite{AlZamo} and \cite{LSS}, allow 
us to obtain the exact free energy and the correlation functions
correspondingly. Thus, the FS approach turns out to be very productive.

The paper \cite{GZ} in such a way sets basis for a more systematic
study of the boundary integrable models, marking them out as a subfield
of integrable models. We undertake in this dissertation a close study of
some particularly interesting boundary models, such 
as the boundary sine-Gordon model (chapters 2,3,4,6), 
as well as develop some general techniques (chapters 5,7). We assume that all  
the properties of the bulk models of interest are known to us,
so we can focuse on the peculiar boundary phenomena. One example of
the boundary phenomena are the boundary bound states (chapter 4).
The existence of such states, localized at the boundary of
the crystal lattice, was first pointed
out by I.E.Tamm.  
Finally, in chapter 7 we describe in some detail the physics behind
the impurity and dissipative two-state models and set up the technique 
for calculating the physical observables (correlation functions) in
these models by making heavy use of the boundary factorized scattering.
Quite remarkable, such characteristics as the conductance can be
expressed in terms of the boundary scattering matrices!

                     In chapter 1  
we review the Bethe ansatz technology.
We start with the traditional Hamiltonian
approach and show  how to extract physical observables
(mass spectrum, scattering matrices etc) from a bare Hamiltonian.
Although we do not present any new contributions with respect to
the existing extensive literature on the subject,
this material is
necessary to understand  further chapters. 
We focuse our discussion on the Thirring model, which is the fermionic
analog of the sine-Gordon model.

                     In chapter 2
we consider the sine-Gordon model on a half-line, with
an additional potential term of the form $-M\cos{\beta\over
2}(\varphi-\varphi_0)$ at the boundary. 
We construct the classical solutions  by using
the
bulk sine-Gordon theory and the ``generalized method of images.''  
From the classical solutions in hand we extract the time delay 
(\ref{phasedel}).
From the time delay we  reconstruct
the semi-classcial phase-shift using the method of Jackiw and Woo
\cite{JW}. We establish the agreement with the semi-classical limit $\beta\to 0$
of the exact boundary reflection matrix, (\ref{semiI}).
The exact expressions for
the boundary reflection matrices  are known up to CDD ambiguities \cite{GZ}.
They were obtained as a ``minimal'' solution to the general set of constraints
for the integrable boundary field theory.

The purpose of 
                           chapter 3 
is to investigate the non-relativistic, $\theta\to 0$,
limit of the boundary sine-Gordon model and 
to determine the quantum-mechanical potential
induced by the presence of boundary.
 We show that the generalized Calogero-Moser model with boundary potential
of the P\"oschl-Teller type describes the non-relativistic
limit in question.

                          In chapter 4 
we address the {\it exact} quantum field theory solution of the 
boundary sine-Gordon model,
which we obtain by means of the Bethe ansatz technique. Among other things,
this solution allows to re-derive the boundary reflection matrices of \cite{GZ},
(\ref{reflamp})-(\ref{EqnII}),
and to relate them to the physical parameters in the Hamiltonian \cite{FS}.
The present chapter includes a complete study of boundary bound states
and related boundary S-matrices for the sine-Gordon model with Dirichlet
boundary condition. Our analysis is based on the solution of the
boundary bootstrap equations, representing the integrability constraints,
together with the explicit Bethe ansatz solution of the inhomogeneous XXZ model
in a boundary magnetic field -- a
lattice regularization of the boundary sine-Gordon model.
 We identify boundary bound states
with new {\it boundary strings} in the Bethe ansatz.

The main  purpose of 
                           chapter 5 
is to study the ground state energy of 1+1
integrable relativistic quantum field theories with boundaries. This involves
several questions. One is the energy associated with a boundary for an
infinite system, the other is the way the energy of the theory on an interval
varies with its length - the ``genuine'' Casimir effect.
The elegant method of Destri and de Vega \cite{DdVg} for the periodic
systems leads directly to the expression for the ground state energy
from which the infinite size contribution and the finite size
correction can be easily extracted. The heart of the DDV method is
a non-linear integral equation (\ref{renormbethe}) being derived 
from the Bethe equations.
We generalize the Destri-de-Vega method to the systems with boundaries
and apply it to compute the ground state energy for the boundary
sine-Gordon model.

                              In chapter 6
We study an open $XXZ$ 
 chain in the regime $\Delta>1$ with a boundary magnetic field $h$
and discuss some of its peculiar features due to the presence of boundary.
In the Bethe ansatz formalism, boundary bound states are represented by
the ``boundary strings'' as described in chapter 4. We find
that for certain values of $h$
 the ground state wave function contains
 boundary strings, and from this infer the existence of two ``critical''
fields in agreement with \cite{miwa}.
 An expression
for the vacuum surface energy in the thermodynamic limit is derived and found
to
be an analytic function of $h$.
We argue that boundary excitations appear only in pairs
with ``bulk'' excitations or with boundary excitations at the other end of the
chain. The case where
the magnetic fields at the left and the right boundaries are antiparallel
has non-trivial differences with the case of the parallel fields.
The Ising ($\Delta=\infty$) and isotropic ($\Delta=1$)
limits are discussed thoroughly and found helpful for the intuitive 
understanding
of the behavior of the boundary $XXZ$ chain at arbitrary $\Delta$. 

                            In chapter 7
we show how to compute analytically time and space dependent correlations in
one-dimensional quantum integrable systems with an impurity. Our approach is
based on a description of these systems in terms of massless scattering 
of quasiparticles \cite{FS:less}. 
Correlators follow then from matrix elements of local
operators between multiparticle states -- the  massless form-factors.
Although, in general an infinite sum of these form-factors has to be considered,
we find that for the current, spin and energy operators only a few 
(two or three) are necessary to obtain an accuracy of more than 1\%. 
Our results hold for {\tt arbitrary impurity strength}, in contrary
to the perturbative expansions in the coupling constants. As an example,
we compute the frequency dependent condunctance, 
at zero temperature, in a Luttinger liquid
with an impurity, and also discuss the succeptibility in the Kondo model
and the time-dependent properties of the two-state problem with dissipation.

 This review is based mostly 
 on the published papers \cite{SSW}, \cite{KapSk}, 
 \cite{Boot:SS}, \cite{LMSS}, \cite{XXZ:KS}, \cite{LSS}.

\chapter{Introduction to the Bethe Ansatz}
 
One of the efficient approaches to quantization of  interacting fields
is based on the conformal field theory  (CFT), while yet another on
the factorized scattering theory (FST).
In both cases the knowledge of the Hamiltonian is not required.
Rather, the data is encoded into the particular representation 
of the Virasoro algebra on the space of fields and the dimension of
the perturbing operator, or in the exact scattering matrices as in the
case of FST. We start, however, with the traditional Hamiltonian
approach and show in this chapter how to extract physical observables
(mass spectrum, scattering matrices etc) from a bare Hamiltonian
using the Bethe ansatz technology.
Although we do not present any new contributions with respect to
the existing extensive literature on the subject 
\cite{Thack:Rev,AFL:Rev}, 
this material is
necessary to understand further chapters. 

This chapter is by no means
the complete review of the Bethe ansatz.
We focuse our discussion on the Thirring model, which is the fermionic
analog of the sine-Gordon model. The relation between the two
models becomes an exact mapping of one onto another due to the bosonization
technique of Coleman and Mandelstam. 

\section{Hamiltonian formulation}

The massive Thirring model is defined by the Hamiltonian
\be
H_T=\int dx [- i (\psi_1^+\partial_x\psi_1-\psi_2^+\partial_x\psi_2)
+m_0(\psi_1^+\psi_2+\psi_2^+\psi_1) + 2g\psi_1^+\psi_2^+\psi_2\psi_1].
\label{Ham:Thir}
\ee
It can be rewritten in terms of the creation and annihilation operators
of the Fock fermionic space
and diagonalized \cite{BThack}. The vacuum of the Fock space $|0\rangle$
is a particular eigenstate
 annihilated by $\psi_1$ and $\psi_2$ and is called {\it bare} vacuum.
The {\it physical} vacuum is the state with the Dirac sea filled which has an
infinite negative energy. Thus, to perform calculations a high-energy
cutoff is required.
Physical excitations are obtained by removing
pseudoparticles from the Dirac sea and placing them above it,
allowing in general some complex combinations called bound states.

The model possesses the conserved charge $N=\int dx (\psi_1^+\psi_1 + 
\psi_2^+\psi_2)$ (total number of pseudoparticles), as well as an infinite
family of other local commuting charges,
 and therefore it is integrable. The latter can be shown by passing
to the discrete Thirring model \cite{Lush}, 
which is known as lattice XYZ chain.\footnote{There exist many lattice
models that have Thirring model as the continuum limit. XYZ chain is, however,
the most studied one. See \cite{Destr_deVeg} for another example.} 
The presence of conserved charges puts tremendous constraints on the
dynamics of the model and, in fact, makes it analagous 
 to the free Dirac fermions.
 The non-trivial difference
comes from the mutual pairwise interaction of pseudoparticles in the vacuum.
 Loosely
speaking, the Dirac sea is sensitive to the removal of one of its pseudoparticles, i.e. a {\it polarization} of vacuum occurs.
In yet another sense the properly regularized Thirring model
is analogous to the quantum-mechanical N-body problem with the pairwise
interaction potential $V\sim\delta(x-y)$.

N-particle bare wave-functions are constructed by glueing up
free solutions to the Dirac equation 
at the boundaries of the domains $x_{i_1}<x_{i_2}<\dots<x_{i_N}$, similar
to the way one proceeds in the quantum-mechanical problem with
 the delta-function potential. 
The solution to the free Dirac equation in two dimensions can be written in
the form
\be
 \vec{\Psi} = \vec{u}(\beta)e^{ipx-iEt}, \label{free:sol}
\ee
where
$$ \vec{u}={1\over \sqrt{2}}\left(\begin{array}{l}e^{-\beta/2} \\ e^{\beta/2}
\end{array}\right), \qquad E^2-p^2=m_0^2. $$
it is convenient to parametrize the energy and momentum in terms of rapidity
$\beta$:
\be
E=m_0\cosh\beta, \qquad p=m_0\sinh\beta. \label{def:rapid}
\ee
Matching solutions (\ref{free:sol}) leads to the pairwise phase shifts
in the wave-function
\be
\Phi(\beta)=- i \log{\sinh{1\over 2}(2 i \mu -\beta) \over
\sinh{1\over 2}(2 i \mu +\beta)} , \qquad \cot\mu=-{1\over 2}g, \label{Phase}
\ee
arising at the hyperplanes $x_i=x_j$,
with the {\it bare} S-matrix being 
\be
S=e^{i\Phi}, \qquad S(0)=1. \label{eq:baresmatr}
\ee
The domain $\mu<\pi/2$ is called a {\it repulsive regime}, while the domain
$\mu>\pi/2$ is called an {\it attractive regime}. The value $\mu=\pi/2$
is the {\it free point}.

\section{Bethe equations, thermodynamic limit and \\ 
solution for the density
of states}

The above
heuristic discussion can be made more rigorous. Let us consider first the model
(\ref{Ham:Thir}) compactified on the circle of length $L$
and then take the limit
$L\to\infty$. 
The wave-functions of the states\footnote{
Notice the peculiar to integrable models factorized structure of the N-particle
states: the wave-function consists of the product of two-particle terms.} 
\begin{eqnarray}
|\beta_1, \beta_2, \dots, \beta_N\rangle = \int\prod_{i=1}^N e^{ip(\beta_i)x_i}
dx_i\prod_{i<j\leq N}&&[1-i\lambda(\beta_i-\beta_j)\theta(x_i-x_j)]\cdot \\
&&\cdot A^+(\beta_1,x_1)\cdots A^+(\beta_N,x_N)|0\rangle, 
\nonumber\label{Thack.wf}
\end{eqnarray}
where $\theta(x)$ is the step function and
$$ A^+(\beta,x)=(2\cos\beta)^{-1/2}[e^{\beta/2}\psi_1^+(x)+e^{-\beta/2}
\psi_2^+(x)] $$
are subject to the periodic boundary conditions, called Bethe equations:
\be
\prod_{j\neq i}[1+i\lambda(\beta_i-\beta_j)]=e^{ip(\beta_i)L}
\prod_{j\neq i}[1-i\lambda(\beta_i-\beta_j)]. \label{eq:Bethe}
\ee
Taking the logarithm of Eq.(\ref{eq:Bethe}) we get
\be
-p(\beta_i)L =\sum_j \Phi(\beta_i-\beta_j)+2\pi I_i . \label{eq:bethe}
\ee
Such equations should be written for each particle's index $i$ and form together
a complex set of coupled transcendental equations. Different solutions are
obtained for different sets of integers $I_i$. For example, the Dirac sea
corresponds to the dense set of integers
from $-I$ to $I$, while excited states correspond
to the sets with $\nu$ of $I_j$'s missing. Such vacancies will be called 
{\it holes}.
Thermodymanic limit is the limit $L\to\infty$, $N\to\infty$.
Class of states for which $\nu$ is fixed and finite when $N\to\infty$
is called {\it scaling states} \cite{FadT}.

Besides the real solutions to (\ref{eq:bethe})
and the Dirac sea solutions with Im$\beta=\pi$, there exist various 
other solutions
with some of $\beta_i$ complex. In the thermodynamic limit it is possible
to show \cite{TakahSuz} that all permissible complex roots form strings.
From the point of view of the wave function, strings are bound states.
They contain the rapidities with common real part Re$\beta$ and equally spaced
imaginary parts and are located symmetrically with respect to Im$\beta=0$
or Im$\beta=\pi$ lines.
(see Figure 1.1). 
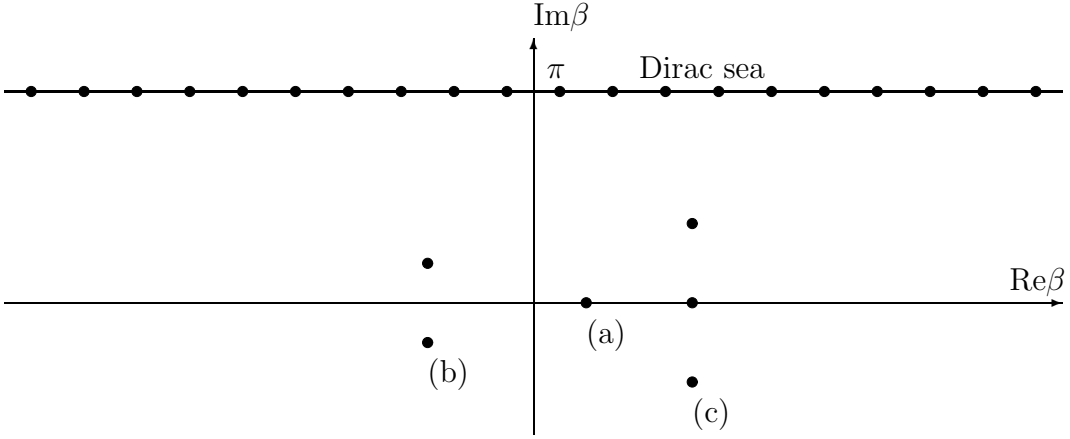
\begin{figure}
\begin{center}
\begin{picture}(400, 220)
\put(0,100){\vector(1,0){400}}
\put(200,50){\vector(0,1){150}}
\put(0,180){\line(1,0){400}}
\put(220, 100){\circle*{4}}
\put(160, 115){\circle*{4}}
\put(160, 85){\circle*{4}}
\put(260, 100){\circle*{4}}
\put(260, 130){\circle*{4}}
\put(260, 70){\circle*{4}}
\put(380,105){Re$\beta$}
\put(200,205){Im$\beta$}
\put(205, 185){$\pi$}
\put(240, 185){Dirac sea}
\put(220, 85){(a)}
\put(260, 55){(c)}
\put(160, 70){(b)}
\put(10, 180){\circle*{4}}
\put(30, 180){\circle*{4}}
\put(50, 180){\circle*{4}}
\put(70, 180){\circle*{4}}
\put(90, 180){\circle*{4}}
\put(110, 180){\circle*{4}}
\put(130, 180){\circle*{4}}
\put(150, 180){\circle*{4}}
\put(170, 180){\circle*{4}}
\put(190, 180){\circle*{4}}
\put(210, 180){\circle*{4}}
\put(230, 180){\circle*{4}}
\put(250, 180){\circle*{4}}
\put(270, 180){\circle*{4}}
\put(290, 180){\circle*{4}}
\put(310, 180){\circle*{4}}
\put(330, 180){\circle*{4}}
\put(350, 180){\circle*{4}}
\put(370, 180){\circle*{4}}
\put(390, 180){\circle*{4}}
\end{picture}
\end{center}
\caption{Various root configurations: (a) 1-string; (b)
2-string; (c) 3-string.}
\end{figure}
The easiest way to determine the spacing
along imaginary axis is to look at the poles of the 
bare S-matrix (\ref{eq:baresmatr}). The latter are given by $\Delta\beta=
2\pi il-2i\mu$. Since by periodicity $-\pi<$Im$\beta<\pi$, it is enough to
consider $l=0,1$. We find that $\Delta\beta=-2i\mu$ in the repulsive regime,
and $\Delta\beta=-2i\omega$ in the attractive regime, where 
$$\omega\equiv \pi-\mu.$$ More rigorous classification of strings
based on the analysis of Bethe equations is given in
\cite{TakahSuz}.

Denote by 
\be
\Phi_{m,n} = \sum_{p=0}^{m-1}\sum_{l=0}^{n-1}\Phi(\beta+i\mu(m-2p-n+2l))
\label{eq:ij_ph}
\ee
the scattering phase of m-string on n-string, and by 
\be
\Phi_{-1,n} = \sum_{p=0}^{n-1}\Phi(\beta+i\pi+i\mu(n-1-2p))=\Phi_{1,n}(\beta+i\pi)
\label{eq:j_ph}
\ee
the scattering phase of the Dirac sea pseudoparticle on the n-string. 
For technical simplicity we restrict $\mu$ and $\omega$ to certain ``rational'' 
values equal to
$\pi/t$, where $t=2,3,\dots$. Then only the strings of the length
$1,2,\dots,t-1$ exist. \footnote{At first glance it seems that the t-string
should be allowed, too. However, it is easy to check that it has vanishing
energy, momentum and scattering phase. So, it is a ``ghost.''}
Note that $(t-1)$-string is qualitatively different from all the shorter
strings and is similar to a hole in the Dirac sea as far as the bare energy,
momentum and scattering phase with other objects are concerned.

In the thermodynamic limit the distribution of roots of Bethe equations
is described by continuous positive functions $\rho_j(\beta)$ and 
$\tilde{\rho}_j(\beta)$, where $\rho_j(\beta)$ is a density of j-strings, 
$\rho_{-1}$ is a density of the Dirac sea pseudoparticles, and 
$\tilde{\rho}_j$ is a density of holes. Summing the Bethe equations
and rearranging, we get the equations for strings:
\be
p_s(\beta_i)L=-\sum_l \sum_j \Phi_{sl}(\beta_i-\beta_j) + 2\pi I_{si},
\label{eq:Bet_str}
\ee
where
$$p_s(\beta_i)=\sum_{a=0}^{s-1}p(\beta_i+i\mu(s-1-2a)),$$
$$p_{-1}(\beta)=p(\beta+i\pi),$$
and $\beta_i$ are now the real numbers. 
Introduce $\sigma_s=$sign$(I_{i+1}-I_i)_s$.
Since the density of roots must be positive, then
\be
 {I_{i+1}-I_i\over L\Delta\beta}\sim\sigma_s(\rho+\tilde\rho)_s.
\label{eq:defro}
\ee
Eq. (\ref{eq:defro}) is the definition of the density functions.
After the thermodynamic limit is taken, the Bethe equations assume
the form:
\be
p_s'(\beta)=-\sum_l\Phi_{sl}'\ast\rho_l + 2\pi\sigma_s(\rho_s+\tilde{\rho}_s).
\label{eq:contBe}
\ee
Introducing
\be
A_{sl}=-{\Phi_{sl}'\over 2\pi} + \sigma_s\delta_{sl}\delta(\beta),
\label{eq:defA}
\ee
we get
\be
{p_s'\over 2\pi} = \sum_l A_{sl}\ast\rho_l + \sigma_s\tilde{\rho}_s,
\label{eq:contBE}
\ee
where the involution operator $\ast$ is defined as:
$$ A\ast\rho(x)\equiv\int A(x-y)\rho(y)dy.$$

The choice of sign of $\sigma_s$ is determined by the behavior of the function
$$p_s(\beta)+L^{-1}\sum_{j,l}\Phi_{sl}(\beta-\beta_j)\equiv y_s(\beta),$$
namely, whether $y_s(\beta)$ increases or decreases.
Thus, we are interested in the sign of $y_s'(\beta)$:
$$p_s'(\beta)+L^{-1}\sum_{j,l}\Phi_{sl}'(\beta-\beta_j)= y_s'(\beta).$$ 
Alas, we cannot compute the infinite sum of the terms whose
values depend on unknown yet $\beta_j$. However, if $p_s'(\beta)$ and
$\sum\Phi'$ have the same sign, then the conclusion regarding the sign of
$y_s'$ can be made without the evaluation of sum. For example,
for the attractive regime $\mu>\pi/2$ we have
$$ p_{-1}'\sim -\cosh\beta <0, \qquad \Phi_{-1,-1}'(\beta)<0$$
$$p_t'(\beta)>0, \qquad \Phi_{-1,t}'(\beta)>0, \qquad t\geq 1.$$
Hence,
$$ \sigma_{-1}=-1, \qquad\qquad \sigma_1,\dots,\sigma_{t-1}=1.$$

Let us compute the density of roots in the ground state in the attractive
regime. We should choose appropriate regularization of the momentum,
since Eqs. (\ref{eq:contBE}) make no sense with $p\sim\sinh\beta$.
In \cite{Korep}, for example, the sharp momentum cutoff is chosen, i.e.
$p=0$ for $|\beta|>\Lambda$ with some large $\Lambda$. Such a choice
makes the function of momentum to be non-analytic, and disregards
large momentum pseudoparticles at all. Although it works in the attractive
regime fine, it gives some problems in the repulsive regime for $\mu<\pi/3$.
We shall choose here a smooth cutoff, which comes naturally from the lattice
regularization of the Thirring model (see Figure 1.2):
\be
p(\beta)=-im_0\ln\left[{\sinh{1\over 2}(\beta+f-i\omega)\over
\sinh{1\over 2}(\beta+f+i\omega)}\right]
-im_0\ln\left[{\sinh{1\over 2}(\beta-f-i\omega)\over
\sinh{1\over 2}(\beta-f+i\omega)}\right],
\label{eq:mom}
\ee
\be
p'(\beta)=m_0\sin\omega\left[{1\over \cosh(\beta+f)-\cos\omega}+
{1\over \cosh(\beta-f)-\cos\omega}\right].
\label{eq:mompr}
\ee
\begin{figure}
\epsfxsize=100truemm
\centerline{\epsfbox{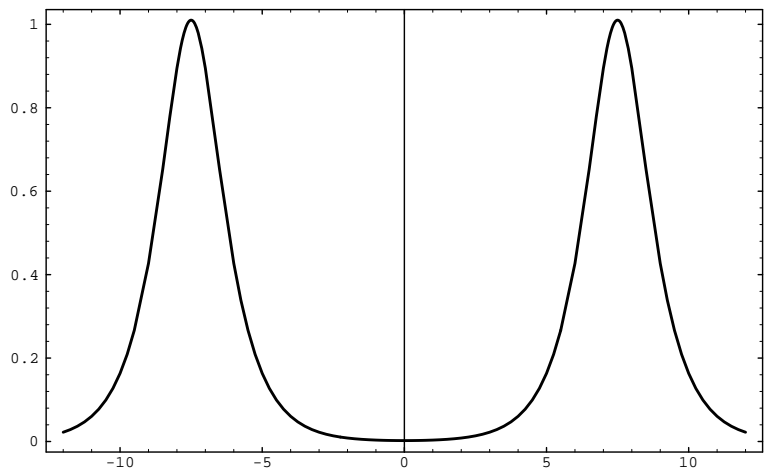}}
\caption{Smoothly regularized momentum $p'(\beta)$.}
\end{figure}

We reproduce the usual relativistic expression in the limit $f\to\infty$:
$$p(\beta)\to 4m_0\sin\omega e^{-f}\sinh\beta=m_1\sinh\beta, \qquad
f\to\infty.$$

Denote the first term in (\ref{eq:mom}) by $f_+(\beta)$ and the second
one by $f_-(\beta)$. Then the regularized energy is
\be
h(\beta)=-f_+(\beta)+f_-(\beta)
\label{eq:ener}
\ee
As expected, $h(\beta)\to m_1\cosh\beta$ as $f\to\infty$.
The Fourier image of (\ref{eq:mompr}) is
\begin{eqnarray}
\hat{p}_{-1}'(k)&=&\int e^{ik\beta}p_{-1}'(\beta)d\beta=
\int e^{ik\beta}p'(\beta+i\pi)d\beta \nonumber \\
&=&-4\pi m_0{\sinh\omega k\cos fk\over\sinh\pi k} \label{eq:four_im}
\end{eqnarray}

\topskip=0cm

We need to solve the Bethe equation for the ground state density $\rho_{-1}$:
$${p_{-1}'\over 2\pi}=A_{-1,-1}\ast\rho_{-1}.$$
This can be easily done by passing to the Fourrier transformed equation:
$${\hat{p}_{-1}'\over 2\pi}=\hat{A}_{-1,-1}\cdot\hat{\rho}_{-1},$$
where 
$$ {\hat{\Phi}'\over 2\pi}=-{\sinh(\pi-2\omega)k\over\sinh\pi k}, \qquad
\hat{A}_{-1,-1}=-\left({\hat{\Phi}'\over 2\pi}+1\right)
=-{2\sinh\omega k\cosh(\pi-\omega)k\over\sinh\pi k}. $$
Thus, we get
\be
\hat{\rho}_{-1}={\hat{p}_{-1}'\over 2\pi\hat{A}_{-1,-1}}=
m_0{\cos fk\over\cosh(\pi-\omega)k}, \label{eq:solro}
\ee
Eventually,
\begin{eqnarray}
\rho_{-1}(\beta)&=&{1\over 2\pi}\int e^{-ik\beta}\hat{\rho}_{-1}(k)dk=
\nonumber \\
&=&{m_0\over 4\mu}\left[{1\over\cosh{\pi\over 2\mu}(\beta+f)}+
{1\over\cosh{\pi\over 2\mu}(\beta-f)}\right]. \label{eq:SolRo}
\end{eqnarray}
Upon the cutoff removal, $f\to\infty$, the density (\ref{eq:SolRo})
becomes
\be
\rho_{-1}(\beta)\to {m_0 e^{-f}\over \mu}\cosh{\pi\beta\over 2\mu}.
\label{eq:limRo}
\ee
We see that what we got looks like the density of free pseudoparticles
in the Dirac sea, with the bare mass and bare rapidity renormalized as
a result of interactions. Note that mass renormalization depends upon the
regularization procedure chosen.

\section{Thermodynamic Bethe ansatz}

The thermodynamic Bethe ansatz (TBA) method allows one to obtian
the excitation energies and to compute the exact free energy. We 
employ it
here for the Thirring model and discuss briefly its main concepts.
The idea of TBA dates back to the work of C.N.Yang and C.P.Yang \cite{YY}.
They introduced the temperature into the Bethe ansatz technique
and used Eqs. (\ref{eq:contBE}) to realize the 
statistics of states at temperature $T$.

The basic equation of the TBA method can be written in the form:
\be
\vrp(\beta)=h(\beta)-A+{T\over 2\pi}\Phi'\ast\ln(1+e^{-\vrp/T}).
\label{eq:YY}
\ee
By virtue of its definition, the pseudoenergy of excitations $\vrp(\beta)$ is
\be
e^{\vrp/T}\equiv{\tilde\rho\over\rho}. \label{eq:defpsen}
\ee
Eq. (\ref{eq:YY}) is the result of minimization  of the free energy
$F=E-TS$ over $\rho$ under the condition $\int\rho=$const. Thus, the
Lagrange multiplier $A$ is just the chemical potential. The entropy
of the system $S$ is given in \cite{YY},
and $E$ is the total energy of all pseudoparticles.
 We note here that the expression
for the entropy differs for the ``fermionic'' and ``bosonic'' cases.
We work with the fermionic case where for each quantum number $I_j$ 
corresponds only one pseudoparticle. 

Eq. (\ref{eq:YY}) is a non-linear integral equation that cannot be solved 
analyticaly in general. 
In the limit $T\to 0$ Eq. (\ref{eq:YY}) reduces to 
\be
\vrp(\beta)=h(\beta)-A-{1\over 2\pi}\Phi'\ast\vrp.
\label{eq:Tvanish}
\ee
Integration in the latter formula is over all rapidities where $\vrp(\beta)$
is negative. From (\ref{eq:defpsen}) also follows that for such $\beta$
we have $\tilde\rho(\beta)=0$, and, respectively, for the values of $\beta$
where $\vrp(\beta)$ is positive, $\rho(\beta)=0$. Let us represent
$\vrp(\beta)$ in the form
$$ \vrp(\beta)=\vrp^++\vrp^-=\vrp H(\vrp)+\vrp H(-\vrp),$$
where $H(\vrp)$ is the Heaviside step function. Then from (\ref{eq:Tvanish})
follows:
\be
\vrp^+(\beta)=h(\beta)-A-\left({\Phi'\over 2\pi}+\delta\right)\ast\vrp^-.
\label{eq:Tvanish-}
\ee
In other words, $\vrp^+>0$ corresponds to the particle excitation energy,
and $|\vrp^-|$ to the hole excitation energy. $\vrp^-$ coincides with
$\vrp$ where $\vrp$ is negative, i.e. in the Dirac sea. For example,
if we take a pseudoparticle out of the Dirac sea and put it above the sea,
the excitation energy would be
$$E_{ext}=\vrp(\beta_p)-\vrp(\beta_h)=\vrp^+(\beta_1)-\vrp^-(\beta_2)>0.$$
The total energy
\begin{eqnarray}
E&=&\int h\rho d\beta=\int\rho(\vrp+A+{1\over 2\pi}\Phi'\ast\vrp^-)
d\beta=\label{eq:smysl}\\
&=&\int[\vrp\rho+A\rho+{1\over 2\pi}\vrp^-\cdot(\Phi'\ast\rho)]=
\int(\vrp^+\rho-\vrp^-\tilde\rho)-{1\over 2\pi}\int p_{-1}'\vrp^- + A\int\rho,
\nonumber
\end{eqnarray}
where we used (\ref{eq:Tvanish}), (\ref{eq:contBE}) and the permutation
property of the convolution operator. Eq. (\ref{eq:smysl}) has clear
physical interpretation: the energy of an arbitrary state at $T=0$
can be expanded as the sum $E=E_{vac}+E_{exc}+$const, where
$$E_{vac}=-{1\over 2\pi}\int p_{-1}'\vrp^-<0$$
is vacuum energy, and
$$E_{exc}=\int(\vrp^+\rho-\vrp^-\tilde\rho)>0 $$
is excitation energy.

TBA equations for $T=0$ can be obtained also directly from the
variation of total energy
\cite{JNWieg}. Under small perturbations $\rho_j^{(0)}\to\rho_j^{(0)}+
\delta\rho_j$ of the ground state densities $\rho_j^{(0)}$ the total
energy changes by
$$\delta E= \sum_j\int h_j\delta\rho_j=\sum_j[\vrp^+_j\delta\rho_j-
\vrp^-_j\delta\tilde{\rho}_j].$$
Substituting $\delta\tilde{\rho}_j$ from the Bethe equations 
(\ref{eq:contBE}) we obtian the {\it basic spectral equations}
\be
h_j=\vrp^+_j+A_{jk}\ast\sigma_k\vrp^-_k.   \label{eq:basic_spectral}
\ee
Similarly, we deduce
\be
p_j=\pi^+_j+A_{jk}\ast\sigma_k\pi^-_k.   \label{eq:basic_spectral_mom}
\ee
From (\ref{eq:basic_spectral}) it follows for the Thirring model
\be
\hat{\vrp}^-_{-1}=-\hat{h}_{-1}/\hat{A}_{-1,-1}, \label{eq:vrpsol}
\ee
with
\be
\hat{h}_{-1}'(k)=-4\pi im_0{\sinh\omega k\sin fk\over \sinh\pi k}. \label{nenuzhno}
\ee
We obtain 
$$ \hat{\vrp}^-_{-1}={2\pi m_0\over k}{\sin fk\over \cosh\mu k}.$$
In the limit $f\to\infty$ we get for the energy of hole in the Thirring model:
\be
\vrp^-_{-1}\to -4m_0e^{-\pi f/2\mu}\cosh{\pi\beta\over 2\mu}.
 \label{eq:hole_ener}
\ee
The hole of the Thirring model corresponds to 
a soliton of the sine-Gordon model.
For the rest of the excitations we have
\be
\hat{\vrp}^+_j=\hat{h}_j+\hat{A}_{j,-1}\hat{\vrp}^-_{-1}
=\hat{h}_j-{\hat{A}_{j,-1}\over\hat{A}_{-1,-1} }\hat{h}_{-1}, 
\label{eq:rest_exc}
\ee
where we have substituted (\ref{eq:vrpsol}) to get the latter.
Using the Fourier transforms $\hat{A}_{jk}$ together with
$$\hat{h}_j'(k)=4\pi im_0{\sinh(\pi-j\omega)k\sin fk\over\sinh\pi k},$$
 we obtain
$$\vrp^+_j(\beta)\to 8m_0e^{-\pi f/2\mu}\sin\left[{(\pi-\mu)\pi j\over 2\mu}
\right]
\cosh{\pi\beta\over 2\mu}, \qquad j<t-1$$
and
$$\vrp^+_{t-1}(\beta)=-\vrp^-_{-1}(\beta).$$
$t-1$ string in the Thirring model correspond to antisoliton of the sine-Gordon
model, while the other j-strings correspond to the breathers.

\section{Scattering matrices}

 The method to compute  elastic S-matrices from the Bethe ansatz equations
was developed in 
\cite{Korep}.  We describe the basic idea briefly, taking as an example
scattering of holes (solitons in the sine-Gordon model).

By definition, the phase shift for scattering of two holes is given by
\be
\delta_h(\beta_1-\beta_2)\equiv{1\over i }\log S = \varphi_1 -
                                                                \varphi_2,
\label{eq:def_scat}
\ee
where $\varphi_1$ is the phase gained by a hole when  going around the system and $\varphi_2$ the same phase but in the presence of another hole. The $\varphi_{1,2}$
are composed of a sum of two-particle bare phase shifts; for example 
$$ \varphi_2 = Lp_{-1}(\beta_1) + \sum_j \Phi_{1,1}(\beta_1-\beta_j)
$$
Here the sum is taken over all the solutions $\beta_j$
of the Bethe equations (\ref{eq:bethe})
 with two holes at positions $\beta_1$ and 
$\beta_2$. The result of subtraction $\varphi_1 - \varphi_2$
is proportional to  the
{\it backflow function} of vacuum: $\delta_h = 2\pi F(\beta_2|\beta_1)$.
The latter is defined using the difference of the two  solutions of
the Bethe equations obtained  with and without the hole  at $\beta_0$,
$$F(\beta_0|\beta)\equiv(\beta-\tilde\beta)L\rho(\beta).$$
 One can show using (\ref{eq:bethe}) 
that $F$ satisfies the following integral equation:
\be
\Phi(\beta-\beta_0)=\dot{\Phi}\ast F + 2\pi F \label{eq:backfl}
\ee                
Equation (\ref{eq:backfl}) describes the backflow caused by a hole at 
$\beta=\beta_0$
on  Im$\beta=\pi$ axis.  
Taking a derivative with respect to $\beta$ and applying the Fourier transform
to both sides of (\ref{eq:backfl})
 we arrive at the following solution (in Fourier space):
\be
\hat{F}'(k)={\sinh(\pi-2\mu)k\over 2\cosh\mu k \sinh(\pi-\mu)k}
\ee
 From  (\ref{eq:def_scat}) we obtain:
\be
{1\over i }{d\over d\theta}\log S(\theta)=
\int_{-\infty}^{+\infty} 
e^{- i  k\theta}{\sinh(\pi-2\mu)k\over 2\cosh\mu k \sinh(\pi-\mu)k}dk,
\label{eq:S-matr}
\ee
where $\theta\equiv \beta_1 - \beta_2$. Note that this method 
does not fix the constant
normalization factor in the matrix element,
which should be fixed by other constraints (e.g. unitarity). 
Expression (\ref{eq:S-matr}) is
in agreement with the results of \cite{ZZ}, 
which enables us to identify coupling
constants in the Thirring model with those of the SG model:
$\mu=\pi - \beta_{SG}^2/8$.

The conserved charge equals to the total number of pseudoparticles 
$Q=\int\rho d\beta$.
The physical (renormalized) 
charge is obtained after the subtraction of the charge
of vacuum. It can be easily calculated from the Bethe equations
using the backflow functions. E.g., for a hole $Q_{hole}=-1+\hat{F}'(0)$,
while for n-string $Q_n=n+\hat{F}_n'(0)$. Thus, we obtain
the following values: $Q_{hole}=-\pi/2\omega$ for the hole, $Q_{t-1}=
\pi/2\omega$ for the $t-1$ string, and $Q_n=0$ for the other strings.

\section{Remarks}

There are many interesting issues of the Bethe ansatz technology left 
beyond the scope of this chapter. We want merely to mention some of them
here. 

Somewhat more accurate analysis of the permissible solutions of the Bethe
equations for the XXZ chain, based on the counting arguments, shows that 
there are certain constraints on the space of physical states. For example,
the holes can exists only in pairs in this model \cite{FadT}. Another
interesting example is so-called imaginary mass Thirring model discussed
in \cite{ImThir}, where the pseudoparticles are paired in the Dirac sea.

The important questions are the norm of the Bethe states \cite{Gaudin},
 and their completeness \cite{Kirill}. The latter issue for the XXZ chain
is equivalent to showing that the number of the physical states obtained
from the Bethe ansatz equations equals to $2^N$, where $N$ is the number of
sites on the chain. 

Finally, it is worth to mention that the thermodynamic Bethe ansatz
technique can be applied also to the {\it physical} excitations, and
in this context it
allows to find the exact free energy as a function of scaling parameter
$mL$, as well as the central charges
of the fixed point theories \cite{AlZamo}. 


\chapter{Classical and semi-classical analysis of the boundary
sine-Gordon model}


We consider the sine-Gordon model on a half-line, with
an additional potential term of the form $-M\cos{\beta\over
2}(\varphi-\varphi_0)$ at the boundary. 
We construct the classical solutions in the next section by using
the
bulk sine-Gordon theory and the ``generalized method of images.''  
From the classical solutions in hand we extract the time delay 
(\ref{phasedel}) as follows.
We send a soliton (anti-soliton) which lives in the semi-infinite
world governed by
the boundary sine-Gordon model from some large position $x_0$ at time
$t_0$ and measure the time $t_1$ 
it takes to bounce off the boundary and come back
to $x_0$.   At the same time $t_0$ in some other, infinite, world 
governed by the bulk sine-Gordon model we send a soliton
with the same speed but in the opposite direction from the position $-x_0$
and measure the time $t_2$ it takes to arrive at $x_0$. Now, 
$\Delta t=t_1-t_2$ is our time delay. From the time delay one can reconstruct
the semi-classcial phase-shift using the method of Jackiw and Woo
\cite{JW}. We establish the agreement with the semi-classical limit $\beta\to 0$
of the exact boundary reflection matrix, (\ref{semiI}).
The exact expressions for
the boundary reflection matrices  are known up to CDD ambiguities \cite{GZ}.
They were obtained as a ``minimal'' solution to the general set of constraints
for the integrable boundary field theory.
   This chapter is based on \cite{SSW}.

\section{The boundary sine-Gordon model}

The Lagrangian of the boundary sine-Gordon model is given by:
\be
{\cal{L}}_{SG}={1\over 2}\int_0^{+\infty}
\left[\left(\partial_t\varphi\right)^2 -
\left(\partial_x\varphi\right)^2 +
{m_0^2\over\beta^2}\cos\beta\varphi\right]dx ~+~ M\cos
{\beta\over 2}(\varphi(x=0)-\varphi_0) \label{lagr}
\ee
Vanishing of the variation of Lagrangian (\ref{lagr})
under a small perturbation $\varphi\to\varphi+\delta\varphi$
is equivalent to two conditions for the field $\varphi$. 
One is the sine-Gordon equation  on the interval $[0,\infty)$
\be
\phi_{tt} - \phi_{xx} = -\sin(\phi) , \label{sgeqn}
\ee
where  $\beta\varphi\equiv\phi$, whereas another is the boundary condition
\be
\partial_x \phi |_{x=0} ~=~ M \sin \half\left(\phi - \phi_0\right)
 \big |_{x=0} \ . \label{intbcs}
\ee
The condition (\ref{intbcs}) is the most generic boundary condition
preserving the integrability of the bulk problem \cite{GZ}. It has
 two arbitrary paramenters, boundary mass $M$ and a phase $\phi_0$.

\section{The classical solutions}

The initial-boundary value problems  compatible with integrability
for the classical
non-linear equations 
is a field itself withing the non-linear science. Such problems were
studied extensively by mathematicians.
One of the first results on the initial-boundary value problem  for the 
classical sine-Gordon equation were obtained by Sklyanin \cite{Skl}.
In \cite{Skl} the particular form of the most generic boundary condition
was reported, $\partial_x\phi=M\sin(\phi/2)$ at $x=0$.
Later, this condition was widened in \cite{BikTar}
to include the Dirichlet $\phi(x=0)=const$ and 
$\partial_x\phi=M\cos(\phi/2)$ cases. Some solutions
to the posed boundary value problems were obtained using the inverse
scattering method and the Backlund transformations \cite{Habibul,Tar}.
Still, the most generic boundary condition was missed in \cite{BikTar}. 
The condition (\ref{intbcs}) first appeared in the physics litearure in the 
seminal work of Ghoshal and Zamolodchikov \cite{GZ}.

In this section we construct explicitly the solution to 
(\ref{lagr})-(\ref{intbcs}) that replicates the  soliton (antisoliton)
scattering from
a boundary, as well as the  solution localized at the boundary, called
{\it boundary breather}. The idea that we use to construct such solutions
is a remniscent of the method of images (and therefore we refer to it as
the ``generalized method of images.'') We employ the known solutions
on the full line -- kinks and anti-kinks.
 To provide the reflection of the
kink from a boundary (and thus the correct
assymptotic behavior at infinity), 
we send another kink of the same kind towards it  --
the ``mirror image''. Such a solution automatically satisfies (\ref{lagr}).
Besides, in order to satisfy also (\ref{intbcs}), we place the third,
stationary kink at the origin. We check by a direct calculation that
condition (\ref{intbcs}) is satisfied. To obtain the three-kink solution,
we need the knowledge of the  $\tau$-function for sine-Gordon equation.

\subsection{The $\tau$-functions}

On the infinite interval, $(-\infty,\infty)$,
the classical multi-soliton solution to the sine-Gordon equation
is well known \cite{ABL}.
  It is usually expressed as:
\be
\phi(x,t) ~=~ 4 ~ \hbox{arg} (\tau) ~\equiv~ 4~
\hbox{artan}\bigg(
{{\cal I}m(\tau)  \over {\cal R}e (\tau) }  \bigg) \ , \label{sgsoln}
\ee
where the $\tau$-function for the $N$-soliton
solution is:
\begin{eqnarray}
\tau &=& \sum_{\mu_j = 0,1} ~  e^{{i \pi \over 2}
( \sum_{j=1}^N \epsilon_j \mu_j)} ~\exp ~\bigg [ - \sum_{j=1}^N
\half
\mu_j ~ \Big[  \Big(k_j  +  \frac{1}{k_j} \Big)   x   +
 \Big (k_j - \frac{1}{k_j} \Big) t ~-~  a_j \, \Big ]  \nonumber \\
 &+&  2
\sum_{1 \le i<j \le N} \mu_i \mu_j ~ \log\bigg( {{k_i - k_j}
\over {k_i + k_j} } \bigg) \bigg]  \ .  \label{taufn}
\end{eqnarray}
The parameters $k_j$, $a_j$ and $\epsilon_j$ have the following
interpretations. The velocity of the $j^{\rm th}$ soliton is given
by:
\be
v_j ~=~ \bigg( {{k_j^2 - 1} \over {k_j^2 + 1} } \bigg) \ . \label{vels}
\ee
(Note that $v_j$ is positive for a left-moving soliton.)
The $a_j$ represent the initial positions of each of the solitons,
and
$\epsilon_j = +1$ if the $j^{\rm th}$ soliton is a kink, while
$\epsilon_j = -1$ if it is an anti-kink.  The rapidity, $\theta$,
of the soliton is defined by $k = e^\theta$, and we have normalized
the
soliton masses to unity (in further discussion, the words ``soliton''
 and ``kink'' will be used synonymously).

It is fairly obvious how to get a single soliton solution on
$[0,\infty)$
with either $\phi |_{x=0} = 0$ or $\partial_x \phi |_{x=0} = 0$.  One
exploits
the symmetry of (\ref{sgeqn}) under $\phi \to -\phi$ and $x \to -x$, and
simply
takes a two soliton solution on  $(-\infty,\infty)$ where one soliton
is a mirror image of the other through $x=0$ \cite{Trull}.
If one does this with a double-kink solution then it satifies the
foregoing Dirichlet condition, while the kink-anti-kink solution
satisfies
the Neumann condition.  It is also not hard to guess how one can go
beyond
this solution:  For $M = \infty$, the boundary condition (\ref{intbcs})
reduces
to $\phi |_{x=0} = \phi_0$.  The only way that this can be obtained
from
a multi-soliton solution on $(-\infty,\infty)$ is to put a third,
{\it stationary  } soliton at the origin.

We therefore consider the three soliton solution with $k_1 = k$, $k_2
= 1/k$
and $k_3 = 1$.  That is, we consider:
\begin{eqnarray}
\tau &=& 1 ~-~  \epsilon v^2 e^{ -{1\over k}
(k^2 +1) x - a} ~-~ \epsilon_0 \left({k-1\over k+1}
\right)^2 e^{- {1\over 2k}  (k +1)^2 x - b} ~ F(t)
\label{bdrytau} \\
{}&+& i \Big\{ ~ e^{- {1\over 2k}  (k^2 +1) x }
{}~ F(t)
{}~+~ \epsilon_0 e^{- x - b}
{}~-~ \epsilon \epsilon_0 v^2 \left( {k-1\over k+1} \right)^4
e^{- {1\over k} (k^2 + k +1) x - a - b} ~\Big\}
 \ , \nonumber 
\end{eqnarray}
where we have introduced the shorthand:
\be
\epsilon = \epsilon_1  \epsilon_2 \ , \qquad
\epsilon_0 = \epsilon_3\ , \qquad a = a_1 + a_2\ , \qquad b = a_3 \
. \label{shorth}
\ee
The function $F(t)$ is defined by:
\be
F(t) ~\equiv~ \epsilon_1 e^{ -{1\over 2k}
(k^2 -1) t - a_1} ~+~ \epsilon_2 e^{ {1\over 2k}
(k^2 -1) t - a_2} \ . \label{Fdefn}
\ee
This solution has $\phi = 0$ at $x = \infty$, and for $k>1$ it
describes a
left-moving soliton moving from $x = \infty$ with a right-moving
``image''
starting at $ x = -\infty$.   There is a stationary soliton with
center
located at $x =-b$.  Viewing this as scattering off a boundary at
$x=0$ one can easily see that $a$ is the phase delay of the returned
soliton.
To make this more explicit, observe that $\tau$ has the following
asymptotic behaviour:
\begin{eqnarray}
\tau(x,t) ~&\sim&~ 1 ~+~ i \epsilon ~ e^{-
{1\over 2k}
 [(k^2 +1) x + (k^2 -1)t] - a_1} \qquad x, -t \to
\infty
\ \ {\rm with} \ \ {x \over t} = -
{k^2 -1 \over k^2 +1} \ ; \nonumber \\
\tau(x,t) ~&\sim&~ 1 ~+~ i \epsilon ~ e^{- {1\over 2k}
 [(k^2 +1) x - (k^2 -1)t] - a_2} \qquad x,t \to
\infty,
\ \ {\rm with} \ \ {x \over t} =
+{k^2 -1 \over k^2 +1} \ . \nonumber
\end{eqnarray}
The problem now is to first show that the $\tau$-function given by
(\ref{bdrytau})
provides a solution to the boundary value problem on $[0,\infty)$
defined\
by (\ref{sgeqn}) and (\ref{intbcs}).
Our second purpose is to relate the parameters $a$ and $b$ of
(\ref{bdrytau}) to
the parameters $M$ and $\phi_0$ of (\ref{intbcs}), thereby obtaining the
classical phase delay, $a$, in terms of $M$ and $\phi_0$.

\subsection{The classical phase delay for $M=\infty$ case}

In the center of mass reference frame the solution to (\ref{sgeqn}) 
obtained by means of the $\tau$-function method reads:
\be
\phi=\mp 4\arctan{2e^{x{\rm ch}\theta - a_1} {\rm ch}
(t{\rm sh}\theta) \pm e^{x - a_3} \mp e^{x(1+2{\rm ch}\theta) - 2a_1
-
a_3 +
2\log(u^2v)}\over  2e^{x(1+{\rm ch}\theta) - a_1 - a_3 +
2\log(u)}{\rm ch}
(t{\rm sh}\theta) \pm  e^{2x{\rm ch}\theta - 2a_1 + 2\log(v)} \mp 1},
\label{A1}
\ee
where $u=\tanh({\theta\over 2})$, $v=\tanh(\theta)$, and we 
have set
$a_1=a_2$, which means that the solution is invariant under
the
transformation $t\rightarrow-t$.
 The upper (resp. lower) sign corresponds to the situation when
a stationary soliton (resp. anti-soliton) is used to adjust the value
of field at
the boundary. Solution (\ref{A1}) refers to the case when the incoming
and outgoing particle is the soliton with asymptotic value
 $\phi=2\pi$
at $x=+\infty$.

Let us represent (\ref{A1}) in the form of rational function of variable
${\rm ch}(t{\rm sh}\theta)$:
$$ \phi(x, t) = 4\arctan {r_2 + s_2{\rm ch}(t{\rm
sh}\theta)\over r_1 + s_1 {\rm ch}(t{\rm sh}\theta) }. $$
The condition $\phi(x=0, t)=\phi_0$ implies that
\be
 {r_2\over r_1}={s_2\over s_1}=\tan\left({\phi_0\over
4}\right),\label{A2}
\ee
from which follows immediately
$$ a_3 = \log\left(\mp u^2\tan{\phi_0\over 4}\right).$$
For the argument of logarithm to be positive one should take
$\phi_0>0$
with the stationary anti-soliton and $\phi_0<0$ with the stationary
soliton.
This is illustrated in figure 2.1. Further, we obtain from (\ref{A2})
\be
 2a_1=\log \left[ u^2v^2 { u^2 + \tan^2({\phi_0\over 4}) \over
1 + u^2\tan^2({\phi_0\over 4})}\right]. \label{A3} 
\ee
 Note that
the time delay, obtained by (\ref{A3}),
is in fact always  a time {\it advance} in both the attractive and
repulsive
cases. For the same value of $\phi_0$ the
time delay
for the soliton that lives on the left half-line $x<0$ is not the
same as
that of the right half-line soliton
(except for $\phi_0=\pm\pi$).  The position of the ``left'' soliton
is not the exact
 mirror image of the ``right'' soliton for generic $\phi_0$.

\subsection{The classical phase delay for generic case}

To summarize the computation briefly, one substitutes (\ref{bdrytau}) into
(\ref{intbcs}),
and obtains the constraint:
\begin{eqnarray}
& &\left[ {\cal R}e (\tau) \partial_x {\cal I}m
(\tau) -
{\cal R}e (\tau) \partial_x  {\cal I}m (\tau)\right]_{x=0}
=\label{midbc} \\
{}&=& M~ \left[ 2  \cos(\half \phi_0) ~{\cal R}e (\tau)  {\cal I}m
(\tau) ~-~
  \sin(\half \phi_0) ~ \left({\cal R}e (\tau)^2 -  {\cal I}m (\tau)^2
\right)  \right]_{x=0} \ . \nonumber
\end{eqnarray}
One can solve this by brute force substitution for the real and
imaginary
parts of $\tau$, but it is somewhat simpler to find constants
$\alpha$,
$\beta$, $\gamma$ and $\delta$ such that:
\begin{eqnarray}
 \partial_x {\cal R}e (\tau)~ \big |_{x=0} &=&
\left[
\alpha~ {\cal R}e (\tau)  ~+~ \beta~{\cal I}m (\tau) \right]\big|_{x=0}
\ , \nonumber \\
 \partial_x {\cal I}m (\tau)~ \big |_{x=0} &=& \left[ \gamma~
{\cal R}e (\tau)  ~+~ \delta ~{\cal I}m (\tau) \right]\big|_{x=0}\ \
. \label{simpler}
\end{eqnarray}
One then finds that (\ref{midbc}) can be satisfied if and only if  one has
$\alpha = \delta$, which indeed turns out to be true.
Using this one arrives at:
\begin{eqnarray}
M~ \cos(\half \phi_0) &=& \nonumber \\
- { (k^2 +1) \over k\Delta} &&
\left\{ \left( 1 ~+~ \epsilon~v^2 e^{-a} \right) ~-~ e^{-2b} \left(
{k-1 \over k+1}
\right)^2 \left[ 1 ~+~ \epsilon~v^2 ~ \left({k-1 \over k+1}
\right)^4 ~e^{-a} \right] \right\} \nonumber \\
M~ \sin(\half \phi_0) &=& - 2~\epsilon_0 ~ {(k-1)^2
\over k} ~ e^{-b} ~ {1 \over \Delta} \left\{  1 ~+~ \epsilon~v^2
{}~ \left({k-1 \over k+1} \right)^2 ~e^{-a} \right\}, \label{Mphireln}
\end{eqnarray}
where
\be
\Delta ~\equiv~   \left( 1 ~-~ \epsilon~v^2  ~ e^{-a}
\right)
{}~+~ e^{-2b} \left( {k-1 \over k+1} \right)^2
\left[ 1 ~-~ \epsilon~v^2 ~ \left( {k-1 \over k+1}
\right)^4 ~e^{-a}  \right] \ . \label{Deldefn}
\ee

It is algebraically very tedious to invert this relationship.  One
proceeds
by eliminating $e^{-b}$, and then  solving for $a$.  It is very
convenient to
introduce a new parametrization of $M$ and $\phi_0$:
\begin{eqnarray}
\mu ~\equiv~ M~ \cos(\half \phi_0) &\equiv&
2~\cosh(\zeta) \cos(\eta) \  \nonumber \\  \nu ~\equiv~ M~ \sin(\half
\phi_0)
{}~&\equiv&~ 2~\sinh(\zeta) \sin(\eta) \ , \label{newparam}
\end{eqnarray}
where $0 \le \zeta < \infty$ and $ -\pi < \eta \le \pi$.
(In the $(\mu,\nu)$-plane the curves of constant $\zeta$ are
ellipses,
while the curves of constant $\eta$ are hyperbolae whose asymptotes
make an angle of $\half \phi_0$ with the $\mu$-axis.)

\newpage
\topskip=18cm
 \includegraphics{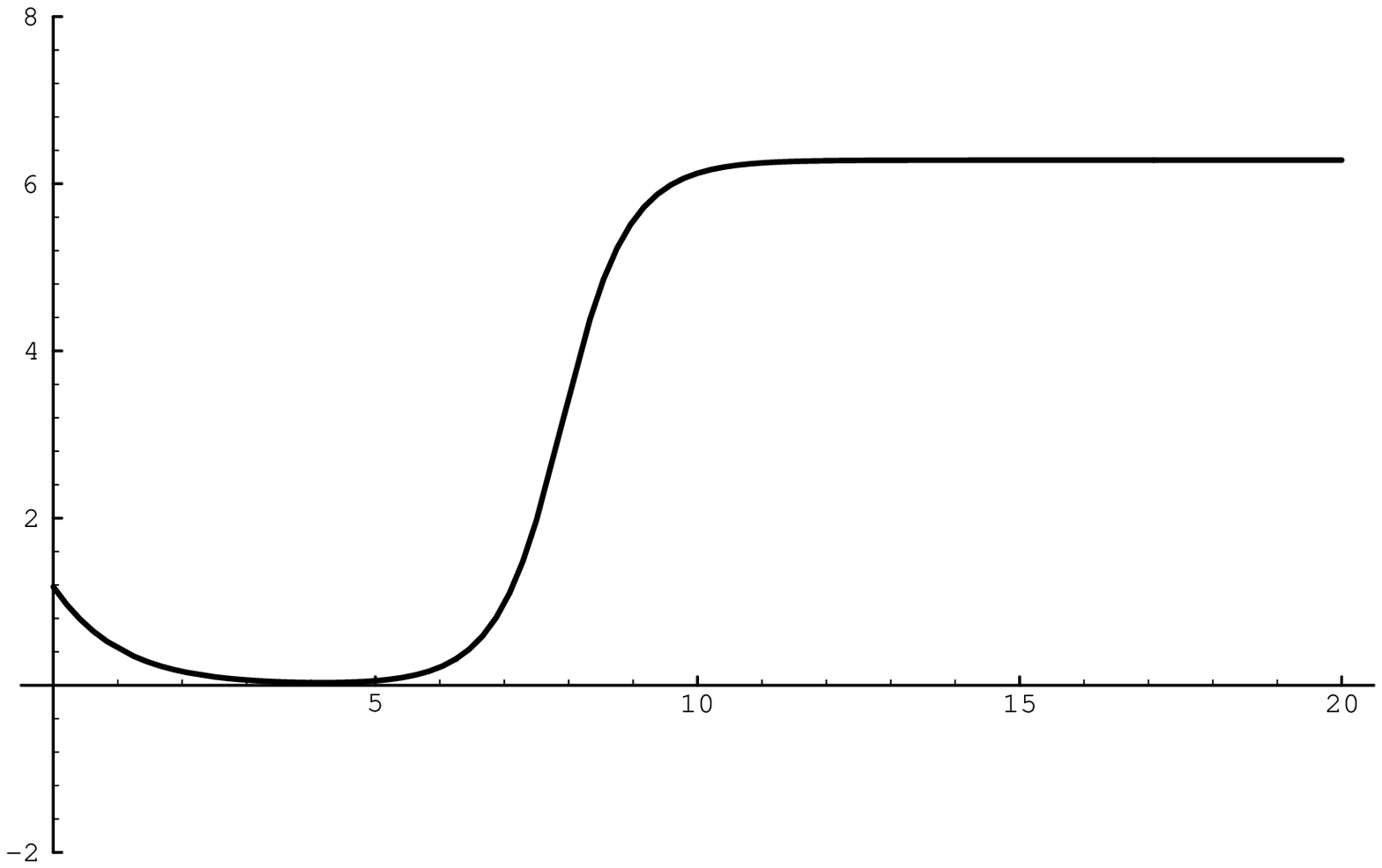}
 \includegraphics{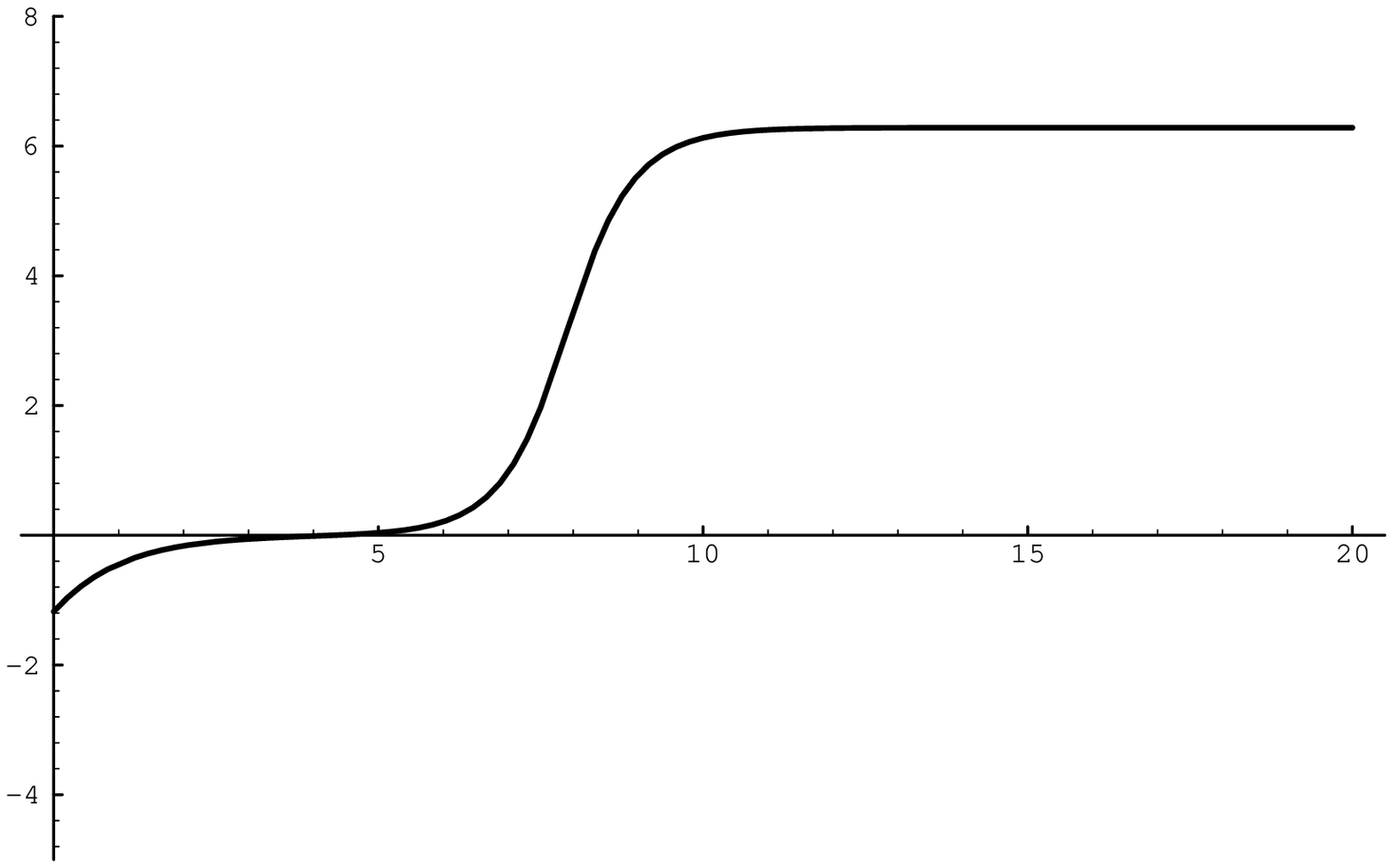}

\noindent Figure 2.1: a solution of the classical Sine-Gordon equation 
 with the fixed boundary
 conditions. For  $\phi|_{x=0}=\phi_0={3\pi\over 8}$ 
 solution is 
 constructed
 out of a left-moving soliton, its right-moving image at $x<0$ and 
 the stationary
 antisoliton in the middle (upper graph). 
 For $\phi|_{x=0}=\phi_0=-{3\pi\over 8}$ 
 the solution is built out of three solitons (lower graph).
\addcontentsline{lof}{figure}{\protect\numberline{2.1}{
Solution of the classical Sine-Gordon equation 
 with the fixed boundary
 conditions.}}
 \newpage
\topskip=0cm

We then find that the phase delay is given by:
\be
a ~=~ \log\left\{ -\epsilon ~ \tanh^2
(\theta/ 2)   \tanh^2\theta   \left[
{{\tanh\half(\theta +
i \eta) ~ \tanh\half(\theta - i \eta) } \over {\tanh\half(
\theta+\zeta )
{}~ \tanh\half(\theta-\zeta) }} \right]^{\pm1} \right\} \  . \label{phasedel}
\ee

There are several things to note about this formula.

\begin{description}

\item[(i)] The ambiguity
of the $\pm1$ power comes from solving a quadratic equation, and is
a direct reflection of the fact that (\ref{intbcs}) is not invariant under
$\phi \to \phi + 2\pi$ (whereas (\ref{sgeqn}) {\it is} invariant under this
shift).
Equivalently, one can flip between the $+$ and $-$ powers by sending
$\phi_0 \to \phi_0 + 2\pi$.

\item[(ii)]  The argument of the log is {\it always} real and
positive.
The discrete parameter, $\epsilon = \epsilon_1 \epsilon_2$, must be
chosen to arrange this.  Hence:
\be
\epsilon = +1 \ \ {\rm for} \ \ - \zeta < \theta < \zeta
\qquad \qquad \epsilon = -1 \ \ {\rm for} \ \ |\theta| > \zeta \ .
\label{epschce}
\ee
This means that a kink reflects into kink for $- \zeta < \theta <
\zeta $,
and reflects into an anti-kink for $|\theta| > \zeta$.  This is
consistent
with the fact that Dirichlet boundary conditions ($M = \infty$)
cause
a kink to reflect as a kink, whereas Neumann boundary conditions
($M = 0$) cause a kink to reflect as an anti-kink.  Note that
these two domains of parameter space (in which a kink reflects
differently)
are separated from one another by a logarithmic singularity in the
classical phase delay.

\item[(iii)]  The choice of the power $\pm 1$ in (\ref{phasedel})
is correlated with the parameter $\eta$ and whether there is a kink,
or anti-kink at the
origin.  Specifically, we have:
\be
\epsilon_0 ~=~ \pm ~sign(\tan({\eta\over 2})) \ , \label{epszero}
\ee
where the $\pm$ is the same as  that of (\ref{phasedel}).

\item[(iv)]  In the $M \to \infty$, or Dirichlet, limit we see that:
\be
\zeta ~\sim~ \log(M) \ , \qquad \eta ~\sim~ \half
\phi_0 \qquad {\rm and} \qquad \epsilon  ~=~ -1 \ , \label{Mtoinf}
\ee
and the phase delay collapses to (\ref{A3}):
\be
a ~=~ \log\left\{
\tanh^2\left({\theta\over 2} \right)   \tanh^2( \theta )
  \left[ \tanh\half(\theta + i {\phi_0\over 2}) ~
\tanh\half(\theta -
i {\phi_0\over 2})   \right]^{\pm1} \right\} \  .\label{simpphasedel}
\ee
In this limit (\ref{intbcs}) enforces Dirichlet boundary conditions.
It is, however, important to note that
there are {\it two} possible distinct
boundary values: $\phi|_{x=0} = \phi_0$ and $\phi|_{x=0} = \phi_0 +
2\pi$.
 Since the boundary potential is
$-M \cos(\half(\phi - \phi_0))$, one sees that $\phi|_{x=0} =
\phi_0$
is stable,  while $\phi|_{x=0} = \phi_0 + 2\pi$ is unstable.
\end{description}

{}From now on we consider only the stable boundary value that
corresponds to the
positive sign in (\ref{simpphasedel}). Then one has from (\ref{epszero})
\be
\epsilon_0 ~=~ - ~sign(\tan({1\over 4}
\phi|_{x=0})) \ . \label{epszerosimp}
\ee

It is essential to observe  that we have taken
$\phi_{x=\infty} = 0$,
{\it ab initio}. For different boundary conditions at $x=\infty$,
one should replace $\phi_0$ by $\phi|_{x=0}-\phi|_{x=\infty}$.
This  physical
parameter is defined  mod $4\pi$.  

\subsection{Boundary breather solutions}

To fully understand the semi-classical scattering computation one
also needs another class of classical solutions,
which we call here {\it boundary breathers}.   It is well-known that
breathers
represent bound states in the soliton-anti-soliton channel in the
bulk
sine-Gordon theory.  In the same way that the classical bulk
breathers  can be obtained from the appropriate solution by analytic
continuation of $\theta$ to
imaginary axis, one might expect that the same procedure would give
boundary breathers on the half-line. To see this, we set
$\theta=i\vartheta$
($0 < \vartheta < {\pi\over 2}$) in the formula (\ref{A1}).
Next, we impose the following conditions for a solution to be
boundary breather:

\begin{description}

\item[a)] it should be a real function,

\item[b)] it should have finite energy and

\item[c)] the asymptotic value at $x=+\infty$ must be fixed and equal
to
$2\pi n$
($n$ -- integer number).
\end{description}

The three-soliton  (resp. soliton-anti-soliton-soliton) configuration
satisfies
the first condition provided that $\vartheta<-{\phi_0\over 2}$
(resp. $\vartheta<{\phi_0\over 2}$). However, the other conditions
are satisfied by the soliton-anti-soliton-soliton configuration only,
which
one could have foreseen from the analogy with the bulk theory.
The boundary breather
solution has the form (see figure 2.2):
\begin{eqnarray}
 & \phi_b & = 2\pi - \label{breath} \\
 & &-4\arctan
{2\cot\vartheta\cot{\vartheta\over 2}\sqrt{K}\cot{\phi_0\over 4}
e^{x+x\cos\vartheta}\cos(t\sin\vartheta) + e^{2x\cos\vartheta}K\cot^2
{\vartheta\over 2} + 1 \over 2\cot\vartheta\cot{\vartheta\over
2}\sqrt{K}
e^{x\cos\vartheta}\cos(t\sin\vartheta)+e^x\cot{\phi_0\over 4}(
e^{2x\cos\vartheta}K +  \cot^2{\vartheta\over 2}) }, \nonumber
\end{eqnarray}
where
$$ K=\cot({\phi_0\over 4} - {\vartheta\over 2})
\cot({\phi_0\over 4} + {\vartheta\over 2}) .$$

So, we have a continuum of classical boundary bound states when
$0<\vartheta<
{\phi_0\over 2}<{\pi\over 2}$ and for other $\phi_0$
according to the
$2\pi$-periodicity. In the quantum theory this continuum shrinks into
a
discrete set of bound states (see next section). Note that in the
limit
$\vartheta\to{\phi_0\over 2}$  the boundary breather (\ref{breath}) reduces
to
the ground state, figure 2.3, and
the phase delay has singularities at $\theta = \pm{1\over 2}
i\phi_0$.
An analogous picture of bound states occurs
for the anti-soliton scattering with fixed boundary conditions.
\subsection{General solutions, integrability and  \\ B\"acklund 
transformations}

Thus far we have only applied the method of images to obtain certain
special classical solutions of the boundary sine-Gordon problem
(\ref{lagr}).  It is natural to suggest that general solutions of 
(\ref{lagr}) can be
obtained by similar methods.  This in turn would establish the
classical integrability the boundary problem (\ref{lagr}).  It is, in fact,
rather straightforward to show that both of these conclusions are
true, at least for the problem (\ref{lagr}) with $\phi_0 =0$. The method we
will employ \cite{SSW}
can almost certainly be generalized to problems with
$\phi_0 \ne 0$, and also
has the virtue that it can be used to construct the integrable
boundary potentials for the more general Toda models. A related
approach has
been followed in \cite{Tar, HabibI}. The basic
idea is to use the fact that any integrable hierarchy has B\"acklund
transformations: that is, solutions can be mapped into one another by
 non-trivial gauge transformations constructed from the affine Lie
algebra action on the corresponding LAX system.   For the
sine-Gordon equation, the requisite B\"acklund transformation can be
cast in the following form:
\begin{eqnarray}
 \partial_u (\phi ~+~ \psi) ~=~ e^\zeta ~
\sin \left({\phi - \psi \over 2} \right) \ ;  \nonumber \\
\partial_v (\phi ~-~ \psi) ~=~ e^{-\zeta} ~
\sin \left({\phi + \psi \over 2} \right) \ , \label{Backlund}
\end{eqnarray}
where $u = x - t$, $v = x + t$, and $\zeta$ is an arbitrary constant
parameter.  The point is that $\phi$ satisfies the sine-Gordon
equation (\ref{sgeqn}) if and only if $\psi$ does so as well.  Suppose that
$\psi$ is a solution to sine-Gordon on $[0,\infty)$ satisfying a
Dirichlet boundary 
condition: $\psi|_{x=0} = 2 \eta$, where $\eta$ is
a
constant. It follows immediately from (\ref{Backlund}) that 
$\phi$ is a
solution to sine-Gordon satisfying (\ref{intbcs}), where $M$ and $\phi_0$
are given in terms of $\zeta$ and $\eta$ by (\ref{newparam}).  Thus, if one
can solve the Dirichlet problem, one can solve the more general
problem by a B\"acklund transformation.

Observe that if $\eta =0$, or equivalently $\phi_0 =0$, then the
Dirichlet problem can be solved trivially by method of images: one
gets the solution on the half-line by extending it as an odd
function on the full line.  Thus, the B\"acklund transformation
essentially defines the generalized  method of images.  It is also
by no means an accident that the parameters entering into the
B\"acklund transformation ($\eta$ and $\zeta$) are precisely the
rapidity parameters that turn up in the phase delay (\ref{phasedel}).   One
should also note that the form of the integrable boundary potential
is given directly by the B\"acklund transformation.  This fact should
easily generalize to Toda systems.

B\"acklund transformations, in general, are invertible
transformations on the solution space of an integrable hierarchy.
The simplest forms of them modify the constants of the motion of
solution, and possibily add or subtract a soliton.   One can
certainly find a B\"acklund transformation that will introduce a
stationary soliton into the soliton-soliton solution of sine-Gordon.
As a result, the general three soliton solution employed above can be
obtained from the two soliton solution that is appropriate for the
``trivial'' Dirichlet problem with $\phi_0 = 0$.   We therefore
expect that any solution of the trivial Dirichlet problem can be
mapped onto the generic problem (\ref{lagr}), thus establishing the
classical integrability.  

\newpage
\topskip=9.5cm
 \includegraphics{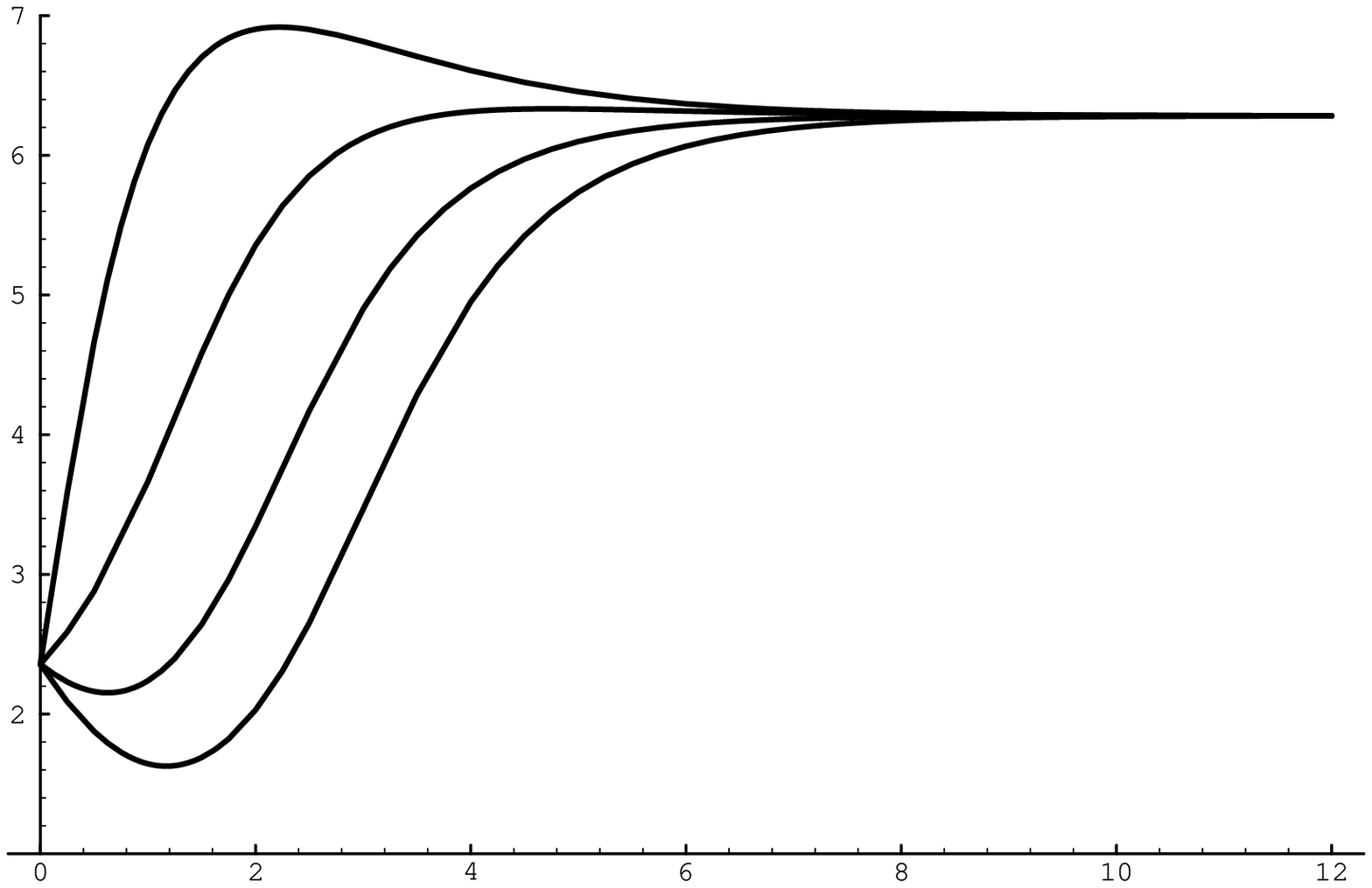}
 \includegraphics{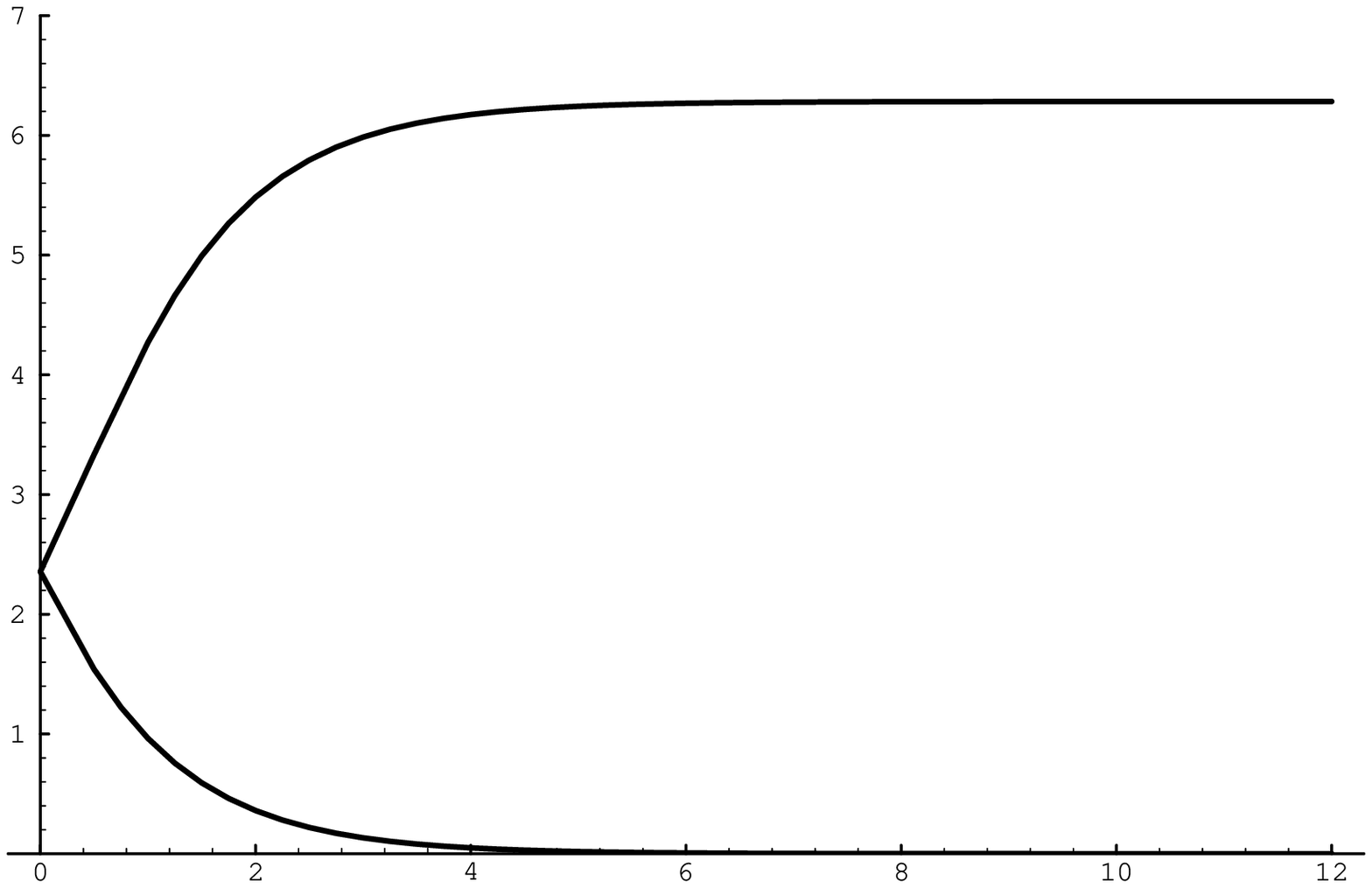}

\noindent Figure 2.2: a boundary bound state (boundary breather) for the values
 $\theta={i\pi\over 6}$, $\phi_0={3\pi\over 4}$ at four distinct 
 instants of time.
\addcontentsline{lof}{figure}{\protect\numberline{2.2}{
A boundary bound state (boundary breather).}}

\vspace{9cm}

\noindent Figure 2.3: two possible ground state configurations for
 $\phi_0={3\pi\over 4}$. 
 The configuration with asymptotic 
 behaviour $\phi\to 0$ at infinity has lower energy than the other one.
\addcontentsline{lof}{figure}{\protect\numberline{2.3}{
Two possible ground state configurations for
 $\phi_0={3\pi\over 4}$.}} 
\newpage
\topskip=0cm

\section{The semi-classical analysis}

\subsection{The exact boundary reflection matrix} 

Let us recall some of the results of \cite{GZ}. 
The {\it exact} (quantum) boundary S matrix
elements are 
\begin{eqnarray}
S_+^+(\th)\equiv
P_+(\th)&=&\cos\left[\xi-(t-1)i\th\right]
R(\th)\nonumber \\
S_-^-(\th)\equiv P_-(\th)&=&\cos\left[\xi+(t-1)i\th\right]R(\th),
\label{bsI}
\end{eqnarray}
and
\be
S_+^-   \equiv Q_+(\th)=S_-^+=Q_-(\th)=
{k\over 2}\sin\left[2(t-1)i\th\right]R(\th) \ , \label{bsII}
\ee
where  $\theta$ is the rapidity of an incoming particle.  The
parameter
t is defined by:
$$
t ~=~ {8\pi\over \beta^2}.
$$
The function $R(\th)$ decomposes as

\be
R(\th)=R_0(\th)R_1(\th), \label{bsR}
\ee
where $R_0$ is a normalization factor ensuring unitarity and crossing
symmetry that does not depend on the boundary conditions. The
dependence upon boundary conditions appears in $R_1$, which reads
\be
R_1(\th)={1\over\cos\xi}\sigma(\eta,\th)\sigma(i\Theta,\th). \label{bsRI}
\ee
Two of the four parameters  $k,\xi,\eta,\Theta$ are  independent, and
we have
the relations
\be
\cos\eta\cosh\Theta={1\over k}\cos\xi,\quad
\cos^2\eta+\cosh^2\Theta=1+{1\over k^2}. \label{rela}
\ee
These parameters are related with $M$ and $\varphi_0$ in an unknown
way. An expression
for the functions $R_0$ and  $\sigma$ involving infinite products of
$\Gamma$-functions is given below (see next chapter). There are also simple integral
representations:
\be
\sigma(\xi,\th)={\cos\xi\over\cos[\xi-i(t-1)\th]}
\exp\left\{\int_{-\infty}^\infty \quad{dx\over
x}\quad{\sinh(t-1+{2\xi\over\pi})x\over
2\cosh(t-1)x\sinh x}\ e^{i{2\over \pi}(t-1)\th x}\right\}, \label{intrep}
\ee
 and
\be
R_0(\th)=\exp\left\{-\int_{-\infty}^\infty\quad{dx\over
x}\quad{\sinh{3\over 2}(t-1)x\sinh({t\over 2}-1)x\over \sinh{x\over
2}\sinh 2(t-1)x}\ e^{i{2\over \pi}(t-1)\th x}\right\}.\label{intrepI}
\ee
One has  $\sigma(\xi,\th)=\sigma(-\xi,\th)$. The only difference
between the scattering of solitons and anti-solitons arises therefore
from the pre-factor $\cos[\xi\mp i(t-1)\th]$ in (\ref{bsI}).

For simplicity we consider only the limit in which
$M\rightarrow\infty$. One can identify
the corresponding values $k=0$ and $\Theta=\infty$ easily since,
at these values,
the topological charge 
$$Q={\beta\over 2\pi}\int_0^{\infty}
\partial_x\varphi dx$$
 is conserved
and therefore the amplitudes $Q_\pm$ must vanish. One has also
$\eta=\xi$ so
\be
R_1(\th)={1\over\cos\xi}\sigma(\xi,\th).\label{fix}
\ee

 Consider now the quantum boundary S matrix at the leading order in
${1\over\beta^2}$ as
$\beta\rightarrow 0$. The computation is most easily  done by  using
the integral representation given above, and evaluating the integrals
explicitly by the residue theorem. This provides convergent
expressions where the $\beta\rightarrow 0$ limit can be investigated
term by term. To get non-trivial results one must scale $\xi$ as
$$
{\beta^2\over 8\pi}\xi\rightarrow\hat{\xi}
$$
We  find then

\be
 P_{\pm}(\th)=\exp\left(\pm  {8i\pi
\hat{\xi}\kappa\over\beta^2} + {8i\pi
|\hat{\xi}|\kappa\over\beta^2}\right)
{\s(\th;0)[\s(2\th;0)]^{1/2}\over
[\s(\th;\hat{\xi})\s(\th;-\hat{\xi})]^{1/2}}, \label{semiI}
\ee
where
\be
\s(\theta; y)=\exp\left({8i\over\beta^2}\int_0^\th
dv\ln\tanh^2{v+iy\over 2}
\right), \label{sdef}
\ee
$\s(\th;0)$ being the semi-classical approximation to the bulk soliton-soliton
S-matrix \cite{JW} (in the following we denote it also as $\s(\th)$),
and $\kappa={\rm sign }\theta$ (in the following  we  assume  that
$\theta>0$).

Before discussing the relation between formula (\ref{semiI}) and the
classical
computations of the preceding section, it is useful to comment on the
 bound states of the quantum theory. Poles of the $R_0$
term are
associated with breathers and do not correspond to the boundary
(new)
bound states.  The latter correspond rather to poles  of the
$\sigma(\xi,\th)$ term which are located in the physical strip ${\rm
Im}\th\in [0,\pi/2]$. By inspection of the
 $\Gamma$-product expression  \cite{GZ}
one finds two families  of  poles of $\sigma$:
\begin{eqnarray}
\th_{n,l}^{(1)}&=&-i{\pi\over t-1}\left(n+{1\over
2}\right)\pm i{\xi\over t-1}-2il\pi\nonumber \\
\th_{n,l}^{(2)}&=&i{\pi\over t-1}\left(n+{1\over 2}\right)\pm
i{\xi\over t-1}+(2l+1)i\pi, \label{poles}
\end{eqnarray}
where $n,l$ are integers. Let us restrict to $\xi>0$. Then the only
physical poles are those that correspond to $+$ sign in $\th^{(1)}$
and $-$ sign in
$\th^{(2)}$. The  first pole that enters the physical strip
(from the bottom) is
$\th^{(1)}_{0,0}$ for $\xi\geq {\pi\over 2}$. The  number of poles of
$\sigma$ in the physical strip increases monotonically with $\xi$ for
$\xi$ small enough, and as $t$ gets large it becomes simply of the
form $\xi/\pi$. These poles are cancelled by the cosine term in $P_-$
and therefore appear only in the $P_+$ amplitude. As $\xi$ reaches
the value $\xi={4\pi^2\over\beta^2}$ the poles densely fill the
interval
$0<{\rm Im}\theta<{\pi\over 2}$ and a pole at $\theta=i\pi/2$
appears corresponding to the emission of a zero momentum soliton.  As
argued in \cite{GZ} this corresponds to a change in the ground state of
the system. For $0<\varphi_0<{\pi\over \beta}$ the ground state is
$\varphi\rightarrow 0$ at infinity but for
${\pi\over\beta}<\varphi_0<3{\pi\over\beta}$ it is
$\varphi\rightarrow {2\pi\over\beta}$ at infinity (see figure 2.3).
 Therefore the
value
$\xi={4\pi^2\over\beta^2}$ corresponds to $\varphi_0={\pi\over\beta}$
and it is the upper  physical value for $\xi$: as
$\varphi_0$ varies arbitrarily, $\xi$ never gets larger than
${4\pi^2\over\beta^2}$ and the set of poles $\th_{n,l}^{(2)}$ never
enters the physical strip.
Observe that there are bound states in the quantum theory when in the
classical
case the kink sitting in the middle and the incoming one are of
opposite topological charges in complete agreement with the
discussion of boundary
breathers in section 2.1.4.

\subsection{Classical time delay and quantum phase shift}

To establish the relation between (\ref{semiI}) and
(\ref{simpphasedel}), first recall \cite{JW} the general relation between the
quasi-classical scattering phase shift, $\delta(\theta)$,
and the classical phase shift, $a(\theta)$~:
\be
\delta(\theta)={\rm constant }+{m\over 2}\int_0^\theta
a(\eta)d\eta,\label{sergei}
\ee
Here, $m$ is the  classical mass of the particles involved in the
scattering.
Using the ``semi-classical Levinson
theorem'' to determine the constant in this formula we deduce,
using (\ref{simpphasedel}), the relation
\be
P_+(\th)\equiv e^{2i\delta(\th)}, \label{levin}
\ee
where
\be
2\delta(\theta)=2n_B\pi+{8\over\beta^2}
\int_0^{\theta}d\eta\log
{\tanh^2\eta \tanh^2{\eta\over 2} \over \tanh{1\over 2}(\eta+i{\beta
\varphi_0\over 2}) \tanh{1\over 2}(\eta-i{\beta\varphi_0\over 2})}, 
\label{levini}
\ee
and $n_B$ is the number of bound states. Since according to the
preceeding discussion $n_B={\xi\over\pi}={8\hat{\xi}\over\beta^2}$
for
$0<\xi<{4\pi^2\over\beta^2}$ as $\beta\rightarrow 0$ we see that
formula (\ref{semiI}) is in complete agreement with the classical phase
shift
(\ref{simpphasedel}) and that $\xi$ is proportional to $\phi_0$. In the
region $-{4\pi^2\over\beta^2}<\xi<0$ there are
no physical poles for $P_+$ and $n_B=0$.
 A similar  discussion can be carried out for
$\xi<0$ and $P_-$. As $\beta\rightarrow 0$ the number of physical
poles of $P_-$ varies as
$n_B=-{\xi\over\pi}=-{8\hat{\xi}\over\beta^2}$ and again we have
agreement with the semi-classical Levinson theorem.

 From comparison of (\ref{semiI}) and the classical phase shift we see that
$\xi$ and $\varphi_0$ are related linearly as
$\xi={4\pi\over\beta}\varphi_0$. This leads correctly to  the
emission of a zero momentum soliton with the ground state
degeneracy as discussed in \cite{GZ}. This linear relation  can hold only
for $|\xi|<{4\pi^2\over\beta^2}$. Beyond that value the ground state
changes. To  compare the quantum result with the classical phase
shift we must then  correlate appropriately  the value of $\phi$ at
infinity in the  latter.  The net effect is to
replace $\phi_0$ by $\phi_0-2\pi$. Eventually, the variation of $\xi$
with
$\varphi_0$ is
therefore
\be
\xi={4\pi\over\beta}\left(\varphi_0-{2\pi\over\beta}{\rm
Int }\left[{\beta\varphi_0\over 2\pi}+{1\over 2}\right]\right),\label{variat}
\ee
as illustrated in figure 2.4. It is very likely that (\ref{variat})
is exactly true for finite $\beta$ as well.

\begin{figure}[htbp]
\centering
\setlength{\unitlength}{0.0125in}
\begin{picture}(478,357)(0,-10)
\path(220,10)(220,320)
\path(222.000,312.000)(220.000,320.000)(218.000,312.000)
\path(10,160)(460,160)
\path(452.000,158.000)(460.000,160.000)(452.000,162.000)
\thicklines
\path(220,159)(281,281)
\path(220,160)(161,42)
\path(341,160)(400,278)
\path(341,159)(282,41)
\path(101,159)(161,279)
\path(100,158)(41,40)
\thinlines
\dashline{4.000}(161,278)(161,42)
\dashline{4.000}(282,280)(282,42)
\path(40,166)(40,154)
\path(400,166)(400,154)
\path(216,279)(224,279)
\put(391,178){$3\pi$}
\put(25,174){$-3\pi$}
\put(165,169){$-\pi$}
\put(291,170){$\pi$}
\put(230,280){${4\pi^2\over\beta^2}$}
\put(218,329){$\xi$}
\put(455,169){$\beta\varphi_0$}
\put(40,0){Figure 2.4: Variation of $\xi$ as a function of $\varphi_0$.}
\end{picture}
\end{figure}

\addcontentsline{lof}{figure}{\protect\numberline{2.4}
{Variation of $\xi$ as a function of $\varphi_0$.}}

We can finally recover (\ref{semiI}) without appealing to our knowledge of
quantum boundary bound states. For this one has to evaluate the
action for the classical configuration.
Following the discussion in \cite{JW} we
have,
\be
2\delta(\theta)=C(\varphi_0)+{8\over\beta^2}
\int_0^{\theta}d\eta\log
{\tanh^2\eta \tanh^2{\eta\over 2} \over \tanh{1\over 2}(\eta+i{\beta
\varphi_0\over 2}) \tanh{1\over 2}(\eta-i{\beta\varphi_0\over
2}) },\quad {\rm if} \quad
|\beta\varphi_0|<\pi, \label{phase}
\ee
where we used the soliton mass
$m={8\over\beta^2}$, and $\delta$ satisfies the differential equation
\be
\delta(\theta) ~-~ \tanh\theta ~\delta'(\theta)
~=~ \int_0^{+\infty}dt
\left( \int_0^{+\infty}dx\dot{\varphi}^2 -
8\sinh\theta\tanh\theta\right).\label{alta}
\ee
We restrict to the situation where we send in a kink, and $\varphi_0$
is
positive and  so there is an anti-kink at the origin (see figure 4).
In this case there are quantum boundary bound states.
It is difficult to
perform the double integral for the three-soliton solution
explicitly because of the cumbersome expression for the integrand (\ref{A1}).
One might hope that in order to find the $\theta$-independent piece of the
phase shift it is sufficient to evaluate both sides of (\ref{alta}) in the
limit as  $\theta\rightarrow +\infty$ or $\theta\to 0$. However both of these
limits do not seem to be helpful, because due to the non-uniform 
convergence of the
right hand side  of (\ref{alta})
 it is not allowed to interchange such a  limit   
with the integration. We evaluated the right hand side  of (\ref{alta})
 (with fixed $\th$)
numerically for
several different values of $\phi_0$   using  Mathematica. 
Combining (\ref{phase}) and (\ref{alta}),
 we obtain an estimate for $C(\varphi_0)$, in 
 agreement with  $C(\varphi_0)={8\pi\over\beta}\varphi_0$ 
with accuracy $0.1\%$. This result
was checked for different values of $\th$. We therefore obtain agreement 
with the semi-classical Levinson theorem.

\section{Remarks}

Although the method of images works nicely in the classical theory,
it
does not seem to extend to the quantum case: we have not been able to
recast the boundary S matrix of \cite{GZ} as a product of bulk S-matrix
elements. A
related phenomenon is the non-trivial structure of the boundary
S-matrix with
one-loop corrections (the semi-classical expressions being the tree
approximation). Recall
that in the bulk one has
$${\cal
S}_1(\theta;y)=\exp\left({8i\over\beta'^2}\int_0^\theta dv\ln\tanh^2
{v+iy\over 2}\right),
$$
where
$$
\beta'^2\equiv \beta^2{8\pi\over
8\pi-\beta^2}={8\pi\over t-1}.
$$
Scaling
$$
{\beta'^2\over 8\pi}\xi\rightarrow\hat{\xi}_1,
$$
one finds the following next-to-leading correction to the boundary
S-matrix in the Dirichlet problem:
$$
P_\pm(\theta)\approx\exp\left(\pm \kappa
{8i\pi\hat{\xi}_1\over
\beta'^2}+\kappa {8i\pi|\hat{\xi}_1|\over
\beta'^2}\right){{\cal S}_1(\theta)[{\cal S}_1(2\theta)]^{1/2}\over
[{\cal S}_1(\theta;-\hat{\xi}_1){\cal
S}_1(\theta;\hat{\xi}_1)]^{1/2}}
\left[\tanh\left({\theta\over 2}-i{\pi\over 4}\right)\right]^{1/2}.
$$
In addition to the usual replacement of $\beta^2$ by $\beta'^2$ we
see the appearance of an entirely new factor involving the
square-root of a
hyperbolic tangent.

The massless limit $m_0=0$ of the boundary sine-Gordon model (\ref{lagr})
was studied in \cite{FSW}, where, in particular,
 the massless classical  solutions were obtained. 

\chapter{Non-relativistic limit of the quantum sine-Gordon model
 with  Dirichlet boundary condition}

In the previous chapter we studied the quantum sine-Gordon model on a
half-line (\ref{lagr}) in the semi-classical limit $\beta\to 0$. The
purpose of this chapter is to investigate the non-relativistic, $\theta\to 0$,
limit of this model and to determine the quantum-mechanical potential
induced by the presence of boundary.
 We show that the generalized Calogero-Moser model with boundary potential
of the P\"oschl-Teller type describes the non-relativistic
limit of (\ref{lagr}).
 The discussion is based on \cite{KapSk}.

\section{Introduction}

When $\theta\to 0$, the energy and momentum of relativistic particles --
in our context solitons and anti-solitons -- degenerates to 
$$ E=M_s\cosh\theta \to M_s - {M_s\theta^2\over 2}, $$
$$ p=M_s\sinh\theta \to M_s\theta. $$
So, $\theta$ is a speed of particles measured in the units of
$c$ (the speed of light). 

Since such properties as integrability and factorized scattering survive 
in the above limit, we expect to get some quantum-mechanical {\it integrable}
model on a half-line. Moreover, since a large class of such models
based on the Lie-algrebraic classification is known \cite{PeOlsh}, it is natural
to look first for the appropriate candidate withing this class. Indeed, we
show that the boundary Calogero-Moser model with boundary potential
of the P\"oschl-Teller type, whose integrability was established
in \cite{IM,VanD}, describes the non-relativistic limit of (\ref{lagr}).

Let us give an idea how one can come up with the boundary Calogero-Moser model
by means of the heuristic arguments. It was known since long ago
\cite{ZZ,Kor} that the solitons and anti-solitons of the bulk sine-Gordon model
interact via $\sinh^{-2}(x)$ (particles of the same kind) and $\cosh^{-2}(x)$
(particles of different kinds) potentials in the non-relativistic limit.
\footnote{The quantum breathers correspond to the bound states in the 
$-\cosh^{-2}(x)$ potential.} Now, the question is to find out the form of the
potential induced by the boundary. But from the semi-classical picture
of the previous chapter we infer that the ``boundary'' can be represented
by a stationary kink or an anti-kink
and a moving ``mirror image''. So, we obtain the
 superposition of $\cosh^{-2}(x)$ and $\sinh^{-2}(x)$ potentials as a boundary
potential, i.e. the P\"oschl-Teller potential \cite{PT}. The $\sinh^{-2}(x)$
term is a hard-core reflecting potential, whereas the term $-\cosh^{-2}(x)$
is necessary to provide the existence of the boundary bound states.

\section{The exact quantum field theory solution}

In this chapter we consider the sine-Gordon model on a half-line (\ref{lagr}),
with the fixed value of field at the boundary: $\varphi(x=0,t)=
\varphi_0$, or $M=\infty$ in (\ref{lagr}). 
In \cite{GZ}, the quantum integrability and  the exact S-matrix were
conjectured for (\ref{lagr}). The boundary scattering matrix is diagonal and, 
according to \cite{GZ}, the reflection amplitude of the soliton $P_+$ 
(resp. $P_-$ for anti-soliton) reads:
\be
P_{\pm}(\theta)=\cos(\xi\pm \lambda u) R(u,\xi) = \cos(\xi\pm
\lambda u)R_0(u)R_1(u,\xi), \label{reflamp}
\ee
where $\theta=iu$ is the rapidity, $\xi={4\pi\over\beta}\varphi_0$
and $\lambda = {8\pi\over\beta^2}-1$;
 \begin{eqnarray}
R_{0}(u)&=&{\Gamma\left(1-{2\lambda u\over\pi}\right)
\Gamma\left(\lambda+{2\lambda u \over\pi}\right)
\over
\Gamma\left(1+{2\lambda u \over\pi}\right)
\Gamma\left(\lambda-{2\lambda u\over\pi}\right)}
\prod_{k=1}^{\infty}{{\Gamma\left(4\lambda k - {2\lambda u\over\pi}\right)
\over
\Gamma\left(4\lambda k+{2\lambda u \over\pi}\right)}}
\times \nonumber \\
&\times &{\Gamma\left(1 + 4\lambda k -{2\lambda u\over\pi}\right)
\Gamma\left(\lambda(4k+1)+{2\lambda u \over\pi}\right)
\Gamma\left(1+\lambda(4k-1)+{2\lambda u \over\pi}\right)
\over
 \Gamma\left(1+4\lambda k+{2\lambda u \over\pi}\right)
\Gamma\left(\lambda(4k+1) -{2\lambda u\over\pi}\right)
\Gamma\left(1 + \lambda(4k-1) -{2\lambda u\over\pi}\right)} \label{EqnI}
\end{eqnarray}
 \begin{eqnarray}
R_1(u,\xi)&=&{1\over\pi}\prod_{l=0}^{\infty}{{
\Gamma\left({1\over 2}+2l\lambda+{-\xi+u\lambda\over\pi}\right)
\Gamma\left({1\over 2}+2l\lambda+{\xi+u\lambda\over\pi}\right)
\over
\Gamma\left({1\over 2}+2l\lambda+\lambda+{-\xi+u\lambda\over\pi}\right)
\Gamma\left({1\over 2}+2l\lambda+\lambda+{\xi+u\lambda\over\pi}\right)
}}\times \nonumber \\
&\times &{\Gamma\left({1\over
2}+2l\lambda+\lambda+{\xi-u\lambda\over\pi}\right)
\Gamma\left({1\over 2}+2l\lambda+\lambda-{\xi+u\lambda\over\pi}\right)
\over
\Gamma\left({1\over 2}+2l\lambda+2\lambda+{\xi-u\lambda\over\pi}\right)
\Gamma\left({1\over 2}+2l\lambda+2\lambda-{\xi+u\lambda\over\pi}\right)}
\label{EqnII}
\end{eqnarray}
The  poles of $P_{\pm}$ located in the physical domain $0<u<\pi/2$ at
$u_n=\pm{\xi\over\lambda}-{2n+1\over 2\lambda}\pi$
correspond to the ``boundary'' bound states of the theory. The latter exist in
the
soliton (resp. anti-soliton) scattering channel if $\xi>0$ (resp. $\xi<0$), and
their energy is $E_n=M_s\cos u_n$. Note that the
``physical'' values of the parameter $\xi$ are bounded:
$|\xi|<4\pi^2/\beta^2$ (see previous chapter).
In  the semi-classical limit
of the quantum field theory (\ref{lagr}), $\beta\to 0$,  the
principal (``tree'') approximation to the amplitudes
(\ref{reflamp}) has the form (\ref{semiI}).

\section{The Calogero-Moser model on a half-line}

Our purpose is
to show that the non-relativistic dynamics of quantum sine-Gordon solitons
in the presence of a boundary is described by the generalized Calogero-Moser
Hamiltonian:
\begin{eqnarray}
\hat{H} =~-~{1\over 2M_s} \sum_{i=1}^{N} {d^2\over dx_i^2}~-~{1\over 2M_s}
 \sum_{j=1}^{M} {d^2\over dy_j^2}&+&\sum_{i<i'}^{N}\left( V_{AA}\left(
 x_i-x_{i'}\right) +V_{AA}\left( x_i+x_i'\right)\right) \nonumber \\
+~\sum_{j<j'}^{M}\left( V_{AA}\left( y_j-y_{j'}\right) +V_{AA}\left( y_j+y_{j'}
\right)\right) &+&\sum_{i=1}^{N}\sum_{j=1}^{M}\left( V_{A\bar{A}}\left( x_i-y_j
\right)+V_{A\bar{A}}\left( x_i+y_j\right)\right)
\nonumber \\
 &+&\sum_{i=1}^{N} W_A\left( x_i\right) ~+~
\sum_{j=1}^{M} W_{\bar{A}}\left( y_j\right) \label{cmoser}
\end{eqnarray}
Here $V_{AA}$ and $V_{A\bar{A}}$ are bulk nonrelativistic potentials obtained
long ago in \cite{ZZ, Kor}:
\be
V_{AA}\left( x\right)~=~{\alpha_0^2\over{M_s}}
{\rho(\rho -1)\over\sinh^2\alpha_0 x} ,\quad
V_{A\bar{A}}\left( x\right) ~=~-{\alpha_0^2\over{M_s}}{\rho(\rho-1)\over
\cosh^2\alpha_0 x},
\label{bulkV}
\ee
with
\be
\rho =  {8\pi \over{\beta^2}} , \label{idI}
\ee
\be
\alpha_0 = {m_0\over 2} , \label{idII}
\ee
 and $W_{A}$ and $W_{\bar{A}}$ are boundary potentials of the \ptp \cite{PT}
 type
\begin{eqnarray}
 W_A\left( x\right)&=&{\alpha_0^2\over{2M_s}}\left(
{\mu(\mu -1)\over\sinh^2\alpha_0 x} - {\nu(\nu-1)\over\cosh^2\alpha_0
x}\right),
 \nonumber \\
  W_{\bar{A}}\left( x\right)&=&{\alpha_0^2\over{2M_s}}\left({\nu(\nu
 -1)\over\sinh^2\alpha_0 x} -{\mu(\mu-1)\over\cosh^2\alpha_0 x}\right),\qquad
\mu >1, \nu >1 . \label{ptell}
\end{eqnarray}

Let us comment on the properties of the one-particle Schr\"odinger
 equation with the \ptp potential $W_A(x)$.
The energy of the bound states, which appear when $\nu>\mu+1$, is given by
$E_n=-{\alpha_0^2\over 2M_s}(\nu-\mu-1-2n)^2$, where $n=0,1,2...$ For a fixed
value of $\nu-\mu$ there are in total $\left[{\nu-\mu-1\over 2}\right]$  bound
states. The reflection coefficient, which is a pure phase, can be obtained
to be equal to
\be
S_A(k)={
\Gamma\left({ik\over\alpha_0}\right)
\Gamma\left({1\over 2}+{\mu-\nu\over 2}-{ik\over 2\alpha_0}\right)
\Gamma\left({\mu+\nu\over 2}-{ik\over 2\alpha_0}\right)
\over
\Gamma\left(-{ik\over\alpha_0}\right)
\Gamma\left({1\over 2}+{\mu-\nu\over 2}+{ik\over 2\alpha_0}\right)
\Gamma\left({\mu+\nu\over 2}+{ik\over 2\alpha_0}\right) } \label{smatr}
\ee
This expression has ``physical'' poles  on
the upper imaginary half-axis
in the complex momentum plane which correspond to the bound states. Besides, it
has poles at the points $k_n=(1+n)\alpha_0$
that come from the first $\Gamma$-function in the
numerator of (\ref{smatr}). The latter set of poles is infinite and does not
correspond to any bound states of the theory. The $S$-matrix for the potential
$W_{\bar{A}}$ can be obtained from (\ref{smatr}) by the substitution $\mu
\leftrightarrow \nu$. One can see that $S_A$ and $S_{\bar{A}}$ satisfy
the boundary Yang-Baxter equation of \cite{GZ}:
\be
S_A \cos \left({\pi\over 2}(\nu-\mu)-\lambda u\right) = S_{\bar{A}}
 \cos\left({\pi\over 2}(\nu-\mu)+\lambda u\right).      \label{noref}
\ee

\section{Relativistic vs non-relativistic \\ integrable models}

To establish the equivalence we will show 
in particular that the $S$-matrices of the quantum
 sine-Gordon theory and the model (\ref{cmoser}) coincide in the appropriate
 limit. The system (\ref{cmoser}) is integrable both at the classical and
quantum levels \cite{IM,VanD}. To see this one takes the hyperbolic-type
 Calogero-Moser Hamiltonian for $N+M$
 particles based on the $BC_{N+M}$ root system \cite{PeOlsh}
 and shifts the  coordinates of the particles $N+1,...,N+M$ by $i\pi/ 2$.
The result is (\ref{cmoser}). Integrability means that the system admits a Lax
 representation and has $N+M$ integrals in involution. Moreover, since as
 $t\to\pm\infty$ these integrals reduce asymptotically to symmetric
 polynomials in particles' momenta, one can use the standard argument
\cite{Kul}
 to show that the $S$-matrix is factorized. A small modification arises due to
 the presence of the boundary; namely, one can consider the particles'
 collisions both very far from the boundary where the problem is reduced to the
 bulk one, and near the boundary where  the colliding particles have enough
time
to reflect and go to $x= +\infty$. In the first case factorization
gives the nonrelativistic Yang-Baxter equation for the bulk $S$-matrix
 \cite{ZZ}, while in the second case we get exactly the boundary Yang-Baxter
equation of \cite{GZ}, which in the Dirichlet case allows to express the
boundary $S$-matrix of the anti-kink through that of the kink (\ref{noref}). The
 unitarity requires the latter to be a pure phase, but otherwise leaves it
undetermined. So in both theories the $S$-matrix is factorized and fully
determined by the bulk two-particle $S$-matrix and the boundary $S$-matrix.
Thus to establish the equivalence it is sufficient to show that these
 $S$-matrices coincide when the nonrelativistic limit is taken.

Note that the translational invariance of the Hamiltonian (\ref{cmoser})
is broken not only
by the boundary potentials, but also by the interaction of particles with their
mirror images. This is very natural from the point of view of the underlying
sine-Gordon theory, since, as was shown above, that the one soliton
 problem on a half-line is equivalent to the three-soliton bulk problem, with
one of the particles staying at $x=0$, and the other two being ``generalized
mirror images'' of each other. The analogy becomes exact if
 we take $\varphi_0=0$. Then the ordinary method of images works, and it is
 obvious that the system of $N+M$ solitons on a half-line is equivalent to the
 system of $2(N+M)$ solitons on a line with symmetric initial conditions. Hence
 the corresponding nonrelativistic Hamiltonian can be obtained from the known
\cite{ZZ,Kor} nonrelativistic bulk Hamiltonian. One can easily see that the
result is just (\ref{cmoser}) with $\mu=\nu\, ,\, \mu(\mu-1)=\rho(\rho-1)/4$.

We will show below that $\mu$ and $\nu$ are related to the parameter $\xi$
of the sine-Gordon model (\ref{lagr}) as follows:
\be
{\nu - \mu \over 2} = {\xi\over\pi} . \label{idIII}
\ee

\section{Taking the non-relativistic limit}

The nonrelativistic limit of (\ref{reflamp}) corresponds to the values
 $\theta\ll 1 .$ Simultaneously we must take the limit $\beta\ll 1,$ so that
 $M_s={{8m_0}\over\beta^2}\gg m_0$ (otherwise the $S$-matrix becomes $1$ and we
 do not get anything interesting.) In general, these two limits are to be
taken without any further assumptions on the relative magnitude of $\theta$
with respect to $\beta$. However, it is worth to note that the region
$\beta^2\ll\theta\ll 1$ corresponds to the quasiclassical limit
${k\over\alpha_0}\gg 1$, where
\be
k\equiv M_s\theta \label{bydef}
\ee
According to (\ref{idII}) ${k\over{\alpha_0}} ={
16\theta\over{\beta^2}}$, which in general is not necessarily large. 
  In what follows we assume that
$\xi$ is positive and $\beta\varphi_0\sim 1$, so that $\xi$ scales as
 $1/\beta^2$. Then $P_+$ has poles corresponding to the boundary bound states,
and the energy of the bound states lying close to the edge of the continuous
spectrum in the theory (\ref{lagr}) becomes
$E_n\simeq M_s - {m_0\pi\over 8\lambda}\left({2\xi\over\pi}-1-2n\right)^2$.
 $P_-$ does not have poles in the physical region. One can easily see then that
 (\ref{ptell}),(\ref{smatr}) describe correctly the spectrum of the boundary
 bound states provided  that equations (\ref{idII}),(\ref{idIII}) are
 fulfilled. To complete the identification of the boundary $S$-matrices we have
 to compare the phase shifts. Since $\xi$ scales as $1/\beta^2$, by virtue
 of (\ref{idIII}) the expression (\ref{smatr}) can be rewritten as
 $S_{NR}(k,\nu-\mu) f(\theta )$, where
\be
S_{NR}(k,\nu-\mu)={
\Gamma\left({ik\over\alpha_0}\right)
\Gamma\left({\mu-\nu\over 2}-{ik\over 2\alpha_0}\right)
\over
\Gamma\left(-{ik\over\alpha_0}\right)
\Gamma\left({\mu-\nu\over 2}+{ik\over 2\alpha_0}\right) } \label{slim}
\ee
is meromorphic and contains the poles located in the arbitrarily small
 neighbourhood of $\theta=0$ as $\beta\to 0$, whereas the factor $f(\theta)$
can
 be expanded into the power series $f(\theta)=1+\sum_{l=1}^{\infty}a_l\theta^l$
with all the coefficients and a radius of convergence $\sim 1$ as $\beta
\rightarrow 0$. In the same limit $P_{\pm}$ can be
factorized analogously: $P_{\pm}=S_{NR}(k,\pm 2\xi/\pi) f_{\pm}(\theta)$ with
$f_{\pm}$ admitting expansions of the form
 $f_{\pm}(\theta)=1+\sum_{l=1}^{\infty}a_l^{\pm}\theta^l$ with all the
coefficients and the radius  of convergence $\sim 1$ in the limit $\beta
\rightarrow 0$. Therefore the boundary $S$-matrices agree when
 $\theta\ll 1$, and $S_{NR}$ represents the nonrelativistic limit of the
boundary $S$-matrix of the sine-Gordon theory.
One can check that the same statements are true for the bulk two-particle
sine-Gordon $S$-matrix of \cite{ZZ} and the $S$-matrix of the particles
 interacting via the potentials (\ref{bulkV}), provided that (\ref{idI}),
(\ref{idII}) are satisfied 
(this result was first established in \cite{Kor} in
the quasiclassical approximation and later confirmed in \cite{ZZ}; note that
 our approach allows to give an exact sense to the statement that the
 nonrelativistic limit of the bulk sine-Gordon theory is the hyperbolic-type
Calogero-Moser model.)
 Thus the equivalence of (\ref{cmoser}) and the
nonrelativistic limit of (\ref{lagr}) is established.

\section{Remarks}

It is instructive
to compare also the combined nonrelativistic/quasiclassical limit of
the S-matrix of  (\ref{lagr})
 with the \ptp S-matrix in the regime  $\beta^2\ll\theta\ll 1$
 (i.e. $k\gg m_0$) without appealing to the exact formula  (\ref{reflamp}),
similar to how it was first done in \cite{Kor} for the bulk
 soliton scattering. This means that one should first take the limit
 $\beta\to 0$ and then, using (\ref{semiI}), pass to the limit
$\theta\to 0$. Expanding the integrals in (\ref{semiI}) and using the
 asymptotic formulas for the $\Gamma$-functions in (\ref{smatr}) we get
 for  (\ref{semiI}) and (\ref{smatr}) in the principal order the
following result:
\be
P_+(k)=e^{2i\xi sign(k)+{4ik\over m_0}\ln{k\beta^2\over m_0}},\qquad
P_-(k)=e^{{4ik\over m_0}\ln{k\beta^2\over m_0}}, \label{asymp}
\ee
once again confirming the equivalence of the two theories.
Note that it is impossible to determine $\mu$ and $\nu$ separately, since
in our limit the boundary $S$-matrices in both theories depend only on the
difference $\mu-\nu$.

The identification of the
 nonrelativistic limit in the case of the most general integrable boundary
condition (\ref{intbcs}) requires a nontrivial generalization of the
 Calogero-Moser Hamiltonians. Indeed, if in (\ref{lagr}) $M<\infty$, then the
 boundary $S$-matrix does not conserve the topological charge, and we are not
 aware of any integrable nonrelativistic model which allows such a process.

\chapter{Boundary bound states and boundary bootstrap}

In the previous chapters we discussed classical, semi-classical
and non-relativistic limits of the boundary sine-Gordon model. 
Now we address the {\it exact} quantum field theory solution of this model,
which we obtain by means of the Bethe ansatz technique. Among other things,
this solution allows to re-derive the boundary reflection matrices of \cite{GZ},
(\ref{reflamp})-(\ref{EqnII}),
and to relate them to the physical parameters in the Hamiltonian \cite{FS}.
The present chapter includes a complete study of boundary bound states
and related boundary S-matrices for the sine-Gordon model with Dirichlet
boundary condition. Our analysis is based on the solution of the
boundary bootstrap equations, representing the integrability constraints,
together with the explicit Bethe ansatz solution of the inhomogeneous XXZ model
in a boundary magnetic field -- a
lattice regularization of the boundary sine-Gordon model.
 We identify boundary bound states
with new {\it boundary strings} in the Bethe ansatz.
This chapter is based on \cite{Boot:SS}.

\section{Introduction}

In the seminal work \cite{GZ} it appeared clearly that the boundary 
sine-Gordon model
presents an extremely  rich structure of boundary bound states, which
was further explored  in \cite{Ghosh}. Our first purpose here is to
study this structure thoroughly in the particular case of Dirichlet
boundary conditions, that is
the model
\be
{\cal L}_{SG}={1\over 2} \int_0^{\infty}\left[(\partial_t\varphi)^2-
(\partial_x\varphi)^2 + {m_0^2\over\beta^2}\cos\beta\varphi\right]dx \label{SG}
\ee
with a fixed value of the field at the boundary: $\varphi(x=0,t)=\varphi_0$.

Also, the consideration of boundary problems poses interesting
challenges from the point of view of lattice models, here lattice
regularizations of (\ref{SG}). In \cite{FS} and also in \cite{GMN}
 it
was shown
in particular how to derive the S-matrices of \cite{GZ} from the Bethe
ansatz. Our second purpose is to complete these studies by
investigating which new types of strings correspond to boundary bound
states, and by deriving as well the set of S-matrices necessary to
close the bootstrap.
Observe that   lattice regularizations
are useful to define what one means
by creating a bound state
at the boundary. Indeed, some bound states (showing up as the pols of
S-matrices) have no
straightforward interpretation, and although they are easy to study
formally using the Yang Baxter equation and the bootstrap,
their meaning in the field theory is unclear.

In section 4.2 we consider the bootstrap problem directly in the
continuum theory. We identify boundary bound states and  we compute the related
boundary $S$ matrices.
In section 4.3 we write the Bethe ansatz equations for
the inhomogeneous six-vertex model with boundary magnetic field,
which is believed \cite{FS} to be a regularization of (\ref{SG}). We  show that
these equations are also the bare equations for the Thirring model
with $U(1)$-preserving boundary interaction,
which is the fermionized version of (\ref{SG}). In section 4.4 we discuss in
details new solutions (``boundary strings'') to the Bethe ansatz
equations made possible by the appearance of boundary terms. In
section 4.5 we study the physical properties of the model, in
particular the masses and S-matrices corresponding to these boundary
strings, and we partially complete the identification with the bootstrap
results of section 4.2. Several remarks, in particular formula for the
boundary energy of the
boundary sine-Gordon model, are collected in sections 4.6-4.7.

\section{Boundary bootstrap}

\subsection{Solving the boundary bootstrap equations}

The S-matrices for the scattering of a soliton  ($P^+$)
and an  anti-soliton ($P^-$) on the ground state
$|0\rangle_B$ of the sine-Gordon model with  Dirichlet
boundary conditions  (\ref{SG})
  read according to \cite{GZ}:
\be
{P^{\pm}(\th)= \cos(\xi\pm\lambda u)R_0(u)R_1(u,\xi),} \label{smatrx}
\ee
where $\th=iu$ is the rapidity, $\xi=4\pi\varphi_0/\beta$ and
$\lambda=8\pi/\beta^2-1$. The explicit form of $R_0, R_1$ is
given
in (\ref{EqnI})-(\ref{EqnII}).
Since the theory is invariant under the simultaneous transformations
$\xi\rightarrow
-\xi$,
and soliton$\rightarrow$anti-soliton, we choose hereafter
$\xi$ to be a generic number in the interval $0<\xi<4\pi^2/\beta^2$
(see the discussion in chapter 2 about the value of the upper bound).

The function $R_0$ contains poles in the physical strip
$0<{\rm Im}\theta<\pi/2$ located at $u=n\pi/
2\lambda, n=1,2,\ldots<\lambda$. These poles come from the
corresponding breather pole in the soliton-antisoliton bulk scattering,
and should not be interpreted as boundary bound states \cite{GZ}.

When $\xi>\pi/2$, the function $P^+(\theta)$ has additional poles in the
physical
strip, located at  $u=v_n$ with 
\be
{0<v_n={\xi\over\lambda}-{2n+1\over 2\lambda}\pi<{\pi\over
2},} \label{pols}
\ee
$(n=0,1,2,\ldots)$ corresponding to  a first  set of boundary bound
states which we denote by $|\beta_n\rangle$, with
masses
\be
{m_n=m\cos v_n=m\cos\left({\xi\over\lambda}-{2n+1\over
2\lambda}\pi\right),} \label{addi}
\ee
where $m$ is the soliton mass. These bound states are easy to interpret
\cite{GZ,SSW}.
For $0<\varphi_0<\pi/\beta$ the ground state of the theory is
characterized by the asymptotic behaviour $\varphi\to 0$ as $x\to\infty$, but
other states, whose energy differs from the ground state by a boundary
 term only,
can be obtained with $\varphi\to \{\hbox{ a multiple of }{2\pi/\beta}\}$
as $x\to\infty$.
Since the $\beta_n$ appear as bound states for soliton scattering,
they all have the same topological charge as the soliton, which
we take equal to unity by convention, so they are all associated with the same
classical solution,   a
 soliton
sitting next to the boundary and performing a motion
periodic in time (``breathing"), with $\varphi(x=0)=\varphi_0$ and
$\varphi\to{2\pi/\beta}$ as $x\to\infty$ (see section 2.2.4).

To deduce the scattering matrices on the boundary bound states we use
the {\it boundary bootstrap equations} as given in \cite{GZ}. We assume that
these S-matrices are diagonal, which is true if all the boundary
bound states
have different energies.  In this case the bootstrap equations read:
\be
{R^b_{\beta}(\theta)=\sum_{c,d}R^d_{\alpha}(\theta)
S_{cd}^{ab}(\theta+iv^{\beta}_{\alpha a})
S_{ba}^{dc}(\theta-iv^{\beta}_{\alpha a}).} \label{BBE}
\ee
These equations allow us to find the scattering matrix of any
particle
$b$  on
the boundary bound state $\beta$ provided that  the latter appears as
a virtual state
in the scattering  of the particle  $a$    on the boundary state
$\alpha$. 
\begin{figure}
\epsfxsize=85truemm
\centerline{\epsfbox{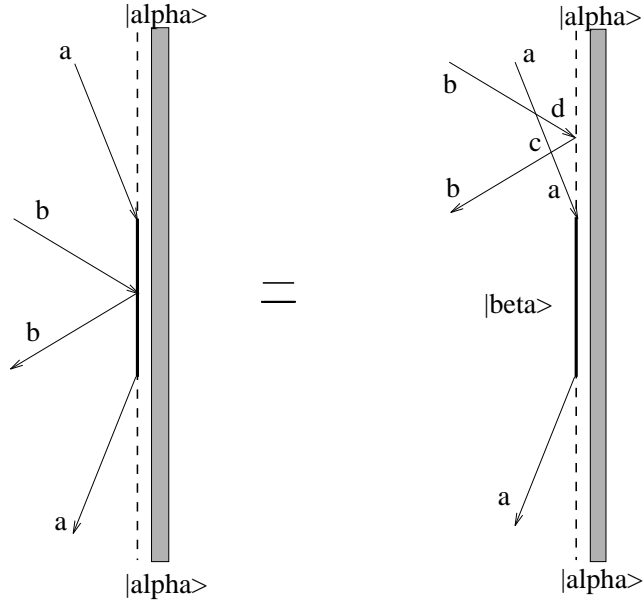}}
\caption{
 A diagrammatic representation of the bootstrap equations (\ref{BBE}).}
\end{figure}


\noindent The masses of the corresponding boundary states
are related through
\be
{m_{\beta}=m_{\alpha}+m_a\cos v^{\beta}_{\alpha a},} \label{mass}
\ee
where $iv^{\beta}_{\alpha a}$ denotes the position of the pole,
corresponding
to  the bound state $\beta$.

Let $\beta_n$ stand for the $n$-th boundary bound state
corresponding to
the pole $v_n$  in $P^+$ (\ref{pols}). Then (\ref{BBE}) gives:
\be
{P^+_{\beta_n}(\theta)=P^+(\theta)a(\theta-iv_n)a(\theta+iv_n)
,} \label{EQI}
\ee
\be
{P^-_{\beta_n}(\theta)=b(\theta-iv_n)b(\theta+iv_n)P^-(\theta
)
+ c(\theta-iv_n)c(\theta+iv_n)P^+(\theta),} \label{EQII}
\ee
where the well known bulk S-matrix elements
$a(\theta)=S_{++}^{++}=S_{--}^{--}$
(kink-kink scattering),
 $b(\theta)=S_{+-}^{+-}=S_{-+}^{-+}$  (kink-anti-kink transmission)
and $c(\theta)=S_{+-}^{-+}=S_{-+}^{+-}$ (kink-anti-kink reflection)
can be  found in \cite{ZZ}.

It is easy to  check that the matrix elements (\ref{EQI})-(\ref{EQII}) satisfy
general
requirements  for the boundary S-matrices, such as boundary unitarity
and boundary crossing-symmetry conditions \cite{GZ}, e.g.
$$ P^-_{\beta_n}({i\pi\over
2}-\theta)=b(2\theta)P^+_{\beta_n}({i\pi\over 2}+
\theta) + c(2\theta)P^-_{\beta_n}({i\pi\over 2}+\theta) ,$$
$$ P^{\pm}_{\beta_n}(\theta)P^{\pm}_{\beta_n}(-\theta)=1.$$
Finally  we obtain from (\ref{EQI})-(\ref{EQII}) by direct calculation:
\be
{P^+_{\beta_n}(\theta)={\cos(\xi-\lambda\pi-i\lambda\theta)
\over\cos(\xi-\lambda\pi+i\lambda\theta)}P^-_{\beta_n}(\theta).}\label{BYBE}
\ee
Hence the boundary Yang Baxter equation  is satisfied since the ratio of the
above two
amplitudes
has a form similar to (\ref{smatrx}) with $\xi\rightarrow\xi-\lambda\pi$,
$\xi$ being a free parameter.

The analytic structure of $P^{\pm}_{\beta_n}(\theta)$ is as follows.
The function $P^+_{\beta_n}(\theta)$ has
simple poles in the physical strip located at  $u={\xi\over\lambda}
+{2N+1\over 2\lambda}\pi$ ,  $N=0,1,2...$, and at $u=v_n$.
It has
double poles at  $u=iv_n+i{k\pi\over\lambda}$, $k=1,2,...n$.
The function $P^-_{\beta_n}(\theta)$  possesses
in the physical strip the same singularities as
$P^+_{\beta_n}(\theta)$
plus the set of simple poles at $u=w_N$ with
\be
{w_{N}=\pi - {\xi\over\lambda} - {2N-1
\over 2\lambda}\pi, \qquad \lambda+{1\over 2}-{\xi\over\pi}
>N>{\lambda+1\over 2} - {\xi\over\pi}. }\label{newpols}
\ee

Interpreting  these poles in terms of boundary bound states
requires some care. First, due to the relation (\ref{BBE}), one sees that if $\beta$
appears as a boundary bound state for scattering of $a$ on $\alpha$, then
the poles of the amplitude for scattering of $b$ on $\alpha$ are also
in general  poles of the amplitude for scattering of $b$ on $\beta$. It seems
unlikely that these poles correspond to new bound states, although in our case
they would have a natural physical meaning, for example one could try to
associate them with classical solutions where $\varphi\to{4\pi/\beta}$ as
$x\to\infty$. Indeed there are strong constraints coming
from statistics that we should not forget. For instance at the
free fermion point $\beta^2=4\pi$, there is a bound state $\beta_1$, but
although $P^+_{\beta_1}$ has again a pole at $\beta_1$, the state of mass
$2m_{\beta_1}$ is not allowed from Pauli exclusion principle, as can easily
be checked on the direct solution of the model (see below section 4.3.3).
Therefore we take the point
of view that the poles already present in the scattering on
 an ``empty boundary'' are ``redundant''.
The only poles we interpret as 
 new boundary bound states are (\ref{newpols}) (the additional poles
 in $P_{\beta_n}^+$ are related to them by crossing). We denote these
boundary bound states $|\delta_{n,N}\rangle$, and their masses,
according to (\ref{mass}) and (\ref{addi}), are given by
\bea
m_{n,N}&=&m(\cos v_n + \cos w_{N})
=m\cos\left({\xi\over\lambda}-{2n+1\over 2\lambda}\pi\right)-
m\cos\left({\xi\over\lambda}+{2N-1\over2\lambda}\pi\right) \nonumber \\
&=&m^b_{N+n}\sin\left({\xi\over\lambda}+{N-n-1\over
2\lambda}\pi\right), \label{massnew}
\eea
where $m^b_p=2m\sin\left({p\pi\over 2\lambda}\right)$ is the mass of
the $p$-th breather, $p=1,2,\ldots<\lambda$.

To understand
the physical meaning of these new boundary bound states it is helpful
to consider the semi-classical limit $\lambda\to\infty$ of the
sine-Gordon
model.  As discussed above,
the boundary bound states $\beta_n$, corresponding to (\ref{pols}), are
associated with solutions where a soliton is sitting next to the boundary and
``breathing''. An
incoming anti-soliton can couple to this soliton, and
together
they form  a breather sitting next to the boundary and performing again some
(rather complicated) motion periodic in time\footnote{To compute this
solution explicitly requires using a  bulk five-soliton
configuration \cite{SSW}, an expression which is very cumbersome.}. The 
WKB quantization
of this solution \cite{DHN}
shoud lead to $|\delta_{n,N}\rangle$. The
topological charge of the states $|\delta_{n,N}\rangle$
 is equal to 0 in our units,
or, equivalently, to the charge of a free breather in the theory (\ref{SG}).

One can in principle continue to solve the bootstrap equations (\ref{BBE})
recursively. For example, for the scattering of solitons or
antisolitons on the boundary bound
states
$|\delta_{n,N}\rangle$ (\ref{newpols}) one obtains the following S-matrices:
\be
{P^-_{\delta_{n,N}}(\th)=P^-_{\beta_n}(\th)a(\th-iw_N)a(\th+iw_N),}\label{FIN}
\ee
\be
{P^+_{\delta_{n,N}}={\cos(\xi-i\lambda\th)\over\cos(\xi+i\lambda\th)}
P^-_{\delta_{n,N}}.} \label{FINI}
\ee
$P^-_{\delta_{n,N}}$ has only one simple pole in the physical strip at
$u=w_N$,  while
$P^+_{\delta_{n,N}}$ has also simple poles at $u=v_k$, $k=n+1,
n+2,...,[{\xi\over
\lambda}-{1\over 2}]$. According to  the discussion below (\ref{newpols}), we
do not consider these poles as associated with new boundary bound states.
Therefore, the boundary bootstrap is closed
for
solitons and antisolitons in the sense
that  further recursion will not generate new boundary bound
states.

So far we have obtained  two sets of boundary bound states (\ref{addi}) and
(\ref{massnew})  by
considering all the poles in the physical strip of  amplitudes for
scattering a soliton and anti-soliton on a boundary with
 or without a boundary bound state.
Of course we should also consider the scattering of breathers off the
boundary.
The scattering of breathers on the ``empty'' boundary was studied in 
\cite{Ghosh},
and we
refer the reader to this work for the explicit boundary S-matrices. By
interpreting  the poles
of the amplitudes in \cite{Ghosh}  as boundary bound states, we find a
spectrum of masses that  look like (\ref{massnew}) 
but with a slightly different range of parameters. Considering then
scattering of breathers off a boundary with a bound state does not give rise to
any new poles beside (\ref{pols}) and (\ref{newpols}), with in the latter case
 an extended
range of values of $N$ (for simplicity we do not
give the relevant boundary S-matrices here). Therefore the complete boundary
bootstrap
is closed in principle.

\subsection{Integral representations of various S-matrices}

For comparison with results obtained from regularizations of the
sine-Gordon model it is useful to write integral representations of
the  boundary $S$-matrices
(\ref{smatrx}), (\ref{EQI}) and (\ref{EQII}) using the well-known formula
\be
{\log\Gamma(z)=\int_0^{\infty}{dx\over x}e^{-x}
\left[z-1+{e^{-(z-1)x}-1\over 1-e^{-x}}\right], \qquad {\rm Re} z >
0. } \label{main}
\ee
Suppose first that $1<2\xi/\pi<\lambda+1$ and denote
\be
{n_{\ast}=\left[{\xi\over\pi}-{1\over 2}\right],}\label{Nn}
\ee
where the square brackets mean the integer part of the number.
For such values of $\xi$
there are $n_{\ast}+1$ poles (\ref{pols}) in the physical strip, i.e. the
spectrum
of excitations contains boundary bound states. Correspondingly, there
is a
finite  number of $\Gamma$-functions in (\ref{smatrx}), (\ref{EQI}), 
(\ref{EQII}) whose
arguments have negative real part so that formula (\ref{main}) is not
applicable.
Treating such  $\Gamma$-functions separately, we obtain the following
results:
\bea
-i{d\over d\theta}\log \left[{P^+(\theta)\over R_0(\theta)}
\right]&=&{2\lambda\over\pi}
\int_{-\infty}^{+\infty}dx \cos\left({2\lambda\theta
x\over\pi}\right) \label{repI} \\
&\times&\left[{\sinh(2\xi/\pi-2n_{\ast}-2)x\over\sinh x} +
{\sinh(\lambda-2\xi/\pi)x\over 2\sinh x \cosh \lambda x}\right], \nonumber
\eea
\bea
-i{d\over d\theta}\log \left[ {P^+_{\beta_n}
(\theta)\over R_0(\theta)}\right]& = &{2\lambda\over\pi}
\int_{-\infty}^{+\infty}dx \cos\left({2\lambda\theta
x\over\pi}\right)
\label{repII} \\
&\times&{\sinh(\lambda-2\xi/\pi)x - 2\cosh x
\sinh(\lambda+1+2n-2\xi/\pi)x
\over 2\sinh x \cosh \lambda x},  \nonumber
\eea
\bea
-i{d\over d\theta}\log \left[{P^-_{\beta_n}(\theta)\over
R_0(\theta)}\right]&=&{2\lambda\over\pi}
\int_{-\infty}^{+\infty}dx \cos\left( {2\lambda\theta
x\over\pi}\right)
\left[ {\sinh(2n_{\ast}+2-2\xi/\pi)x\over\sinh x} \right. \label{repIII} \\
&+&\left.{\sinh(\lambda-2\xi/\pi)x - 2\cosh x
\sinh(\lambda+1+2n-2\xi/\pi)x
\over 2\sinh x \cosh \lambda x}
\right] .\nonumber 
\eea
In the derivation of analogous representation for $P^-$ there are no
subtleties
because the ``dangerous" $\Gamma$-functions cancel. We get
\be
{-i{d\over d\theta}\log \left[{P^-(\theta)\over
R_0(\theta)}
\right]={2\lambda\over\pi}
\int_{-\infty}^{+\infty}dx \cos\left({2\lambda\theta
x\over\pi}\right)
{\sinh(\lambda-2\xi/\pi)x\over 2\sinh x \cosh \lambda x}.  }\label{repIIII}
\ee
In the region $0<2\xi/\pi<1$
where there are no poles and no  boundary bound states in the
spectrum,
formula (\ref{repIIII}) is valid, too. The expression for $P^+$ can be
obtained
 from
(\ref{repI}) by setting formally $n_B\equiv n_{\ast}+1=0$, which gives
\be
{-i{d\over d\theta}\log \left[{P^+(\theta)\over R_0(\theta)}
\right]={2\lambda\over\pi}
\int_{-\infty}^{+\infty}dx \cos\left({2\lambda\theta
x\over\pi}\right)
{\sinh(\lambda+2\xi/\pi)x\over 2\sinh x \cosh \lambda x}. }\label{repV}
\ee
Note that if $2\xi/\pi>1$, the integral in (\ref{repV}) diverges.
Finally, we complete this list by the following two expressions:
\bea
&&-i{d\over d\theta}\log \left[{P^{\pm}_{\delta_{N,n}}(\theta)\over
R_0(\theta)}\right]=-i{d\over d\theta}\log
\left[{P^{\pm}_{\beta_n}(\theta)\over
R_0(\theta)}\right]  + 
 {2\lambda\over\pi}
\int_{-\infty}^{+\infty}dx \cos\left( {2\lambda\theta
x\over\pi}\right) \nonumber \\
&\times&\left[{\sinh({2\xi\over\pi}-2n_{\ast}-2)x\over\sinh x}
-{2\cosh x\sinh({2\xi\over\pi}+2N-\lambda-1)x\over 2\sinh x\cosh\lambda
x}\right]. \label{reepr}
\eea
For the
integral representation
of $R_0$ see \cite{FS}.

\section{Exact solution of the regularized boundary sine-Gordon
model}

\subsection{The XXZ chain with boundary magnetic field}

The XXZ model in a boundary magnetic field
\be
{{\cal H}= {\pi-\gamma\over 2 \pi\sin\gamma}
\left[\sum_{i=1}^{L-1} \left(
\sigma^x_i \sigma^x_{i+1} + \sigma^y_i
\sigma^y_{i+1}+\Delta(\sigma^z_i
\sigma^z_{i+1}-1)\right) +h(\sigma_1^z-1)+h^{\p}(\sigma_L^z-1)
\right],} \label{XXZ}
\ee
was discussed in \cite{SKLYANIN} and in \cite{ABBBQ}, 
where its eigenstates were constructed using
the Bethe ansatz.  As usual, these  eigenstates ${\cal H}|n\rangle=E|n\rangle$
are linear combinations of
the states with $n$ down spins, located at
$x_1,...,x_n$ on the chain:
$$|n\rangle=\sum
f^{(n)}(x_1,...,x_n)|x_1,...,x_n\rangle.
$$

Consider for simplicity the case $n=1$. The wave-function  $f^{(1)}(x)$ reads
\cite{ABBBQ}:
\bea
f^{(1)}(x)&=&[e^{-ik}+(h^{\p}-\Delta)]e^{-i(L-x)k} -
(k\to-k) =\label{wf} \\
&=&\left[{\sinh{1\over 2}(i\gamma+\alpha)\over
\sinh{1\over 2}(i\gamma-\alpha)}\right]^{L-x} {\sin\gamma\sinh{1\over
2}
(\alpha+i\gamma H^{\p})\over \sinh{1\over 2}(i\gamma-\alpha)
\sin{1\over 2}(\gamma+\gamma H^{\p})} -  (\alpha\to-\alpha), \nonumber
\eea
where we defined the new variables  as in
\cite{ABBBQ}: $\Delta=-\cos\gamma$, $k=f(\alpha, \gamma)$,
\be
{\g H=f(i\g, -i\ln(h-\Delta))=-\g-i\ln{h-i\sin\g\over
h+i\sin\g}} \label{Hh}
\ee
(and similarly for $H^{\p}$), and
\be
{f(a,b)=-i\ln\left[{\sinh{1\over 2}(ib-a)\over
\sinh{1\over 2}(ib+a)}\right].}\label{phaseI}
\ee
 When $h$ varies from  $0$ to $+\infty$, $\g H$ increases
monotonically
 from $-\pi-\g$ to $-\g$ according to (\ref{Hh})  if we take the
main branch
of the logarithm.

Denote $h_{th}=1-\cos\g$. This
``threshold'' value of $h$ corresponds to $\g H=-\pi$;  its meaning
will become clear below. When $h$ varies from $-\infty$ to $0$, $\g
H$
increases monotonically  from $-\g$ to $\pi-\g$. For the purposes of
the
present work we confine our attention to the region $h, h^{\p}>0$ and
choose
$\g\in(0,{\pi\over 2})$. Other regions in the parameter space
can be obtained using the discrete symmetries of the Hamiltonian
(\ref{XXZ}):
$\sigma^z\to -\sigma^z$ on each site or on the odd sites only.
The parameter $k$ in (\ref{wf}) is not arbitrary, but satisfies the Bethe
equation  \cite{ABBBQ}:
\be
{
e^{i(2L-2)k}{(e^{ik}+h-\Delta)(e^{ik}+h^{\p}-\Delta)\over(e^{-ik}+h-
\Delta)(e^{-ik}+h^{\p}-\Delta)}=1,}\label{BAEV}
\ee
or
\be
{\left[{\sinh{1\over 2}(\alpha-i\gamma)\over
\sinh{1\over 2}(\alpha+i\gamma)}\right]^{2L}
{\sinh{1\over 2}(\alpha-i\gamma H)\sinh{1\over 2}(\alpha-i\gamma
H^{\p})
\over
\sinh{1\over 2}(\alpha+i\gamma H)\sinh{1\over 2}(\alpha+i\gamma
H^{\p})}=1.}\label{BAEI}
\ee
Note that the wave-function  (\ref{wf}) depends on $H$ implicitely through
the solution of the Bethe equation (\ref{BAEI})
$\alpha(H,H^{\p})$. Besides, one can multiply the amplitude (\ref{wf}) by
any
overall scalar factor depending on $\alpha$, $L$, $H$ and $H^{\p}$.
The Bethe equations in the sector of arbitrary $n>1$ can be found in 
\cite{ABBBQ}.

\subsection{The Bethe equations for the inhomogeneous \\ XXZ chain}

The real object of interest for us is actually the  inhomogeneous
six-vertex model with boundary magnetic field
on an open strip. The inhomogeneous six-vertex model is obtained by
giving an alternating
imaginary part $\pm i\Lambda$  to the spectral parameter on
alternating vertices of the six-vertex model
\cite{RS,DDV}.  It was argued in \cite{FS}, generalizing known
results for the periodic case \cite{RS},  that this  model
on an open strip provides in the scaling limit  $\Lambda,
L\to\infty$, lattice spacing $\to 0$
a lattice regularization of (\ref{SG}),
with $\beta^2=8\g$ and  a value of $\varphi_0$ at the boundary
related to
the magnetic field. The reader can find more details on the model in
the
references; it is actually closely related to the XXZ chain we
discussed above. In particular,
the wave function can be
expressed in
terms of the  roots  $\alpha_j$
 of the Bethe equations  \cite{ABBBQ,RS}:
\bea
&&\left[{\sinh{1\over 2}(\alpha_j+\Lambda-i\g)\over
\sinh{1\over 2}(\alpha_j+\Lambda+i\g)} 
{\sinh{1\over 2}(\alpha_j-\Lambda-i\g)\over
\sinh{1\over 2}(\alpha_j-\Lambda+i\g)}\right]^L
{\sinh{1\over 2}(\alpha_j-i\g H)\over\sinh{1\over 2}(\alpha_j+i\g H)}
{\sinh{1\over 2}(\alpha_j-i\g H')\over\sinh{1\over 2}(\alpha_j+i\g H')} 
\nonumber \\
& = &\prod_{ m\neq j}
 {\sinh{1\over 2}(\alpha_j-\alpha_m-2i\g)\over
\sinh{1\over 2}(\alpha_j-\alpha_m+2i\g)}
 {\sinh{1\over 2}(\alpha_j+\alpha_m-2i\g)\over
\sinh{1\over 2}(\alpha_j+\alpha_m+2i\g)}.    \label{origbe}
\eea
By construction of the Bethe-ansatz wave function, ${\rm Re}\alpha_j>0$.
Note that the solutions of (\ref{origbe}) $\alpha_j=0, i\pi$ should be
excluded
  because the wave function
vanishes identically in this case.  The analysis of solutions of eq.
(\ref{origbe})
is very similar to the case of the XXZ chain in a boundary magnetic
field.  We consider the regime $0<\g<\pi/2$, which falls
into the
attractive regime $0<\beta^2<4\pi$ in the sine-Gordon model (\ref{SG}).
We set hereafter $\g=\pi/t$ and for technical simplicity restrict
$t$
to be positive integer. In the limit $L\to\infty$ this  constraint
implies that
in the bulk
only the strings of length from 1 to $t-1$ are allowed, along with the
anti-strings (see chapter 1).

Taking the logarithm  of eq. (\ref{origbe}), one obtains:
\bea
L\left[f(\alpha_j+\Lambda,\gamma)+
f(\alpha_j-\Lambda,\gamma)\right]+
f(\alpha_j,\gamma H)+f(\alpha_j,\gamma H^{\p})\nonumber \\
=2\pi l_j + \sum_{ m\ne
j} \left[f(\alpha_j-\alpha_m,2\gamma) +
f(\alpha_j+\alpha_m,2\gamma)\right],\label{foral}
\eea
where $l_j$ is an integer.
We also recall the formula for the
eigen-energy associated with the roots $\alpha_j$ \cite{ABBBQ,RS},
\be
{E={\pi-\gamma\over\pi}\sum_{\alpha_j}[f^{\p}
(\alpha_j+\Lambda, \gamma) + f^{\p}(\alpha_j-\Lambda, \gamma)].}\label{Energy}
\ee

\subsection{Thirring model with boundary}

Like the bulk sine-Gordon model is a bosonized version
of the bulk massive Thirring model, one can expect that the boundary
sine-Gordon model is a bosonized version of the
Thirring model with  certain boundary conditions.
The quickest way to identify this boundary
Thirring model is to use the Bethe ansatz equations (\ref{origbe}). Write
the most general $U(1)$-invariant boundary interaction
\bea
H_{T}=
\int_0^L & dx & [-i\psi_1^+\psi_{1x} + i\psi_2^+\psi_{2x} +
m_0\psi_1^+\psi_2+m_0\psi_2^+\psi_1+2g_0\psi_1^+\psi_2^+\psi_2\psi_1]
\nonumber \\
&+&\sum_{ij}a_{ij}\psi_i^+\psi_j(0)+\sum_{ij}a^{\p}_{ij}\psi_i^+
\psi_j(L).\label{Thir}
\eea
The entries of the $2\times 2$ matrices $A=\{a_{ij}\},
A^{\p}=\{a^{\p}_{ij}\}$
can be determined up to one arbitrary parameter $\phi$ by the
hermicity
of $H_{T}$ and the consistency of the boundary conditions $({\rm det}
A=0)$.
For the left boundary, the matrix $A$ looks like
\be
{A={1\over 2\sin\phi}\pmatrix{e^{-i\phi}& 1 \cr
                        1& e^{i\phi}\cr}}
\label{lmatr}
\ee
and the boundary condition reads $\psi_1(0)=-e^{i\phi}\psi_2(0)$
(similarly for the right boundary).

To find the eigenvectors of the Hamiltonian (\ref{Thir}), $H_{T}\Psi=E\Psi$,
one can use the same wave-functions as for the bulk Thirring model
(\ref{Thack.wf}), 
and modify them by analogy with the example of XXZ chain in a
boundary
magnetic field \cite{ABBBQ}. This way one gets the equations
for the set of rapidities
$\alpha_j$ :
\bea
 e^{2im_0L\sinh\alpha_j}&=&
{\cosh{1\over 2}(\alpha_j+i\phi)\over\cosh{1\over
2}(\alpha_j-i\phi)}
 {\cosh{1\over 2}(\alpha_j+i\phi^{\p})\over
\cosh{1\over 2}(\alpha_j-i\phi^{\p})}\nonumber \\
 & \times &\prod_{ m\neq
j}
 {\sinh{1\over 2}(\alpha_j-\alpha_m-2i\g)\over
\sinh{1\over 2}(\alpha_j-\alpha_m+2i\g)}
 {\sinh{1\over 2}(\alpha_j+\alpha_m-2i\g)\over
\sinh{1\over 2}(\alpha_j+\alpha_m+2i\g)}, \label{origbeI}
\eea
where $\gamma$ is related to $g_0$ in the usual way \cite{Korep}. These
equations look quite similar to (\ref{origbe}). The mapping can be made
complete by taking in (\ref{origbe}) the limit $\Lambda\rightarrow\infty$
with the identification $m_0=4e^{-\Lambda}\sin\gamma$.

The
derivation of
these equations is rather cumbersome, therefore to illustrate the
procedure
we comment on the simplest case of one-particle sector, which is
nevertheless
sufficient to obtain the form of the boundary terms in (\ref{origbeI}).
We make an ansatz $\Psi=\int_0^L dy
\chi^{\lambda}(y)\psi^+_{\lambda}(y)|0\rangle$,
where $\lambda$ is the spinor index, $\chi(y)$ is the wave-function
and $|0\rangle$ is the unphysical vacuum annihilated by $\psi_{\lambda}$.

The equation $H_{T}\Psi=E\Psi$ reduces to 
\be
{ -i\sigma_3{\partial\over \partial x}\vec\chi
+m_0\sigma_1\vec\chi
+A\vec\chi\delta(x)+A^{\p}\vec\chi\delta(x-L)=E\vec\chi.}\label{IwI}
\ee
where $\sigma_i$ are the Pauli
matrices. We look for the solution of (\ref{IwI}) in the form
\be
{\pmatrix{\chi_1\cr\chi_2\cr}=a(\alpha)\pmatrix{e^{-\alpha/2}
\cr
e^{\alpha/2}\cr}e^{im_0x\sinh\alpha} -
a(-\alpha)\pmatrix{e^{\alpha/2}\cr
e^{-\alpha/2}\cr}e^{-im_0x\sinh\alpha}. } \label{IIwI}
\ee
Substituting it into (\ref{IwI}) we get $E=m_0\cosh\alpha$ and,
besides, two boundary conditions
to be solved. The first one, at $x=0$, determines the form of the
 factor $a(\alpha)=\cosh{1\over 2}(\alpha-i\phi)$, while the
second one
at $x=L$ gives rise to the Bethe equation
$$e^{2im_0L\sinh\alpha}=
{\cosh{1\over 2}(\alpha+i\phi)\over\cosh{1\over
2}(\alpha-i\phi)}
{\cosh{1\over 2}(\alpha+i\phi^{\p})\over
\cosh{1\over 2}(\alpha-i\phi^{\p})},$$
which determines $\alpha$. Comparing the Bethe equation (\ref{origbe}) with
(\ref{origbeI}) and using the relation
between $\xi$ and $H$ obtained below in section 4.5
 we find the relation between the boundary
parameters $\phi$ and $\varphi_0$ in the Hamiltonians (\ref{Thir}) and
 (\ref{SG}) respectively:
$$\phi=\beta\varphi_0-\beta^2/8.$$
Thus, the integrable boundary condition
for the $U(1)$-invariant boundary Thirring model 
equivalent to (\ref{SG}) reads:
\be
{\psi_2(0)=-e^{i\beta^2/8-i\beta\varphi_0}\psi_1(0).} \label{intconI}
\ee

It would be interesting to obtain the result (\ref{intconI}) directly
from the Hamiltonian (\ref{SG}) using
an extension of the Coleman-Mandelstam bosonization technique to
the case with boundary. However, to our knowledge such an extension
has not been developed yet, except for the free point \cite{boot:AKL}.

\section{Solutions of the Bethe ansatz equations with boundary terms}

As is well known in the case of the bulk Thirring model or
equivalently
the periodic XXZ chain, the  bound states are associated with
various types of solutions of the Bethe ansatz equations involving
in general complex roots (see chapter 1). By analogy, we expect the
boundary bound states  to
correspond to new solutions
made possible by the boundary terms.

Consider first the example of the XXZ chain
as given in section 4.3.1. Since our goal is to study
purely boundary effects, we will look for  the solutions of the Bethe equations
that give rise to
a wave-function localized at $x=0$ or $x=L$ and exponentially
decreasing
away from the boundary.

The states described by such wave-functions will be referred to
as the ``boundary bound states'' below.  For this, one should have
$\alpha$ purely imaginary in (\ref{wf}).
We consider here the limit of $L$
large, when the left and the right boundaries can be treated
independently
and the overlap of the corresponding wave-functions is negligibly
small
(for the physical applications it is necessary to take the scaling
limit
anyway).
 In the limit $L\to\infty$ it is easy to check that there are two such
solutions
to (\ref{BAEI}): $\alpha=i\alpha_0=-i\gamma H +
i\varepsilon(L,H,H^{\p}) $ and
$\alpha=i\alpha_0^{\p}=-i\gamma
H^{\p}+i\varepsilon^{\p}(L,H,H^{\p})$,
where $\varepsilon\sim
\exp(-2\kappa L)$ and we defined $\kappa>0$ as
$$ e^{-\kappa}=\left|{\sin{1\over 2}(-\gamma H-\gamma)\over
\sin{1\over 2}(-\gamma H+\gamma)}\right|$$
(similar relations are assumed for $\varepsilon^{\p}, \kappa^{\p}$).  Solution
$\alpha_0^{\p}$ gives a wave-function (\ref{wf}) localized at
$x=L$: $f^{(1)}(x)\sim  e^{-\kappa^{\p}(L-x)}$. Solution
$\alpha_0$ gives a wave-function localized at $x=0$,
$f^{(1)}(x)\sim  e^{-\kappa x}$, provided we renormalize  the
wave-function
(\ref{wf}):
\be
{f^{(1)}\rightarrow f^{(1)}\left[\sinh{1\over
2}(\alpha-i\gamma H)
\sinh{1\over 2}(\alpha+i\gamma H)\right]^{1/2}.}\label{wfI}
\ee
In the special case $H=H^{\p}$ there is  only one proper solution
$\alpha=i\alpha_0=-i\g H +i\varepsilon(L,H)$ with
$\varepsilon\sim\exp(-\kappa L)$.
The  wave-function  (\ref{wf}) behaves as the superposition
of the ``left'' and the ``right'' boundary bound states,
$f^{(1)}\sim(e^{-\kappa x} + e^{-\kappa(L-x)})$.
Note that the boundary bound state appears in the above example
only when the boundary magnetic field is large enough:\footnote{
More generally, the criterion of existence of boundary
 bound
state
solutions
allows us to determine  threshold fields for any anisotropy
$\Delta$.
For this, let us examine (\ref{BAEV}). The parameter $k$ is defined modulo
$2\pi$,
therefore we restrict it to lie withing $k\in (0,2\pi)$. Two
possibilities
$k=ia$ and $k=\pi+ia$, where $a>0$, lead to two different threshold
fields,
determined by the fact that the denominator in (\ref{BAEV}) should vanish:
$$h_{th}^{(1)}=\Delta +1, \qquad h_{th}^{(2)}=\Delta-1,$$
 and the regions where boundary bound states could
in principle exist are $h<\Delta-1, \quad
h>\Delta+1$ (one has to be carefull here and check that these solutions
of BE indeed correspond to the {\it stable} states of the model).
When $\Delta>1$, there are two different threshold fields, in
agreement with the results of Jimbo et al \cite{miwa}.
 In the region of interest,
$|\Delta|<1$,
 there is only one threshold field $h_{th}^{(1)}$.}
namely, $h>h_{th}$. This follows from the fact that $\alpha$ should be such
that
$0<\alpha_0<\pi$.

Now, consider the equations for the inhomogeneous model (\ref{origbe}).
The basic boundary 1-string solution to (\ref{origbe}) is
$\alpha=i\alpha_0=-i\g H+i\epsilon$, provided that $0<\alpha_0<\pi$.
This solution is possible due to an argument very similar
to the one used in the bulk: as $L\to\infty$,  the two first terms
of (\ref{origbe}) decrease exponentially fast, while the third increases
exponentially fast, and $\epsilon\sim \exp(-2\kappa L)$ with
\be
{e^{-\kappa}={ \sinh^2{\Lambda\over 2}+
\sin^2{\alpha_0-\g\over 2}
\over \sinh^2{\Lambda\over 2}+ \sin^2{\alpha_0+\g\over 2}}.} \label{Defkap}
\ee
Recall that for the  bulk problem when there is no boundary term, the right
hand side of (\ref{origbe}) would have to decrease exponentially, forcing the
existence of a ``partner''  root at $\alpha-2i\gamma$.

One can construct similarly  boundary n-strings which  consist of
the points $i\alpha_0$, \newline 
$i\alpha_0+2i\g, ... , i\alpha_0+2i(n-1)\g$
(see  figure 4.2). 
By convention $n=0$ means there is no boundary string, that is
all complex solutions are in the usual bulk strings.
The possible values of $n$ are restricted  by the fact that the upper
point of the complex should be below $i\pi$:
$\max(n)=[{\pi-\alpha_0\over
2\g}]+1$, where the square bracket denotes the integer part.  To show
that
the boundary n-string is indeed a solution to (\ref{origbe}), we introduce
infinitesimal  corrections $\varepsilon_i$ to the positions of the
points of
complex \cite{Gaudin}. Taking the modulus of both sides of (\ref{origbe}) with
$\alpha_j=
i\alpha_0+2ik\g$ and multiplying equations for $k=0,...,n-1$
we obtain $\exp\{-2L(\kappa_1+\kappa_2+...+\kappa_n)\}\sim
\varepsilon_1$, where $\varepsilon_1$ denotes the correction to the
point
$i\alpha_0$. The behavior of the remaining $\varepsilon_k$ follows
from
$\varepsilon_1$ by recursion. For example, for the 2-string
$\varepsilon_2$ is given by  $|\varepsilon_1
-\varepsilon_2|\sim \exp(-2L\kappa_2)$.

\begin{figure}
\begin{center}
\begin{picture}(400, 220)
\put(0, 50){\vector(1,0){400}}
\put(200,40){\vector(0,1){160}}
\put(0,180){\line(1,0){400}}
\put(200, 80){\circle*{4}}
\put(200, 100){\circle*{4}}
\put(200, 120){\circle*{4}}
\put(200, 140){\circle*{4}}
\put(200, 160){\circle*{4}}
\put(205, 80){$i\alpha_0$}
\put(205, 100){$i\alpha_0+2i\g$}
\put(205, 120){$i\alpha_0+4i\g$}
\put(205, 140){$i\alpha_0+6i\g$}
\put(205, 160){$i\alpha_0+8i\g$}
\put(380, 55){Re$\alpha$}
\put(200, 210){Im$\alpha$}
\put(205, 185){$\pi$}
\put(240, 185){Dirac sea}
\put(210, 180){\circle*{4}}
\put(230, 180){\circle*{4}}
\put(250, 180){\circle*{4}}
\put(270, 180){\circle*{4}}
\put(290, 180){\circle*{4}}
\put(310, 180){\circle*{4}}
\put(330, 180){\circle*{4}}
\put(350, 180){\circle*{4}}
\put(370, 180){\circle*{4}}
\put(390, 180){\circle*{4}}
\end{picture}
\caption{ The first type of boundary string. In the ground state
the 
boundary string of maximum allowed length is filled.}
\end{center}
\end{figure}
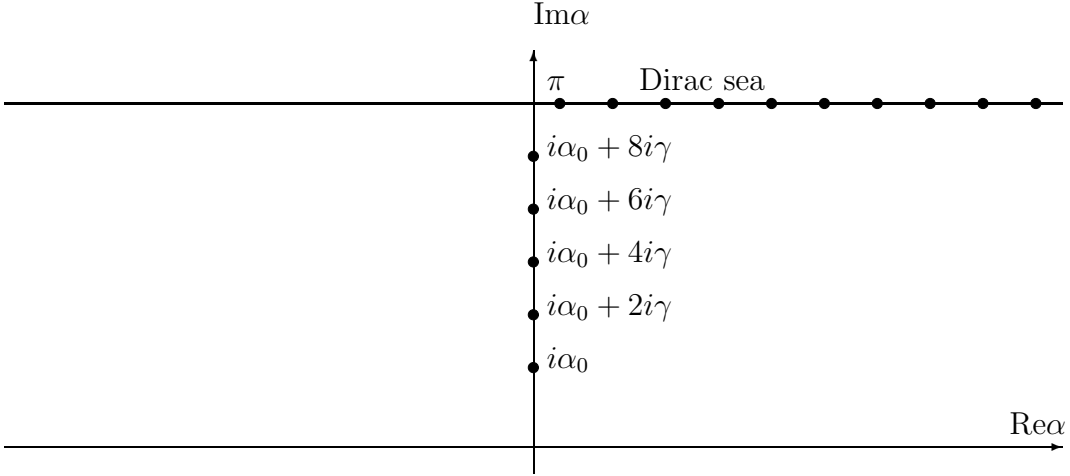

Note that associated  with  each
boundary n-string there
is also the solution to (\ref{origbe}) obtained by complex
conjugation of  all $\alpha'$s.  The existence of such a "mirror
image" is the consequence of the symmetry of equations (\ref{origbe})
and it is of no importance to physics. In the bulk case, it is easy to show
\cite{BdVV} that all solutions are invariant under complex conjugation, 
but this result does not hold here.
In fact, a solution which has both the boundary n-string and its
mirror image would lead to a vanishing wave-function.

Additional boundary strings can be obtained by adding the roots
$i\alpha_s$
below $i\alpha_0$ so that $i\alpha_s=i\alpha_0-2is\g$, with
$s=1,2,...,N$
(see  figure 4.3).
Together with the boundary n-string above $\alpha_0$, they form the
complex
which we call boundary $(n,N)$-string. To analyze the existence of
such
complexes  as the solutions of (\ref{origbe}) we introduce  as before the
infinitesimal
corrections $\varepsilon_s$ to the roots $\alpha_s$, where now
 $s=n, n-1,..., 1,
-1, -2,..., -N$. Then, the equations (\ref{origbe}) with
$\alpha_j=i\alpha_s$ tell us
that the range of $N$ should be
\be
{ {\alpha_0\over 2\g}<N<{\pi+\alpha_0\over 2\g}.}\label{RangeN}
\ee
In other words, the inequality (\ref{RangeN}) states that the lowest root
of the
boundary string should be below the axis ${\rm Im}\alpha=0$ and above
the axis ${\rm Im}\alpha=-i\pi$.
Another constraint  follows if we multiply the equations (\ref{origbe}) for
all the roots of boundary $(n,N)$-string. This gives
$\exp(-2L\sum\kappa_s)
=\varepsilon_1$. So, one should have $\sum\kappa_s>0$. The latter sum
can
be easily evaluated if one uses the expression (\ref{Defkap}) simplified in
the
limit $\Lambda\to\infty$: $\kappa=4e^{-\Lambda}\sin\g\sin\alpha_0$.
The constraint obtained in such a way forces the number of roots
above ${\rm Im}\alpha=0$ axis in the boundary string to be greater
than the
number of roots below ${\rm Im}\alpha=0$.

\begin{figure}
\begin{center}
\begin{picture}(400, 300)
\put(0, 150){\vector(1,0){400}}
\put(200,40){\vector(0,1){240}}
\put(0,260){\line(1,0){400}}
\put(0, 50){\line(1,0){400}}
\put(200, 100){\circle*{4}}
\put(200, 180){\circle*{4}}
\put(200, 200){\circle*{4}}
\put(200, 220){\circle*{4}}
\put(200, 240){\circle*{4}}
\put(200, 160){\circle*{4}}
\put(200, 140){\circle*{4}}
\put(200, 120){\circle*{4}}
\put(205, 180){$i\alpha_0$}
\put(205, 200){$i\alpha_0+2i\g$}
\put(205, 220){$i\alpha_0+4i\g$}
\put(205, 240){$i\alpha_0+6i\g$}
\put(205, 160){$i\alpha_0-2i\g$}
\put(205, 140){$i\alpha_0-4i\g$}
\put(205, 120){$i\alpha_0-6i\g$}
\put(205, 100){$i\alpha_0-8i\g$}
\put(380, 155){Re$\alpha$}
\put(200, 290){Im$\alpha$}
\put(205, 265){$\pi$}
\put(205, 60){$-\pi$}
\put(240, 265){Dirac sea}
\put(210, 260){\circle*{4}}
\put(230, 260){\circle*{4}}
\put(250, 260){\circle*{4}}
\put(270, 260){\circle*{4}}
\put(290, 260){\circle*{4}}
\put(310, 260){\circle*{4}}
\put(330, 260){\circle*{4}}
\put(350, 260){\circle*{4}}
\put(370, 260){\circle*{4}}
\put(390, 260){\circle*{4}}
\end{picture}
\caption{The second type of boundary string.}
\end{center}
\end{figure}
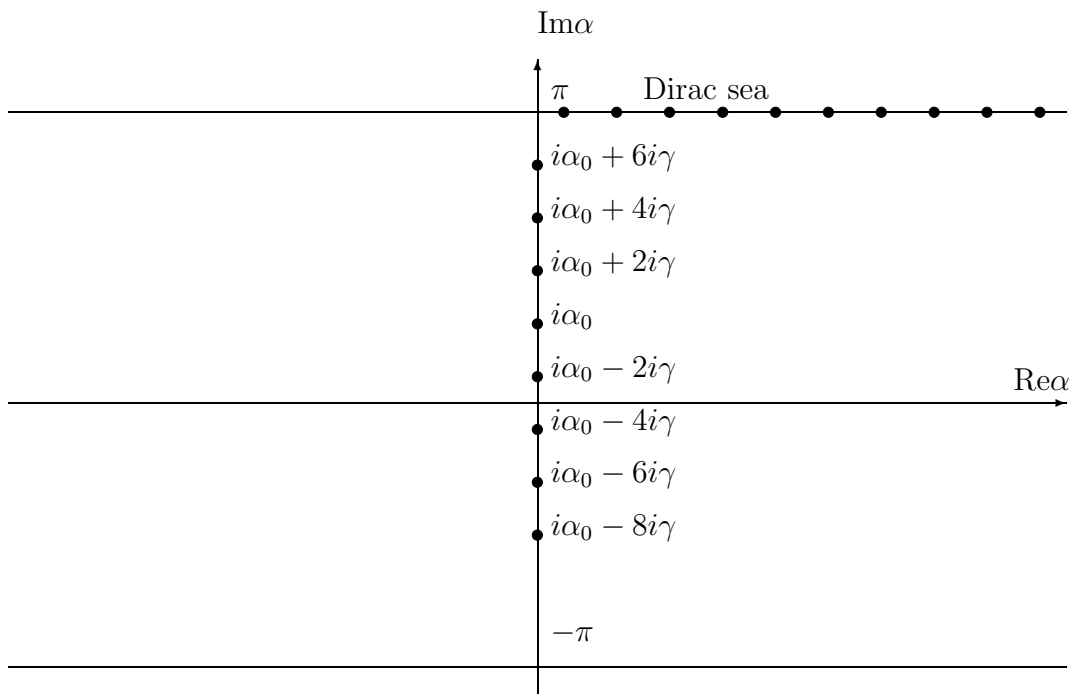

We have not been able to find any reasonable additional solution to
the
Bethe ansatz equations.
The  two sets of
boundary strings we have encountered  appear
to be in one to one correspondence with the boundary bound states
identified in section 4.2 using the bootstrap approach. To clarify
this identification we now compute related  masses and S-matrices.

\section{S-matrices and bound state properties from the exact
solution}

\subsection{Bare and physical Bethe ansatz equations}

The {\it bare} Bethe equations follow from taking the derivative of
(\ref{foral}).
Defining $2L(\rho_k+\rho_k^h)d\alpha$ to be the number of roots
in the interval $d\alpha$, one obtains coupled integral equations
for the densities of strings $\rho_1,...,\rho_{t-1}$ and anti-strings
$\rho_a$:
\bea
2\pi (\rho_k + \rho_k^h)&=&  {1\over 2}
p_k^{\p} - f^{\p}_{ka}\ast\rho_a -
\sum_{l=1}^{t-1}f^{\p}_{kl}\ast\rho_l
+{1\over 2L}(u_k-\omega f^{\p (L)}_{n;k}-
\omega^{\p}f^{\p (R)}_{n^{\p};k})\nonumber \\
2\pi(\rho_{a} +\rho_a^h)&=&-{1\over 2}p_a^{\p}+f^{\p}\ast\rho_a
+\sum_{l=1}^{t-1}f^{\p}_{al}\ast\rho_l+
{1\over 2L}(u_a+\omega f^{\p (L)}_n + \omega^{\p} f^{\p(R)}_{n^{\p}})
\label{bba}
\eea
where $\ast$ denotes convolution:
$$ f\ast g(\alpha)= \int_{-\infty}^{\infty} d\beta f(\alpha-\beta)
g(\beta).$$
These densities are originally defined for $\alpha>0$, but the
equations allow
us to define $\rho_k(-\alpha)\equiv\rho_k(\alpha)$ in order to
rewrite the
integrals to go from $-\infty$ to $\infty$.  If we totally neglect
the boundary
terms (terms $\sim L^{-1}$) in  (\ref{bba}), we will end up with the same
equations
as for the periodic inhomogeneous six-vertex model \cite{RS}.  The various
kernels
and sources in (\ref{bba}) are defined as follows:
$$p_a(\alpha)=f(i\pi+\alpha+\Lambda, \g)+f(i\pi+\alpha-\Lambda,
\g),$$
$$p_k(\alpha)=\sum_{\alpha_i}f(\alpha_i+\Lambda, \g)+
f(\alpha_i-\Lambda, \g),$$
where the sum in the last expression is taken over the rapidities of
the
bulk k-string roots centered on $\alpha$.

The kernels $f_{kl}$
are the phase shifts of bulk k-string on bulk $l$-string obtained by
summing
(\ref{phaseI})
over the rapidities of string roots. The boundary terms are:
$$u_a(\alpha)=-2f^{\p}(2\alpha, 2\g)-f^{\p}(\alpha+i\pi, \g H)
-f^{\p}(\alpha+i\pi, \g H^{\p})-2\pi\delta(\alpha),$$
$$u_k(\alpha)=\sum_{\alpha_i}[2f^{\p}(2\alpha_i, 2\g)+
f^{\p}(\alpha_i, \g H)+f^{\p}(\alpha_i, \g
H^{\p})]-2\pi\delta(\alpha),$$
(the sum above is over the roots of bulk k-string centered on $\alpha$),
$$f_n^{(L,R)}(\alpha)=\sum_{\alpha_i}
f(i\pi+\alpha-\alpha_i, 2\g)
+f(i\pi+\alpha+\alpha_i, 2\g),$$ and $\alpha_i$
denotes the rapidities of the roots in the boundary n-string.
$$f^{(L,R)}_{n;k}(\alpha)=\sum_{\alpha_i}\sum_{\alpha_j}
f(\alpha_j-\alpha_i, 2\g)
+f(\alpha_j+\alpha_i, 2\g),$$
where $\alpha_i$ denotes the roots in the boundary n-string, while
$\alpha_j$ denotes the roots in the bulk k-string centered on $\alpha$.
The parameters
$\omega, \omega^{\p}$ are equal to $1$ or $ 0$, depending on whether
the boundary string is present or not.
In our ferromagnetic case the ground state
of the periodic inhomogeneous XXZ chain is filled with anti-strings.

The physical Bethe equations are obtained \cite{DL,KR}
 by
eliminating the ``non-physical'' density $\rho_a$ from
 the right-hand side of (\ref{bba}). This is done simply
by solving  for
$\rho_a$ in the last equation in (\ref{bba}) and substituting it into
the others. The result is
\bea
2\pi(\rho_k+\rho_k^h)&=&{1\over 2}p^{\p}_k
+{1\over 2}{f^{\p}_{ak}\over 2\pi-f^{\p}}\ast p^{\p}_a +
+ {f^{\p}_{ak}\over 2\pi-f^{\p}}\ast 2\pi\rho_a^h \nonumber \\
&-& \sum_{l=1}^{t-1}
\left(f^{\p}_{kl}+{f^{\p}_{ak}f^{\p}_{al}\over
2\pi-f^{\p}}\right)\ast\rho_l
+{1\over 2L}U_{n,n^{\p}; k}, \label{physbe} \\
2\pi(\rho_a+\rho_a^h)&=&-{1\over 2}{2\pi p^{\p}_a\over 2\pi-f^{\p}}
-{f^{\p}\over 2\pi-f^{\p}}\ast 2\pi\rho_a^h + \sum_{l=1}^{t-1}
{f^{\p}_{al}\over 2\pi-f^{\p}}\ast 2\pi\rho_l + {1\over
2L}U_{n,n^{\p}; a},
\nonumber
\eea
where
\bea
U_{n,n^{\p}; a}&=&2\pi{
u_a+\omega f^{\p (L)}_n + \omega^{\p} f^{\p (R)}_{n^{\p}}
\over 2\pi-f^{\p}}, \label{defI} \\
U_{n,n^{\p}; k}&=&u_k-\omega f^{\p (L)}_{n;k}-
\omega^{\p}f^{\p (R)}_{n^{\p};k}-f^{\p}_{ak}\ast U_{n,n^{\p};a}/2\pi, 
\label{defII}
\eea
and different products  (ratios) of kernels are defined through their
Fourier transforms.

\subsection{The mass spectrum of boundary bound states}

We assume at first
that the ground state is built by filling up the Dirac sea
with anti-strings, as in the case of the periodic XXZ chain.
We will see below that this is not always true.
The presence of the boundary strings in the Bethe equations
deforms the distribution of  roots and modifies
the density of the Dirac sea $\rho_a$ by a term $\delta\rho_a/2L$
of  order $L^{-1}$. With the boundary n-string, the Bethe equation
for the density of the Dirac sea particles $\tilde{\rho}_a$ is
\be
{{1\over 2} p_a^{\p}(\alpha)-{1\over
2L}u_a(\alpha)=\int_{-\infty}^{
+\infty}f^{\p}(\alpha-\beta)\tilde{\rho}_a(\beta)d\beta
-2\pi\tilde{\rho_a}
(\alpha) + {1\over 2L}f_n^{\p}(\alpha),}\label{beI}
\ee
where $f_n$ was defined above.
Subtracting from (\ref{beI}) the equation for the density of the Dirac sea
alone,
\be
{{1\over 2} p_a^{\p}(\alpha)-{1\over
2L}u_a(\alpha)=\int_{-\infty}^{
+\infty}f^{\p}(\alpha-\beta)\rho_a(\beta)d\beta -2\pi\rho_a(\alpha),}
 \label{beII}
\ee
one obtains the equation for $\delta\rho_a$:
\be
{0=-\int_{-\infty}^{
+\infty}f^{\p}(\alpha-\beta)\delta\rho_a(\beta)d\beta +
2\pi\delta\rho_a(\alpha) - f_n^{\p}(\alpha),} \label{beIII}
\ee
where
$$\delta\rho_a\equiv 2L(\tilde{\rho}_a-\rho_a).$$
The solution to (\ref{beIII}) can be written in terms of the Fourier
transform
$$\delta\hat{\rho}_a(k)=\int d\alpha e^{ik\alpha}\delta\rho_a(\alpha)$$
as follows:
\be
{\delta\hat{\rho}_a(k)={\hat{f}_n^{\p}(k)\over
 2\pi-\hat{f}^{\p}(k)}.}\label{delro}
\ee
For the boundary n-string  $i\alpha_0+2i\gamma s$,
$s=0,1,...,n-1,$ we obtain
\bea
\hat{f}_n^{\p}(k)&=&-2\pi{4\cosh \gamma k \sinh n\gamma k
\cosh(\alpha_0 +\gamma n -\gamma)k \over \sinh \pi k},  \label{ftI} \\
\delta\hat{\rho}_a(k)&=&-{2\cosh \gamma k \sinh n\gamma k
\cosh(\alpha_0
+\gamma n -\gamma)k \over \sinh \gamma k \cosh(\pi-\gamma)k}, 
\label{deformI}
\eea
where we used
$$\hat{f}^{\p}(k)=2\pi{\sinh(\pi-2\gamma)k\over\sinh \pi k}.$$
Expressions (\ref{ftI}), (\ref{deformI}) are valid for the n-strings with
$n=1,2,...,
\left[{t+H\over 2}\right]$. For the longest n-string with
$n=\left[{t+H\over 2}\right]+1\equiv n_{\ast}+1$  the Fourier
transforms
$\hat{f}_n^{\p}, \delta\hat{\rho}_a$ differ from (\ref{ftI}), (\ref{deformI}):
\bea
\hat{f}_{\ast}^{\p}(k)&=&2\pi{2\sinh(\pi-2\gamma)k \cosh
(\alpha_0+2\gamma n_{\ast}-\pi)k \over \sinh \pi k} \nonumber \\
&-& 2\pi{4\cosh \gamma k \sinh n_{\ast}\gamma k \cosh(\alpha_0
+\gamma n_{\ast}-\gamma)k\over \sinh \pi k}, \label{ftII}
\eea
\bea
\delta\hat{\rho}_a(k)&=&{\sinh(\pi-2\gamma)k \cosh
(\alpha_0+2\gamma n_{\ast}-\pi)k \over \sinh \gamma k \cosh
(\pi-\gamma)k} \nonumber \\
&-& {2\cosh \gamma k \sinh n_{\ast}\gamma k \cosh(\alpha_0
+\gamma n_{\ast}-\gamma)k\over  \sinh \gamma k \cosh
(\pi-\gamma)k}.  \label{deformII}
\eea
The conserved $U(1)$ charge in the boundary XXZ chain is  the
total projection of the spin on the z-axis. In the thermodynamic limit
the charge of the boundary n-string with respect to the vacuum
is determined by \cite{ABBBQ}:
\be
{ Q_n=n+\int_0^{+\infty}2L\tilde{\rho}_ad\alpha
- \int_0^{+\infty}2L\rho_ad\alpha=n+{1\over
2}\int_{-\infty}^{+\infty}
\delta\rho_a d\alpha=n+{1\over 2}\delta\hat{\rho}_a(0).}
\label{charge}
\ee
Using (\ref{deformI}), we obtain for the n-string  $ Q_n=0$, and for
the longest boundary string Eq. (\ref{deformII}) yields $Q_{\ast}=\pi/
2\gamma$.
Similarly, the mass of the boundary strings
in the thermodynamic limit according to (\ref{Energy}) is given by
\be
m_n=h_n+\int_0^{+\infty}2L\tilde{\rho}_a
h_ad\alpha - \int_0^{+\infty}2L\rho_ah_ad\alpha=h_n+{1\over 2}
\int_{-\infty}^{+\infty}h_a\delta\rho_a d\alpha,
\label{ener}
\ee
where the expression for $h_a$ is
$$ \hat{h}_a(k)={\pi-\g\over\pi}\hat{p}_a^{\p}=
-2(\pi-\g){2\sinh \gamma k \cos
\Lambda k
\over \sinh \pi k}$$
and the soliton mass \cite{RS}
$$ m=2e^{-{\Lambda\pi\over 2(\pi-\gamma)}}.$$

We obtain in the limit $\Lambda\to\infty$
\bea
m_n&=&m\left[\sin{\pi\over 2\lambda}\left(2n-1-H\right)+
\sin{\pi\over 2\lambda}\left(H+1\right)\right],
\label{enerI} \\
m_{\ast}&=&m\sin{\pi\over 2\lambda}\left(H+1\right).
\label{enerII}
\eea
Since the parameter $H$ varies in the interval $-\lambda-1<H<-1$,
the mass of
the longest string $m_{\ast}$ (\ref{enerII}) is always negative, while
the other boundary strings have positive masses (\ref{enerI}). This means
that
the vacuum we have been working with is an unstable one in the region
$-t<H<-1$ $(h>h_{th})$.  To cure the
situation we define a new correct  ground state by attributing the
longest
boundary string  to the Dirac sea.  The boundary excitations are
obtained by
succeessive removing  of particles from the top of the longest
boundary string.
The charge and mass of such excitations with respect to the correct
ground
state
are given by
\be
{Q_n=-{\pi\over 2\gamma}, \qquad m_n=
m\cos{\pi\over 2\lambda}(\lambda+1+H-2n), \qquad n=0,1,...,n_{\ast}.}
\label{correct}
\ee
Note that the number of excitations (\ref{correct}) is equal to the
number of roots in
the longest boundary string, $n_{\ast}+1$. The charge of such
boundary
excitations is equal to the charge of the hole in the Dirac sea. We
identify
a hole with a sine-Gordon soliton, and the boundary excitations
described above, with the boundary bound states  $|\beta_n\rangle$ (\ref{addi}). Their
masses and
charge (\ref{correct})  and the counting coincide provided that
\be
{t+H+1={2\xi\over\pi},} \label{param}
\ee
and the lattice charge $Q$ is properly
normalized. This expression is in fact valid for all values of $h>0$.

In the above discussion we considered the boundary bound states
related
to  one of the boundaries (say, the left one). In principle, one
should include
into the ground state the longest boundary string
$i\alpha^{\p}_0+2i\g l,
\quad l=0,1,...,\left[{t+H^{\p}\over 2}\right]$, corresponding to the
right
boundary as well. The energy of the excitations
due to both boundary strings is a superposition of energies of the
form
(\ref{correct}). When $H=H^{\p}$, these two  boundary strings overlap and
the
usual Bethe wave-function vanishes. However on physical grounds we do not
expect anything special to happen when the boundaries are identical.
So, in such a case
 one should use as a wave function a properly renormalized
version of the
limit $H\to H^{\p}$
of the
usual Bethe wave function.

When the magnetic field varies, the above picture indicates
a   qualitative change in the
structure
of the  ground state at values  $H=-t, -t+2, -t+4,...$. At these values,
the mass of the bound state with the highest mass approaches  the soliton mass
and it becomes unstable.
As discussed  in \cite{McCoy} and \cite{GZ} for the Ising case,
this decay corresponds to large
  boundary fluctuations that propagate deeply into the bulk.

 The mass of the boundary
$(n,N)$-string with respect to the correct vacuum  can be calculated
analogously.
The result is:
\be
{m_{n,N}=m\cos({\xi\over\lambda}-{\pi\over 2\lambda})
+m\cos({\xi\over\lambda}-{2n+1\over 2\lambda}\pi)
-m\cos({\xi\over\lambda}+{2N-1\over 2\lambda}\pi),}
\label{enerNn}
\ee
where we used (\ref{param}) to express $H$ in terms of $\xi$.
This result seems rather confusing, because the above
mass does not correspond in general to one of the bound state masses
found in the bootstrap apporach. It can be considered as a sum of such masses,
hinting that the $(n,N)$ string describes actually coexisting bound states,
but the calculation of 
corresponding boundary S-matrix does not allow such an interpretation.
We  are forced  (but see the conclusion) to consider that only the
$(n,N)$-strings with $n=n_{\ast}+1$
occur, that is
the physical excitations  are built by adding
roots to the ground state configuration below $i\alpha_0$.
The charge and energy of such excitations
with respect to the correct vacuum is given by
\be
{Q_N={\pi\over 2\g}-{\pi\over 2\g}=0, \qquad
m_{N}=m\cos({\xi\over\lambda}-{\pi\over 2\lambda})
-m\cos({\xi\over\lambda}+{2N-1\over 2\lambda}\pi),}
\label{enerNnI}
\ee
These coincide with   the charge and  mass
of the boundary bound states $|\delta_{n=0,N}\rangle$
(\ref{massnew}). The
range
of $N$ (\ref{RangeN}) agrees with the range of corresponding parameter in
(\ref{newpols}).

\subsection{Boundary S-matrices}

It remains to check that the boundary S-matrices obtained above by
the bootstrap approach  coincide with  those of the lattice
model. To extract the boundary
S-matrices from the Bethe equations we will follow the discussion of
\cite{FS}.  Briefly, the idea of the method is the following.  The
physical
excitations of the  lattice model in the limit $\Lambda\to\infty$
can be thought
of as  relativistic quasi-particles with  rapidities $\th_i$. The
integrability
implies
that  the set $\{\th_i\}$ is conserved. Moreover, if the scattering
matrices are
diagonal, each  particle preserves its rapidity. Assuming that this
is the
 case, the quantization of a gas of ${\cal N}$ quasi-particles on
an interval of
lenght $L$ results in  the  integral equations for the set of allowed
rapidities
\cite{FS}:
\be
{2\pi(\rho_b+\rho_b^h)= m_b \cosh\th + \sum_{c=1}^p
\varphi_{bc}*\rho_c+ {1\over 2L}
\Theta_b,}
\label{sba}
\ee
where subscript stands for the type of particle, and
\bea
\varphi_{bc}(\th)&=&-i {d\over d\th}
\ln S_{bc}(\th)\label{forphi} \\
\Theta_{b}(\th)&=&-i {d\over d\th} \ln R_{\beta}^{b(L)}(\th)
-i {d\over d\th} \ln R_{\beta^{\p}}^{b(R)}(\th)
+ i {d\over d\th} \ln S_{bb}(2\th)
-2\pi \delta(\th).\nonumber
\eea
Equations (\ref{sba}) should be compared with the physical BE (\ref{physbe}),
which gives
bulk and boundary S-matrices.  We will confine our attention to the
boundary
S-matrices only, keeping track of those terms  in (\ref{defI}), (\ref{defII}),
(\ref{forphi}),
which depend on the boundary magnetic field (the field-independent
terms
contribute to $R_0$ and their agreement has been shown in \cite{FS}).
The discussion for the left boundary is completely parallel to that
of the
right
one. Also, it is
sufficient to consider only $b=$soliton  and $b=$anti-soliton in
(\ref{sba}). We
identify a hole in the anti-string distribution in (\ref{physbe}) with a
soliton
in (\ref{sba}), and $(t-1)$-string with  an anti-soliton. Below we give
explicit expressions only  for
the
 kernels in (\ref{physbe}) which we need
for our
analysis. The other expressions are listed in \cite{FS}.

Suppose first that $h<h_{th}$ ($-t-1<H<-t$). This corresponds to the
case
without boundary excitations in the spectrum, $\xi<\pi/2$. Choose
$\omega=\omega^{\p}=0$ in (\ref{physbe}). Then
$$\hat{u}_a^{(L)}(k)=2\pi{\sinh(2\pi+\g H)k\over \sinh\pi k}+\ldots$$
(we omitted the $H$-independent terms and $H^{\p}$-dependent ones),
$$\hat{U}_a^{(L)}=2\pi{\sinh(2\pi+\g H)k\over 2\sinh\g k
\cosh(\pi-\g)k}+\ldots.$$
Using (\ref{forphi}) we compare this expression with (\ref{repV})
  (recall that
the rapidity $\alpha$ should be renormalized
$\alpha\to\th=t\alpha/2\lambda$)
and find complete agreement under the identification (\ref{param}).
Similarly,
one can use
$$\hat{u}_{t-1}^{(L)}=-2\pi{\sinh(\pi+\g H)k\sinh(\pi-\g)k\over
\sinh\pi k \sinh \g k}+\ldots, $$
$$\hat{U}_{t-1}^{(L)}=-2\pi{\sinh(2+H)\g k\over  2\sinh \g k
\cosh(\pi-
\g)k}+\ldots$$
to compare $U_{t-1}$ with (\ref{repIIII}) and obtain   agreement as
well.

Next, suppose that $h>h_{th}$ ($-t<H<-1$). To obtain  the boundary
S-matrices
for scattering on the ground state $|0\rangle_B$ set
$\omega=\omega^{\p}=1$ and
choose the boundary string to be the longest string, $n=n_{\ast}+1$
in (\ref{physbe}).
Then, using (\ref{ftII}), $\hat{f}^{\p}_{n_{\ast}+1;
t-1}=-\hat{f}^{\p}_{\ast}$ and
$$\hat{u}_a^{(L)}=2\pi{\sinh\g Hk\over\sinh\pi k}+\ldots,$$
\be
{\hat{u}_{t-1}^{(L)}=\hat{u}_a^{(L)} -
2\pi{\sinh(H+2[{1-H\over 2}])\g k \over \sinh\g k}+\ldots}
\label{defIII}
\ee
we obtain
$${\hat{U}_{n_{\ast}+1; a}^{(L)}\over 2\pi}={\sinh\g Hk\over
2\sinh\g k \cosh(\pi-\g)k}
+{\sinh(\pi-2\g)k\cosh(H+t-2n_{\ast})\g k \over  \sinh\g k
\cosh(\pi-\g)k}-$$
$$ -{2\cosh\g k \sinh n_{\ast}\g k \cosh(H-n_{\ast}+1)\g k \over
\sinh\g k \cosh(\pi-\g)k}+\dots,$$
$$\hat{U}_{n_{\ast}+1; t-1}^{(L)}=\hat{U}_{n_{\ast}+1; a}^{(L)}-
2\pi{\sinh(H+2[{1-H\over 2}])\g k\over \sinh\g k}+\ldots,$$
which agrees with (\ref{repI}), (\ref{repIIII}) under the identification 
(\ref{param}).
Note that the last relation, which follows directly from (\ref{defI}),
(\ref{defII}) and
(\ref{defIII}),
is valid also for  $\hat{U}_{n; a}$ and $\hat{U}_{n; t-1}$ with any
$n$.
In the same manner one can calculate the boundary S-matrices for
scattering
on the boundary n-strings  and
check that they indeed coincide   with (\ref{repII}), (\ref{repIII}) under the
condition
 (\ref{param}).
For this, one needs to take
$\omega=\omega^{\p}=1$ in (\ref{physbe}) and use (\ref{ftI}),
$\hat{f}^{\p}_{n; t-1}=-\hat{f}^{\p}_n$. Finally one
can compute also the boundary S-matrix
for the scattering on the $(n_{\ast}+1,N)$-strings, again in agreement
with the bootstrap results.

\section{Conclusion}

The question of boundary bound states even in the simple  Dirichlet
case appears rather difficult:
using the XXZ lattice regularization or equivalently
the Thirring model, we have only been able to recover
the $|\beta_n\rangle$ and $|\delta_{n=0,N}\rangle$ 
boundary bound states. A way out is
to consider solutions of the Bethe ansatz equations made of an $(n,N)$ string
superposed with the $n_{\ast}+1$ string that describes the ground state. This
is not allowed in principle in the model we consider because the Bethe
wave function vanishes
when two roots are equal. However, putting formally such a solution
in the equations gives the masses of the $|\delta_{n,N}\rangle$ states and the
S-matrix also agrees with the bootstrap results! But the meaning
of this is not clear to us.

\section{Remarks}

Finally let us mention that one can calculate the ground state energy
in the thermodynamic
limit  by solving the equation (\ref{beI}) for the ground state
density and using (\ref{Energy}):
$$E_{gr}=\int_0^{+\infty}2L\tilde{\rho}_a(\alpha)h(\alpha)d\alpha.$$
As a result, we get the combination $E_{gr}=E_{bulk}+E_{boundary}$,
where $E_{bulk}$ is the well-known sine-Gordon ground state energy
\cite{DdVg}:
$$E_{bulk}=-{Lm^2\over 4}\tan{\pi\g\over 2(\pi-\g)}$$
and $E_{boundary}$ is the contribution of the boundary terms
($\Lambda\to\infty$):
$$E_{boundary}
={m\over 2}\left[ {\sin{(H+2)\g\pi\over 2(\pi-\g)}
\over \sin{\pi^2\over 2(\pi-\g)}  } +1 + \cot{\pi^2\over 4(\pi-\g)}
\right].$$
We see that the ground state energy of the boundary sine-Gordon model
is a smooth function of the boundary magnetic field for the whole range
of $h$ in the XXZ regularization, hence of $\varphi_0$. The changes in ground
state structure do not affect $E$, as is expected since in such a unitary model
there is no
(one dimensional) boundary transition.


The finite size corrections to
the ground state energy themshelves (the genuine Casimir effect)
can be computed using the technique developed in \cite{DdVg}. We will
show how such a calculation can be carried out in the next chapter.
 It is also interesting to consider
the inhomogeneous 6-vertex model with an imaginary boundary magnetic field
ensuring commutation with $U_q sl(2)$ \cite{PSAL}. 
This should  lead to a solution of minimal
models
with integrable boundary conditions \cite{LMSS}.

\chapter{Boundary energy  in integrable quantum field theories}

The main  purpose of this chapter is to study the ground state energy of 1+1
integrable relativistic quantum field theories with boundaries. This involves
several questions. One is the energy associated with a boundary for an
infinite system, the other is the way the energy of the theory on an interval
varies with its length - the ``genuine'' Casimir effect.
The elegant method of Destri and de Vega \cite{DdVg} for the periodic
systems leads directly to the expression for the ground state energy
from which the infinite size contribution and the finite size
correction can be easily extracted. The heart of the DDV method is
a non-linear integral equation (\ref{renormbethe}) being derived 
from the Bethe equations.
We generalize the Destri-de-Vega method to the systems with boundaries
and apply it to compute the ground state energy for the boundary
sine-Gordon model. This chapter is a part of a more complete work
\cite{LMSS}.

\section{TBA for a QFT defined on cylinder}

For a Quantum Field Theory defined on a torus, the standard way to
compute its ground state energy is through the Thermodynamic Bethe
Ansatz \cite{AlZamo}. If the theory is defined on a circle of circumference $R$,
one switches to a modular transformed point of view where now the theory is
defined on a circle of very long circumference $L$ and at temperature $T=1/R$.
The free energy $F$ of the theory in the ``$R$ channel'' can be computed
using TBA, and it is simply related to the ground state energy $E^0(R)$ of
the theory in the ``$L$ channel'' by $F=-TLE^0(R)$. The limit $R\to 0$
corresponds to the UV limit and is described by a conformal field theory.
It is known that $E^0(R)$ behaves as $E^0(R)\sim\pi c/6R$ in this limit
\cite{CARDY},
where $c$ is the {\it central charge} of the underlying conformal
field theory \cite{BPZ}.\footnote{ More precisely, one has 
$$ E^0(R)={2\pi\over R}\left(\Delta+\overline{\Delta}-{c\over 12}\right) $$
where $(\Delta,\overline{\Delta})$ are the scaling dimensions of the
ground state.}

Consider now a quantum field theory on a cylinder of finite length $R$
and circumference $L$, with some boundary conditions $(a,b)$ at the ends
of the cylinder.

\begin{figure}
\epsfxsize = 100truemm
\centerline{\epsfbox{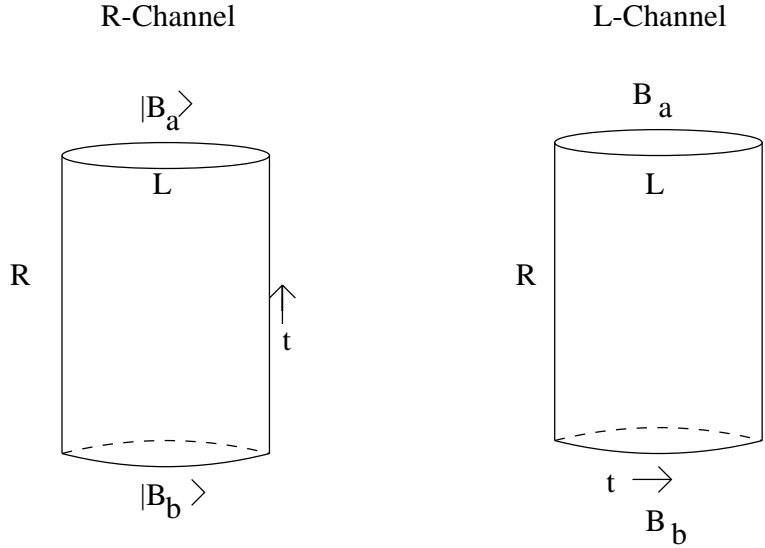}}
\caption{ Cylindrical geometry in the R and L channels.}
\end{figure}


Depending on quantization scheme, we have two possible ways to 
compute the partition function of the system
(see figure 5.1). The first possibility consists
in choosing as direction of time the horizontal axis and therefore the
partition function will be expressed as
\be
{
Z_{ab} = Tr \, e^{-L H_{ab}}\, ,
}\label{part}
\ee
where $H_{ab}$ is the Hamiltonian relative to the system with boundary
conditions $(a,b)$. In the second method the time evolution takes place
along the vertical axis and therefore the partition function is given by
the matrix element of the time evolution operator between the boundary states,
i.e.
\be
{
Z_{ab} = \langle B_a\mid e^{-R H}\mid B_b\rangle=e^{RF}\, ,
}\label{partt}
\ee
where now $H$ is the Hamiltonian of the bulk system. The ground state energy
$E_{ab}(R)$ is, by definition, the leading term arising in the large $L$ limit
of the first expression, eq. (\ref{part}), i.e.
\be
{
Z_{ab} \sim e^{-L E_{ab}^0(R)}\,.}\label{Llim}
\ee
However, in view of the equivalence of the two quantization schemes, we can
compute this quantity by looking at the large $L$ limit
of the second expression, eq. (\ref{partt}).
In the large $R$ limit, $R\to\infty$, the partition function (\ref{partt})
becomes
\be
Z_{ab}\to \langle B_a|0\rangle e^{-RE^0}\langle 0|B_b\rangle
\label{eqn:largeRlim}
\ee
where $|0\rangle$ is the ground state of Hamiltonian $H$ on the circle.
It is the only state that contributes in the limit $R\to\infty$;
other states do not propagate along the cylinder. The scalar products 
$g_a=\langle 0|B_a\rangle$ and $g_b=\langle 0|B_b\rangle$ are called
the {\it ground state degeneracy}. The quantity $S(0)=\ln(g_ag_b)$,
corresponding to the {\it zero-temperature entropy}, is one other universal
term that appears in the expansion of the free energy
\cite{AFFLECK}. For periodic
boundary conditions one must have $g_ag_b=1$.

Since eq. (\ref{partt}) employs the boundary states of the model, let us shortly
recall their basic properties (for more detail see the original
reference \cite{GZ}). In the QFT description of the model,   the information
on boundary conditions is encoded into a boundary state $\mid B\rangle$, 
which for infinite
length $L$ reads \cite{GZ}:
\be{
\mid B\rangle = g\exp\left[\int_0^{\infty} {d\th \over 2 \pi} K(\th)
A^{\dagger}(-\th) A^{\dagger}(\th)\right]\mid 0\rangle .}\label{bounk}
\ee
Here $g$ is an overall normalization related to the boundary entropy.
 For simplicity, we ignore possible additional
contributions to the boundary state from zero momentum particles. From the
point of view of QFT, the boundary state can be therefore regarded as a
particular state of the Hilbert space of the bulk theory, made of a
superposition of pairs of particles of equal and opposite momentum
(``Cooper pairs''). All information relative to a particular boundary
condition is encoded into the function $K(\th)$ which can be seen as the
elementary amplitude to create a virtual pair of particles.

In  this chapter we address the question
of computing the ground state energy of the sine-Gordon model with Dirichlet
boundary conditions. Instead of deriving a TBA in the R channel, we 
adopt the beautiful approach of Destri and de Vega \cite{DdVg}.
 This requires working for a while with the lattice theory, here
chosen to be the XXZ model with boundary magnetic fields.

\section{TBA for the inhomogeneous XXZ model with boundary fields}

like in the previous chapter, we  start from the inhomogeneous
6-vertex model with boundary fields $h$ as a regularization of the boundary
sine-Gordon model with Dirichlet boundary conditions, with the only
difference that now we consider the antiferromagnetic regime.

In the inhomogeneous antiferromagnetic 6-vertex model with anisotropy
parameter $\gamma$,
one gives an alternating  imaginary part
$\pm i\Lambda$ to the spectral parameter on alternating vertices
\cite{RS,DDV}. The scaling limit is given by taking
$\Lambda\rightarrow\infty$, $N\rightarrow\infty$, and the lattice spacing
$\Delta\rightarrow 0$, such that $R\equiv N\Delta$ remains finite. In the bulk,
this provides a regularization of the sine-Gordon model with
Lagrangian
\be
L_{SG}=\int_0^R dx\ \left[ \half (\partial\phi)^2 + \mu^2 \cos \beta_{SG}
\phi\right]\label{sg}
\ee
where  $\mu\propto {1\over\Delta}\exp(-{\rm const}\Lambda)$,
$\beta_{SG}^2=8(\pi-\gamma)$, and the field is fixed at $x=0$ and $x=R$
(Dirichlet
boundary conditions) to a value that is simply related to $h$ (see previous
chapter, eq. (\ref{param})).

The wave function of the inhomogeneous six-vertex model can be expressed in
terms of a set of ``roots''  $\alpha_j$, where $j=1\dots n$. They must be
solutions of the set of equations (\ref{foral}):
\bea
&&N\left[f(\alpha_j+\Lambda,\gamma)+f
(\alpha_j-\Lambda,\gamma)\right]+
2f(\alpha_j,\gamma H)=\nonumber \\
&&2\pi l_j + \sum_{m=1,m\ne
j}^n \left[f(\alpha_j-\alpha_m,2\gamma) +
f(\alpha_j+\alpha_m,2\gamma)\right],\label{DDV:foral}
\eea
where $l_j$ is an integer and all $\alpha_j$ are positive. The function $f$ is
defined as
$$f(a,b)=2\tan^{-1}\left(\cot{b\over 2}\tanh a\right)$$
%
and
\be
{H\equiv{1\over\gamma} f(i\gamma,-i\ln(h+\cos\gamma)).}\label{forH}
\ee
 By construction of the Bethe-ansatz wave
function, $\alpha_j>0$. Even though there is a solution of  (\ref{DDV:foral})
 with one
vanishing root for any $N$ and $n$, we emphasize that $\alpha_j=0$ is
{\bf not} allowed  because the wave function vanishes identically in this
case. Observe that equations (\ref{DDV:foral}) are formally satisfied
as well by the opposite of the roots, $-\alpha_j$. Often in what follows
we shall consider that the roots take both signs in order to rewrite equations
in a way which is similar to the bulk case.

 For simplicity, we restrict to the case $\gamma={\pi\over t}$ where $t$
is an integer, and restrict to the choice $\epsilon=-1$ \cite{RS}.
In the sine-Gordon model, this falls in the repulsive regime. We make the
standard assumption that all the solutions of interest are collections of
``$k$-strings'' for $k=1,2\dots t-1$ and antistrings $a$.
 A $k$-string is a group of $\alpha_j$ in the pattern
$\alpha^{(k)}-i\pi(k-1),  \alpha^{(k)}-i\pi(k-3), \dots, \alpha^{(k)}+
i\pi(k-1)$ where $\alpha^{(k)}$ is real. The antistring has
$\alpha_j=\alpha^{(a)}+i\pi$, where  $\alpha^{(a)}$ is real.

The thermodynamic  limit is obtained by sending $N\to\infty$. In this case,
we can define densities of the different kinds of solutions.
The number of allowed solutions of
(\ref{DDV:foral}) of type $k$ in the interval $(\alpha,\alpha+d\alpha)$ is
$2 N (\rho_k(\alpha) + \rho_k^h(\alpha)) d\alpha$, where $\rho_k$ is the
density
of ``filled'' solutions (those which appear in the sum in the right-hand-side
of (\ref{DDV:foral}) ) and $\rho_k^h$ is the density of ``holes'' (unfilled solutions).
The densities $\rho_a$ and $\rho_a^{h}$ are defined likewise for the
antistring. The ``bare'' Bethe ansatz equations follow from taking the
derivative of (\ref{DDV:foral}). 
For $\gamma=\pi/t$ they can be written in the form:
\bea
2\pi (\rho_k + \rho_k^h)&=&  a_k(\alpha) -
\dot{\phi}_{k,t-1}*\rho_{a} +\sum_{l=1}^{t-1}
\dot{\phi}_{kl}*\rho_l +{1\over 2N}u_k\nonumber \\
2\pi(\rho_{a} +\rho_a^h)&=& 2\pi(\rho_{t-1}+ \rho_{t-1}^h) +
{1\over 2N}(u_a-u_{t-1})\label{new:bba}
\eea
These densities are originally defined for $\alpha>0$, but the equations allow
us to define $\rho_k(-\alpha)\equiv\rho_k(\alpha)$ in order to rewrite the
integrals to go from $-\infty$ to $\infty$. The kernels in these equations are
defined most easily in terms of their Fourier transforms
\be
{\hat{f}(x)=\int_{-\infty}^\infty {d\alpha\over 2\pi}e^{i\alpha
tx/\pi}
f(\alpha),\quad f(\alpha)={t\over\pi}\int_{-\infty}^\infty e^{-i\alpha
tx/\pi}\hat{f}(x)dx.} \label{new:fourier}
\ee
One has
\be
{\hat{\dot{\phi}}_{kl}(x)=
\delta_{ab}-2{\cosh x \sinh(t-k)x
\sinh lx \over \sinh  x\sinh tx},}\label{new:forphi}
\ee
for $k\geq l$  with $\dot{\phi}_{lk}=\dot{\phi}_{kl}$, and
\bea
\hat a_k&=&{\sinh (t-k) x
\over\sinh  tx} \cos\Lambda tx/\pi\nonumber \\
\hat u_k&=&2 {\sinh (t-H)x \sinh kx  \over \sinh x \sinh tx} +
{\sinh (t-2k)x/2\over \sinh  tx/2} - 1\label{forps} \\
\hat u_a&=&2{\sinh Hx \over \sinh  tx}- {\sinh
(t-2)x/2\over \sinh tx/2} - 1, \nonumber
\eea
with in particular $a_1(\alpha)={1\over 2}\left[\dot{f}(\alpha+\Lambda,\gamma)
+\dot{f}(\alpha-\Lambda,\gamma)\right], \phi_{11}(\alpha)=-f(\alpha,2\gamma)$.
The boundary manifests itself in the first term in $u_k$; notice that even for
$h=0$, it still modifies the equations. A few technicalities account for the
other terms (these are relevant here because we are interested in subleading
boundary effects). The second term in $u_k$ arises from the fact that the sum
in (\ref{DDV:foral}) does not include the term $m=j$; the integration over densities
includes such a contribution and so it must be subtracted off by hand. The
third term in $u_k$ arises because $\rho$ and $\rho^h$ are defined for allowed
solutions, while as already explained, $\alpha=0$ is not allowed because it
does not give a valid wavefunction. Since it is a valid solution of 
(\ref{DDV:foral}) but
is not included in the densities, we must subtract an explicit ${2\pi\over
2N}\delta(\alpha)$ (corresponding to $1/2N$ in Fourier space). Explicitly,
one has
\be
{u_1=2\dot{f}(\alpha,\gamma
H)+2\dot{f}(2\alpha,2\gamma)-2\pi\delta(\alpha).}
\label{formi}
\ee

For compactness we rewrite (\ref{new:bba}) as
\be
{2\pi\sigma^{(k)}(\rho_k+\rho_k^h)=a_k(\alpha)+\sum_l
\dot{\phi}_{kl}*\rho_l+{1\over 2N}u_k,}
\label{newbba}
\ee
where $\sigma^{(k)}=-1$ for the antistring.

The energy reads, with proper hamiltonian normalization,
\be
{{E^{latt}\over 2N}=-{1\over t}\sum_k \int_{-\infty}^\infty
a_k(\alpha)\rho_k(\alpha)d\alpha.} \label{DDV:ener}
\ee

It is easy to write the thermodynamic Bethe ansatz  for this model. One
finds that the TBA equations, since they are obtained by a variational method,
do not depend on boundary terms, and read as usual
\be
{-{2\over t}a_k(\alpha)=T\ln\left(1+e^{\epsilon_k}\right)-
T\sum_l\sigma^{(l)}{A_{kl}\over 2\pi}*\ln\left(
1+e^{-\epsilon_l}\right),}\label{TBA}
\ee
where
\be
{A_{kl}(\alpha)=2\pi\sigma^{(k)}\delta_{kl}\delta(\alpha)-
\dot{\phi}_{kl}\, .} \label{Adef}
\ee
The free energy does depend on the boundary term and reads
\bea
F^{latt}=&-&TN\sum_k\int_{-\infty}^\infty
\sigma^{(k)}a_k(\alpha)\ln
\left(1+e^{-\epsilon_k}\right){d\alpha\over 2\pi}\nonumber \\
&-&{T\over 2}\sum_k\int_{-\infty}^\infty
\sigma^{(k)}u_k\ln\left(1+e^{-\epsilon_k}\right){d\alpha\over2\pi}.
\label{freeen}
\eea
In the above formulas the temperature $T$ corresponds in the two-dimensional
point of view to having a cylinder of radius $L=1/T$. We can deduce from this
result the ground state energy. Indeed recall that the ground state is
obtained by $\rho_k=0,k\neq 1$ and $\rho^h_1=0$ so
\be
{E^{latt}=-N\int_{-\infty}^\infty a_1(\alpha)
|\epsilon_1^-|{d\alpha\over 2\pi}-{1\over 2}\int_{-\infty}^\infty
u_1|\epsilon_1^-|{d\alpha\over
2\pi},}
\label{gren}
\ee
where from (\ref{TBA}) we have
\be
{\hat{\epsilon}_1^{\ -}=-{1\over t}{\cos \left({\Lambda
tx/\pi}\right)\over\cosh x}.}\label{epsmin}
\ee
Replacing and  using (\ref{forps}) we find
\bea
E^{latt}_{bulk}&=&-{N\over\pi}\int_{-\infty}^\infty
\cos^2\left({\Lambda
tx\over\pi}\right){\sinh (t-1)x\over\cosh x\sinh tx}\nonumber \\
E^{latt}_{bdry}&=&-{1\over 2\pi}\int_{-\infty}^\infty {\cos(\Lambda tx/
\pi)\over\cosh x}
\left(2{\sinh (t-H)x\over\sinh tx}+{\sinh (t-2)x/2\over\sinh
tx/2}-1\right) \label{DDV:energs}
\eea
where we used the formula
$$
\int_{-\infty}^\infty a(\alpha)b(\alpha)d\alpha=2t\int_{-\infty}^\infty
\hat{a}(x)\hat{b}(-x)dx \, .
$$

In the continuum limit $\Lambda\rightarrow\infty$ the energy contains various
terms. We keep only the finite part which is obtained by closing the
above integrals in the upper half plane and selecting the pole
at $x=i{\pi\over 2}$, leading to
\bea
E_{bulk}&=&R{m^2\over 4}\cot {t\pi\over 2}\nonumber \\
E_{bdry}&=&-{m\over 2}\left(2{\sin (t-H)\pi/2\over\sin t\pi/2}-\cot
{t\pi\over 4}-1\right)\,,\label{contener}
\eea
where $m$ is the soliton mass,
\be
{m=2e^{-t\Lambda/2}\, .}
\label{solmass}
\ee
All these results trivially generalize to the case of two different
boundary fields by splitting the $H$ dependent terms into the sum
of an $H$ and an $H'$ term. The bulk result agrees with what is obtained
by other methods. As in the bulk case, when $t$ is even, there are additional logarithmic terms.
 The boundary entropy is actually the same in the UV and IR limits.

\section{The Destri De Vega equations for the \\ boundary sine-Gordon model}

We now would like to compute the complete Casimir effect in a theory
with boundary. TBA in the R channel is pretty intricate in the non diagonal
case. We adopt an alternative method elaborated by Destri and De Vega (DDV) in
the periodic case.

\subsection{The  DDV equations with boundary conditions}

Consider eq. (\ref{DDV:foral}) which we rewrite as
\be
{2Np(\alpha_j)+p_{bdry}(\alpha_j)
+\sum_{\alpha_m>0}\phi(\alpha_j-\alpha_m)+\phi(\alpha_j+\alpha_m)=
2\pi n_j,}\label{newDDV:foral}
\ee
where the sum  runs over all roots (including $m=j$) and we introduced the
notations
\be
p(\alpha)\equiv{1\over 2}\left[f(\alpha+\Lambda,\gamma)+
f(\alpha-\Lambda,\gamma)\right],\quad p_H(\alpha)\equiv f(\alpha,\gamma
H),\quad \phi(\alpha)\equiv \phi_{11}(\alpha), \nonumber
\ee
\be
p_{bdry}(\alpha)=2p_H(\alpha)-\phi(2\alpha). \label{pdef}
\ee

The ground state is obtained by filling the real positive solutions,
 $\alpha_j=0$ excepted. This corresponds to the choice $n_j=1,2,\ldots$.
Recall that
if $\alpha_j$ is solution of
(\ref{newDDV:foral}) with some $n_j$, so is formally $-\alpha_j$ (with $-n_j$).
Given the set of roots $\{\alpha_j>0\}$ representing the ground state, one can
construct the {\it counting function} as follows:
\be
{f(\alpha)\equiv 2iNp(\alpha)+
ip_{bdry}(\alpha)
+i\sum_{\alpha_m>0}\phi(\alpha-\alpha_m)+\phi(\alpha+\alpha_m),}
\label{deff}
\ee
Define then
\be
{Y(\alpha)\equiv e^{f(\alpha)}.}\label{defy}
\ee
We have $Y(\alpha_j)=Y(-\alpha_j)=1$ for every root $\alpha_j$ from
the ground state, as well as $Y(0)=1$.
 Therefore we
can rewrite (\ref{deff}) as
\be
{f(\alpha)=2iNp(\alpha)+
ip_{bdry}(\alpha)-i\phi(\alpha)-
\int_C\phi(\alpha-\alpha'){\dot{Y}(\alpha')\over
1-Y(\alpha')}{d\alpha'\over 2\pi},} \label{newwDDV:foral}
\ee
where $i\phi(\alpha)$ at the right-hand side takes care of the unwanted
contribution of the pole $\alpha'=0$
and the contour $C$ consists of two parts as shown in figure 5.2, $C_1$ above
 and $C_2$ below the real axis.

\begin{figure}
\epsfxsize = 125truemm
\epsfysize = 80truemm
\centerline{\epsfbox{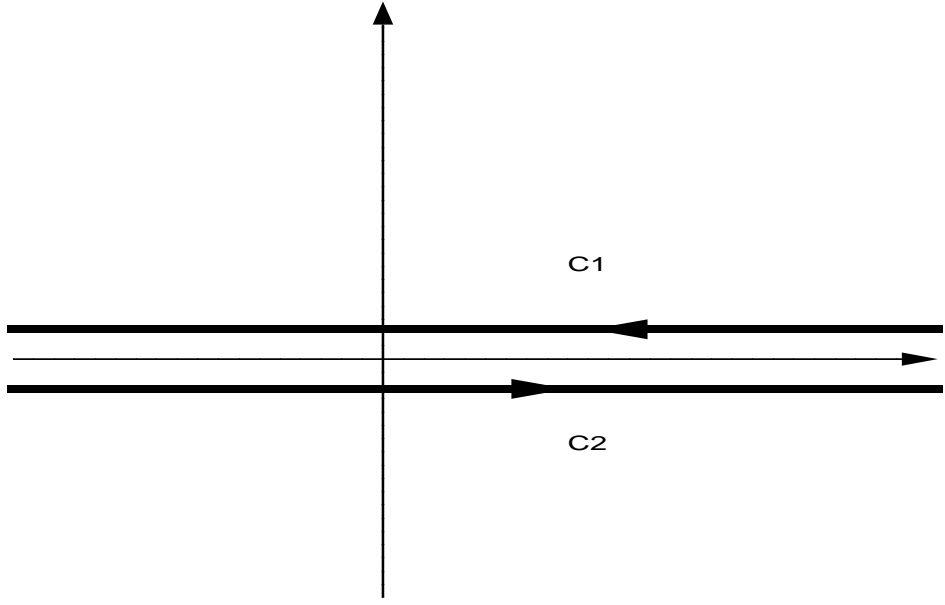}}
\caption{ Contours of integration in the DDV method.}
\end{figure}

 Like in the bulk case \cite{DdVg},
 simple manipulations
allow us to rewrite this non-linear integral equation as
\bea
f(\alpha)&=&2iNP(\alpha)+iP_{bdry}(\alpha)+
\int_{C_1}\Phi(\alpha-\alpha')\ln\left(1-e^{f(\alpha')}\right)
d\alpha'\nonumber \\
&+&\int_{C_2}\Phi(\alpha
-\alpha')\ln\left(1-e^{-f(\alpha')}\right)d\alpha'.\label{ddvi}
\eea
In  equation (\ref{ddvi}) one has
\be
{\Phi(\alpha)=-
{t\over 2\pi^2}\int_{-\infty}^{+\infty}
dx e^{-itx\alpha/\pi}{\sinh(t-2)x\over 2\sinh(t-1)x\cosh x},}
\label{phiren}
\ee
together with
\be
{P(\alpha)=
\int_{-\infty}^{+\infty}dx{e^{-itx\alpha/\pi}-1\over -ix}
{\cos(\Lambda tx/\pi)\over 2\cosh x}}\label{pren}
\ee
and
\bea
P_{bdry}&=&
\int_{-\infty}^{+\infty}dx{e^{-itx\alpha/\pi}-1\over -ix}
\left[{\sinh (t-H)x\over\sinh (t-1)x\cosh x}\right.\nonumber \\
&+&\left.{\sinh (t-2)x/2\cosh tx/2\over\sinh (t-1)x\cosh x}+
{\sinh (t-2)x\over 2\sinh (t-1)x\cosh x}\right]\label{pbdrren}
\eea
Obtaining (\ref{ddvi}) requires some care with the definition of logarithms.
One proceeds as follows. Before integrating by parts in the integral over
$C_2$ one factors out ${d\over dx}\ln Y(x)$:
$$
-{\dot{Y}(x)\over 1-Y(x)}={d\over dx}\ln[1-Y^{-1}] +
{d\over dx}\ln Y(x).
$$
Then both integrals over $C_1$ and $C_2$ can be taken by parts, resulting in
\bea
f(\alpha)&-&\int_{-\infty}^{+\infty}\dot{\phi}(\alpha-\alpha')
f(\alpha'){d\alpha'\over 2\pi}=2iNp(\alpha)+
ip_{bdry}(\alpha)-i\phi(\alpha)\nonumber \\
&-&\int_{-\infty}^{+\infty}\dot{\phi}(\alpha-\alpha'-i0)\ln[1-Y(\alpha'+i0)]
{d\alpha'\over 2\pi} \nonumber \\
&+&\int_{-\infty}^{+\infty}\dot{\phi}(\alpha-\alpha'+i0)
\ln[1-Y^{-1}(\alpha'-i0)]
{d\alpha'\over 2\pi}, \nonumber
\eea
(surface terms from $C_1$ and $C_2$ cancel against each other provided
$N,\Lambda$ are finite).
To make the source term vanish at infinity  we
take the derivative of both sides in the latter equation, after which it
can be Fourier-transformed and ``dressed'' by the factor
$(1-\hat{\dot{\phi}})^{-1}$. Finally, one goes back in the rapidity space
and integrates the equation, using $f(0)=0$, to obtain (\ref{ddvi}).

The energy of the ground state configuration can be expressed as
\bea
E^{latt}&=&-{2\over t}\sum_{\alpha_j>0} \dot{p}(\alpha_j)=
-{1\over t}\sum_{\alpha_j}\dot{f}(\alpha_j-\Lambda, \gamma)+
\dot{f}(-\alpha_j-\Lambda, \gamma)\nonumber \\
&=&
{1\over t}\dot{f}_{\gamma}(\Lambda)+
{1\over t}\int_C \dot{f}_{\gamma}(\alpha-\Lambda)
{\dot{Y}(\alpha)\over 1- Y(\alpha)}
{d\alpha\over 2i\pi},\label{DDV:enerix}
\eea
where $f(\alpha, \gamma)\equiv f_{\gamma}(\alpha)$.
One finds after exactly the same manipulations as above,
\bea
&E&={1\over t}\dot{f}_{\gamma}(\Lambda)-
{1\over t}\int_{-\infty}^{+\infty}f_{\gamma}''(\alpha-\Lambda+i0)
\ln[1-Y(\alpha+i0)]{d\alpha\over 2i\pi}\label{tati} \\
&+&{1\over t}\int_{-\infty}^{+\infty}f_{\gamma}''(\alpha-\Lambda-i0)
\ln[1-Y^{-1}(\alpha-i0)]{d\alpha\over 2i\pi}
+{1\over t}\int_{-\infty}^{+\infty}f_{\gamma}''(\alpha-\Lambda)f(\alpha)
{d\alpha\over 2i\pi}.\nonumber
\eea
Substituting (\ref{ddvi}) instead of $f(\alpha)$ in the last term of the latter,
we obtain
\bea
E^{latt}&=&E^{latt}_{bulk}+E^{latt}_{bdry}
 - {i\over t}\int_{-\infty}^{+\infty}
s(y-\Lambda+i0)\ln[1-Y(y+i0)]{dy\over 2\pi}\nonumber \\
&+&{i\over t}\int_{-\infty}^{+\infty}
s(y-\Lambda-i0)\ln[1-Y^{-1}(y-i0)]{dy\over 2\pi}, \label{DDV:eneriixx}
\eea
where we defined $s(y)$ by
\be
{s(y)=i\int_{-\infty}^{+\infty}{kdk\over 2\cosh\gamma k}
e^{-iky}={t^2\over 4}{\tanh(ty/2)\over\cosh(ty/2)}.}
\label{epsdefx}
\ee
The last two terms in (\ref{DDV:eneriixx}) represent finite-size corrections to
the ground state energy, while
\bea
E^{latt}_{bulk}+E^{latt}_{bdry}&=&{1\over
t}\dot{f}_{\gamma}(\Lambda)
-{1\over t}\int_{-\infty}^{+\infty}{dk\over 2\pi}
e^{-ik\Lambda}\hat{\dot{f}}_{\gamma}(k){2N\hat{\dot{p}}(k)+\hat{\dot{p}}_{bdry}
(k)-
\hat{\dot{\phi}}(k)\over 2\pi-\hat{\dot{\phi}}(k)}\nonumber \\ 
&\equiv&
{1\over t}\dot{f}_{\gamma}(\Lambda) - {2N\over t}\int_{-\infty}^{+\infty}
\dot{f}_{\gamma}(\alpha-\Lambda)\rho(\alpha)d\alpha, \label{kust}
\eea
where the function $\rho(\alpha)$ defined so satisfies the following equation:
\be
{2\pi\rho(\alpha)=\dot{p}(\alpha)+\dot{\phi}\ast\rho(\alpha)
+{1\over 2N}\left(\dot{p}_{bdry}(\alpha)-\dot{\phi}(\alpha)\right),
}\label{kustik}
\ee
which can be checked by solving this linear equation in Fourier space.
Introduce $\rho_1(\alpha)=\rho(\alpha)-\delta(\alpha)/2N$.
Then
\be
{ E^{latt}_{bulk}+E^{latt}_{bdry}=- {2N\over
t}\int_{-\infty}^{+\infty}
\dot{f}_{\gamma}(\alpha-\Lambda)\rho_1(\alpha)d\alpha}\label{dela}
\ee
and, by virtue of (\ref{kustik}),  $\rho_1$ satisfies the equation
\be
{2\pi\rho_1(\alpha)=\dot{p}(\alpha)+\dot{\phi}\ast\rho_1
(\alpha)
+{1\over 2N}\left(\dot{p}_{bdry}(\alpha)-2\pi\delta(\alpha)\right).}
\label{kustiki}
\ee
Hence $\rho_1$ is the density of the ground state configuration (\ref{new:bba})
(see also (\ref{formi})) and $E^{latt}_{bulk}$
and $E^{latt}_{bdry}$
coincide with the quantities computed in the previous section.

\subsection{The continuum DDV equations}

Having checked the correct values of bulk and boundary energies,
we now let the cutoff $\Lambda\to\infty$ according to $\Lambda = {2\over t}
\log (2/m\Delta)$ with  the size of the system $R=N\Delta$ and the   physical
mass $m={2\over\Delta}e^{-t\Lambda/2}$ fixed, and work only with
the renormalized theory. In that limit one has, evaluating all integrals
in Fourier transforms and keeping the leading terms,
\be
{s(\alpha+\Lambda)+s(\alpha-\Lambda)=
-{t^2\over 2}m\sinh\theta,\qquad
NP(\theta)=mR\sinh\theta,} \label{conteqs}
\ee
where we set
\be
{\theta={t\alpha\over 2}.}\label{defrap}
\ee
Recall indeed that when one studies the excitations of the model,
a relativistic dispersion relation is obtained provided
the rapidity in the relativistic theory and the ``bare rapidity'' of the
Bethe excitations are related by (\ref{defrap}). 
We redefine implicitely all functions
to depend on $\theta$ from now on. The energy therefore reads now,
\bea
E&=&E_{bulk}+E_{bdry}+
{1\over 2}\int_{C_1} m\sinh\theta \ln\left(1-e^{
f(\theta)}\right){d\theta\over
2i\pi}\nonumber \\
&+&{1\over 2}\int_{C_2} m\sinh\theta \ln\left(1-e^{
-f(\theta)}\right){d\theta\over
2i\pi},\label{newenrform}
\eea
where $f$ is solution of the integral equation
\bea
f(\theta)&=&2imR\sinh\theta +i
P_{bdry}(\theta)+\int_{C_1}\Phi(\theta-\theta')
\ln\left(1-e^{f(\theta')}\right){d\theta'}\nonumber \\
&+&\int_{C_2}\Phi(\theta-\theta')
\ln\left(1-e^{-f(\theta')}\right){d\theta'},\label{renormbethe}
\eea
and
\be
{\Phi(\theta)=-\int_{-\infty}^\infty
{dx\over 2\pi^2}{\sinh (t-2)x\over
\cosh x\sinh(t-1)x}e^{2ix\theta/\pi},}
\label{kerndef}
\ee
which can be identified with
\be
{\Phi(\theta)=-{1\over 2i\pi}{d\over d\theta}\ln S_{++}(\theta),}
\label{ident}
\ee
where $S_{++}$ is the soliton-soliton $S$ matrix element 
(see eq. (\ref{eq:S-matr})).

\subsection{The Casimir effect}

We define the effective  central charge by the formula
\be
{E=E_{bulk}+E_{bdry}-{\pi c_{eff}\over 24R}.}\label{cdeff}
\ee
 From (\ref{newenrform}) we find
\be
{c_{eff}(mR)=-{6 R\over i\pi^2}\left\{\int_{C_1}m\sinh\theta\ln\left(
1-e^{f(\theta)}\right)d\theta+\int_{C_2}m\sinh\theta\ln\left(
1-e^{-f(\theta)}\right)\right\}d\theta.} \label{cform}
\ee
To study the ultraviolet behaviour of the above expression, let us
use $f(\overline{\theta})=\overline{f(\theta)}$ and rewrite
 (\ref{renormbethe}) as
\be
{f(\theta)=2imR\sinh\theta+iP_{bdry}(\theta)-
2i\int_{-\infty}^\infty \Phi(\theta-\theta')\hbox{ Im}
\ln\left[1-e^{f(\theta'+i0)}\right]d\theta',} \label{rewrit}
\ee
and (\ref{cform}) as
\be
{c_{eff}(mR)={12mR\over\pi^2}\int_{-\infty}^\infty
d\theta\sinh\theta\hbox{ Im}
\ln\left[1-e^{f(\theta+i0)}\right].}\label{rewriti}
\ee
It might be useful to remind the reader of the form of the corresponding
bulk equations \cite{DdVg}:
\be
{c_{eff}(mR)={6mR\over\pi^2}\int_{-\infty}^\infty
\sinh\theta\hbox{ Im}
\ln\left[1+e^{f(\theta+i0)}\right]d\theta,}
\label{rewritibulk}
\ee
with $f$ satisfying
\be
f(\theta)=imR\sinh\theta+i\omega-2i\int_{-\infty}^\infty
\Phi(\theta-\theta'){\rm Im}
\ln\left[1+e^{f(\theta'+i0)}\right]d\theta',
\label{rewritbulk}
\ee
and $\omega$ is the twist of the 6-vertex model. Note the deep similarity
between these two systems; the factors of $2$ can obviously be
absorbed in a redefinition of $mR$ and the minus sign
in the arguments of logarithms in a redefinition of $f$, so the only
essential difference
is that the twist angle $\omega$ is replaced by $P_{bdry}$. It is well known
that the twist corresponds to a soliton fugacity \cite{FS}
 so we see that the boundary acts by some effective, rapidity dependent
fugacity.

In the limit when $R\rightarrow 0$, only the region $|\theta|$ large
contributes to $c_{eff}$. Let us focus on the limit  $\theta\gg 1$, the results
for negative
$\theta$ following by symmetry. Then one finds $f\approx f_K$, where
\be
{f_K(\theta) =imR e^\theta+iP_{bdry}(\infty)
-2i\int_{-\infty}^\infty \Phi(\theta-\theta')\hbox{ Im}
\ln\left[1-e^{f_K(\theta'+i0)}\right]d\theta',}
\label{bethekink}
\ee
together with
\be
{c_K(mR)= {6\over\pi^2}\int_{-\infty}^\infty mRe^{\theta}
{\rm Im} \ln\left[1-e^{f_K(\theta+i0)}\right]d\theta.}
\label{uvc}
\ee
It is now useful to recall some well known results about
dilogarithms \cite{KR}. Define
\be
{L(x)\equiv \int_0^x du\left[{\ln(1+u)\over u}-
{\ln u\over 1+u}\right].}\label{dilogdef}
\ee
Assume
\be
{-i\ln F(x)=\phi(x)+2\int_{-\infty}^\infty dy \ G(x-y)\hbox{ Im}
\ln\left[
1+F(y+i0)\right],} \label{suppi}
\ee
with  $G$ an even function. Then one has
\bea
\hbox{Im}\int_{-\infty}
^\infty dx\phi'(x)\ln\left[1+F(x+i\epsilon)\right]={1\over 2}\hbox{Re}\left\{
L[F(-\infty)]-L[F(\infty)]\right\}\nonumber \\
+{1\over 2}\hbox{Im}\left\{
\phi(\infty)\ln\left[1+F(\infty)\right]-\phi(-\infty)
\ln\left[1+F(-\infty)\right]
\right\},\label{new:main}
\eea
(where we did not write the $i0$ part of some arguments for
simplicity). Set $F=e^{f_K-i\pi}$ and denote
$P_{bdry}(\infty)\equiv\sigma$. Then, according to (\ref{bethekink}):
 $\phi=mRe^\theta+\sigma-\pi, \quad G=-\Phi$.
We have $\phi(-\infty)=\sigma-\pi$, $\phi(\infty)=\infty$.
One has also $F(\infty+i0)=0$, and from this and (\ref{bethekink}) one can get
the value of $F$ at $-\infty$:
\bea
{f_K(-\infty)\over i}&=&
\sigma-2\hbox{ Im}\ln\left[1-e^{f_K(-\infty+i0)}\right]
\int_{-\infty}^\infty \Phi(\theta)d\theta \nonumber \\
&=&\sigma+{t-2\over t-1}\hbox{ Im}\ln\left[1-e^{f_K(-\infty+i0)}\right].
\label{fff}
\eea
So, if $e^{f_K(-\infty+i0)}=e^{i\omega}$ one may use
$\hbox{ Im}\ln(1\pm e^{i\omega})= {1\over 2i}\ln(\pm e^{i\omega})$ to   find
\be
{e^{i\omega}=
-\exp\left\{2i{t-1\over t}\sigma+2i{\pi\over t}\right\}.}\label{omm}
\ee
 From (\ref{new:main}) it follows that
$$\hbox{ Im}\int_{-\infty}^\infty
mRe^{\theta}\ln(1-e^{f_K(\theta+i0)})d\theta={1\over 2}
\hbox{ Re}L(-e^{i\omega})-{1\over 2}\left(\sigma-\pi\right)
\left({t-1\over t}\sigma-\pi+{\pi\over t}\right).$$
In the region $\theta\ll 1$ we have $f\approx f_A$, and similar calculations
yield
$$\hbox{ Im}\int_{-\infty}^\infty
mRe^{-\theta}\ln(1-e^{f_A(\theta+i0)})d\theta=-{1\over 2}
\hbox{ Re}L(-e^{-i\omega})+{1\over 2}\left(\pi-\sigma\right)
\left(-{t-1\over t}\sigma+\pi-{\pi\over t}\right).$$
Collecting both $\theta\gg 1$ and  $\theta\ll 1$ contributions we obtain
\bea
c_{UV}&=&{6\over\pi^2}\left\{{1\over 2}
\left[L(-e^{2i\omega)})+
L(-e^{-2i\omega})\right]+
 {t-1\over t}\left(\sigma-\pi\right)^2\right\}=\nonumber \\
&=&{6\over\pi^2}
\left[{\pi^2\over 6}-{t-1\over
t}\left(\sigma-\pi\right)^2\right].\label{ccc}
\eea
{}From (\ref{pbdrren}) we get
\be
{P_{bdry}(\infty)\equiv\sigma= 2\pi-{\pi\over 2}{H+H'\over t-1}.}
\label{pinf}
\ee
Finally, from this we find
\be
c_{UV}=1-6{t-1\over t}\left(1-{H+H'\over 2(t-1)}\right)^2.\label{cccc}
\ee
In the  case with no boundary field, $H=H'=t-1$ so $c_{UV}=1$ as expected.

\vspace{1cm}

{\it Remark.} So far we tacitly assumed that $0<H<t-1$. In general,
from relation (\ref{forH})
it follows that when $h>0, \quad H$ varies between $-t-1$
and $-1$, while when $h<0, \quad -1<H<t-1$. To generalize our results,
we should use the most general form of $p_H$:
\be
{\hat{\dot{p}}_H(k)=\int_{-\infty}^\infty\dot{p}_H(\alpha)e^{ik\alpha}d\alpha=
2\pi
\hbox{ sign}(H){\sinh(\pi-\omega_H)k\over\sinh\pi k},\quad -\pi<\gamma H<\pi,}
\label{balda}
\ee
where we defined $\omega_H\equiv|\gamma H|$. For $-2\pi<\gamma H<-\pi$
set $\omega_H=2\pi+\gamma H$ and $\hbox{ sign}(H)=1$ in (\ref{balda}).
Then (\ref{pinf}) generalizes to
$$
\sigma=2\pi-{\pi\over 2}{\omega_H+\omega_H'\over \pi-\gamma}
$$
if $H$ and $H'$ are both positive or $-2t<(H, H')<-t$, and
$$
\sigma=2\pi-{\pi\over 2}{4\pi-\omega_H-\omega_H'\over \pi-\gamma}
$$
if they are both negative, but greater than $-t$.
In the case when $0<H<t-1$ and $-t<H'<-1$  (that is, $h<0, h'>0$) we get:
$$
\sigma=\pi+{\pi\over 2}{\omega_H'-\omega_H-2\gamma\over \pi-\gamma}.
$$
So, when $\omega_H'-\omega_H=2\gamma$ we have $\sigma=\pi$ and $c_{UV}=1$,
as in the free case. The condition $\omega_H'-\omega_H=2\gamma$
is equivalent to $h=-h'$, as could be seen from (\ref{forH}). 
That $c=1$ when the
two surface field are real and opposite is well known from lattice studies 
\cite{DDV:BS}.

\section{Remarks}

A particularly interesting case is when the XXZ chain or the inhomogeneous
6-vertex model commutes with the quantum group $U_qsl(2)$. In that
case $h=-h'=2i\sin\gamma$ and the net result is that all $H$ dependent terms
simply disappear from the equations, so, in particular
\be
{E_{bdry}={m\over 2}\left(\cot{t\pi\over 4} +1\right).}\label{equu}
\ee
At the $N=2$ supersymmetric point, $t=3$, the boundary energy vanishes,
a result
well expected from supersymmetric considerations. More generally, it vanishes
if $t=4n+3$, $n$ an integer. Notice that the bulk energy vanishes
for $t$ an odd number (as
a consequence of the generalized fractional $N=2$ supersymmetries
studied in \cite{vafa}).

In the quantum group symmetric case \cite{PSAL}
 one has $H+H'=2t$ so from (\ref{cccc})
\be
{c=1-{6\over t(t-1)},}
\label{cmin}
\ee
the expected result for the restricted sine-Gordon model \cite{DDV:RSm,DDV:LCL,
DDV:BLCL}.

  \chapter{Surface excitations and surface energy
 of the  antiferromagnetic XXZ chain by the Bethe ansatz approach}

We study an open $XXZ$ 
 chain in the regime $\Delta>1$ with a boundary magnetic field $h$
and discuss some of its peculiar features due to the presence of boundary.
In the Bethe ansatz formalism, boundary bound states are represented by
the ``boundary strings'' as described in chapter 4. We find
that for certain values of $h$
 the ground state wave function contains
 boundary strings, and from this infer the existence of two ``critical''
fields in agreement with \cite{miwa}.
 An expression
for the vacuum surface energy in the thermodynamic limit is derived and found
to
be an analytic function of $h$.
We argue that boundary excitations appear only in pairs
with ``bulk'' excitations or with boundary excitations at the other end of the
chain. The case where
the magnetic fields at the left and the right boundaries are antiparallel
has non-trivial differences with the case of the parallel fields.
The Ising ($\Delta=\infty$) and isotropic ($\Delta=1$)
limits are discussed thoroughly and found helpful for the intuitive 
understanding
of the behavior of the boundary $XXZ$ chain at arbitrary $\Delta$. 
This section is based on the work \cite{XXZ:KS}.

\section{Introduction}

In this chapter we study
the $XXZ$ chain with even number of spins $L$ in a boundary magnetic field,
\be
{\cal H}= {1\over 2 }
\left\{\sum_{i=1}^{L-1} \left(
\sigma^x_i \sigma^x_{i+1} + \sigma^y_i
\sigma^y_{i+1}+\Delta\sigma^z_i
\sigma^z_{i+1}\right) +h_1\sigma_1^z+h_2\sigma_L^z \right\}, \label{ham}
\ee
in the regime $\Delta>1$, $h_1\geq 0$, $h_2\leq 0$,
 focusing on the effects peculiar to systems with boundaries
\cite{McCoy}. At $h_1=h_2=0$ this model describes one-dimensional antiferromagnet
 with non-magnetic impurities, accessible experimentally. We exploit the Bethe
 ansatz
solution for this model, first derived  in \cite{ABBBQ}, together with the
well-known results for the periodic chain \cite{Gaudin, GAU}. We find new ``boundary
string'' solutions to the Bethe equations, similar to the boundary strings
existing in the $|\Delta|<1$ regime \cite{Boot:SS}. For certain
 values of the boundary
magnetic field the ground state configuration
contains boundary 1-strings. Boundary excitations are obtained by removing
(or adding, depending on the sign of $h$) boundary strings from the
 ground state wave-function.
 Their energy was first obtained in \cite{miwa} by the algebraic approach.

A peculiar feature of the Bethe ansatz solution of the periodic chain is that
 the excitations (holes in the Dirac sea) appear only in pairs
\cite{FadT}. We argue that similarly the boundary excitations
can appear only in pairs with bulk excitations or with boundary excitations at
the other end of the spin chain. There is no such restriction in the  
soultion of
the semi-infinite chain by the algebraic approach \cite{miwa}.

Using  the Bethe ansatz
solution we calculate the surface energy (see e.g.,\cite{BH}):
\be
E_{surf}(L,\Delta, h)=E_{gr}-E^0_{gr}, \label{surfen}
\ee
 in the thermodynamic limit $L=\infty$. Here $E_{gr}$
is the ground state energy of (\ref{ham}) and  $E_{gr}^0$ is that of the
periodic
chain. We give an interpretation of our results in the
limits $\Delta\to\infty$ and $\Delta\to 1$, corresponding to the 1D Ising
and $XXX$ models respectively. Finally, we comment on the structure of the
ground state when the boundary magnetic fields are parallel.

\section{The Bethe ansatz equations}

Let us first set up the Bethe ansatz (BA) notations and list the relevant
 results about the $XXZ$ chain \cite{ABBBQ,GAU}.
In \cite{ABBBQ} the eigenstates of (\ref{ham}) were constructed
for arbitrary $\Delta$.
 As usual in the BA picture, the $n$-magnon eigenstates $|n\rangle$, satisfying
${\cal H}|n\rangle=E|n\rangle$,
are linear combinations of
the states with $n$ spins down, located at sites
$x_1,...,x_n$:
$$|n\rangle=\sum
f^{(n)}(x_1,...,x_n)|x_1,...,x_n\rangle.
$$
The wave-function
\be
 f(x_1,...,x_n)=\sum_P \ve_P A(p_1,...,p_n)e^{i(p_1x_1+...+p_nx_n)},
\label{new:wf}
\ee
contains $n$ parameters $p_j\in(0,\pi)$ which are subject
to quantization conditions, called Bethe equations (BE):
\be
e^{2iLp_j}\cdot{e^{ip_j}+h_1-\Delta\over 1+(h_1-\Delta)e^{ip_j}}\cdot
{e^{ip_j}+h_2-\Delta\over 1+(h_2-\Delta)e^{ip_j}}=\prod_{l\neq j}^n
e^{i\Phi(p_j,p_l)}. \label{beq}
\ee
The summation in (\ref{new:wf}) is over all permutations and negations
of $p_j$.
The energy and spin of the
$n$-magnon state are given by \cite{ABBBQ}:
\be
E={1\ov 2}\left[(L-1)\Delta+h_1+h_2\right]+2\sum_{j=1}^n(\cos p_j-\Delta),
\qquad S_z={L\ov 2}-n. \label{ene}
\ee

It is convenient to rewrite BE using the following mappings:
\be
 \Delta=\cosh\g \geq 1, \qquad \g\geq 0, \label{mapI}
\ee
\be
p=-i\ln{\cosh{1\over 2}(i\alpha+\gamma)\ov \cosh{1\ov 2}(i\al-\g)},
\label{mapII}
\ee
(our definition of $p(\alpha)$ differs from that of \cite{Gaudin, GAU} by the shift
$\al\to\al+\pi$ and it was chosen in such a way that
$p(\al)$ be an odd function that  maps $-\pi<\al<\pi$ to $-\pi<p<\pi$),
\bea
 h&=&\cosh\g +{\sinh{\g\ov 2}(1-H)\ov\sinh{\g\ov 2}(1+H)}=
\sinh(\g)\cdot\coth{\g\ov 2}\left(H+1\right), \qquad
h_{lim}<|h|<\infty, \label{maphI} \\
 h&=&\cosh\g -{\cosh{\g\ov 2}(1-H)\ov\cosh{\g\ov 2}(1+H)}=
\sinh(\g)\cdot\tanh{\g\ov 2}\left(H+1\right), \qquad |h|<h_{lim}.
\label{maphII}
\eea
The latter two mappings are defined on $H\in(-\infty,\infty)$ and are
necessary to cover the region $-\infty<h<\infty$, with
positive $h$ corresponding to $H\in(-1,\infty)$.
 The value
$h_{lim}\equiv h(\infty)=\sinh\g$ lies between two critical fields
 $h_{cr}^{(1)},h_{cr}^{(2)}$ defined as follows \cite{miwa}:
\be
 h_{cr}^{(1)}=\Delta-1, \qquad h_{cr}^{(2)}=\Delta+1. \label{crfields}
\ee
Both critical fields correspond to $H=0$, and the gap $h_{cr}^{(1)}<h<
h_{cr}^{(2)}$ corresponds to $0<H<\infty$.
In these notations eq. (\ref{beq}) becomes:
\bea
\left[{\cosh{1\over 2}(i\alpha_j+\g)\over
\cosh{1\over 2}(i\alpha_j-\g)}
\right]^{2L}&&B(\al_j,H_1) B(\al_j, H_2) \nonumber \\
&=&\prod_{ m\neq j}
 {\sinh{1\over 2}(i\alpha_j-i\alpha_m+2\g)
\sinh{1\over 2}(i\alpha_j+i\alpha_m+2\g)\over
\sinh{1\over 2}(i\alpha_j-i\alpha_m-2\g)
\sinh{1\over 2}(i\alpha_j+i\alpha_m-2\g)},   \label{be}
\eea
where
\bea
 B(\al,H)&=&{\cosh{1\over 2}(i\alpha+\g H)\over
\cosh{1\over 2}(i\alpha-\g H)}, \qquad h_{lim}<|h|<\infty, \label{defBI} \\
 B(\al,H)&=&{\sinh{1\over 2}(i\alpha+\g H)\over
\sinh{1\over 2}(i\alpha-\g H)}, \qquad |h|<h_{lim}, \label{defBII}
\eea
are called boundary terms.
The energy eq. (\ref{ene}) takes the form:
\be
E={1\ov 2}\left[(L-1)\cosh\gamma+h_1+h_2\right]
-2\sinh\g\sum_{j=1}^n p'(\al_j), \qquad
 p'(\al)={\sinh\g\ov\cosh\g+\cos\al}.
\label{new:ener}
\ee

In the thermodynamic limit $L\to\infty$ the real roots $\alpha_j$ of BE
form a dense distribution in the open interval $(0,\pi)$
with density $\rho(\al)$, $dI=2L(\rho+\rho_h)d\al$ being the number of
roots in the interval $d\al$. The logarithm of eq. (\ref{be}) is:
\be
 2Lp(\al_j)+{1\ov i}\ln B(\al_j,H_1)+{1\ov i}\ln B(\al_j,H_2)+\phi(2\alpha_j)
=\sum_{l=1}^n \phi(\al_j-\al_l)+\phi(\al_j+\al_l) +
2\pi I_j, \label{lbe}
\ee
where $I_j$ form an increasing sequence of positive integers, and
\be
\phi(\al)=-i\ln{\sinh{1\ov 2}(2\g+i\al)\ov\sinh{1\ov 2}(2\g-i\al)},
\qquad \phi(0)=0. \label{new:phase}
\ee
Taking the derivative of eq. (\ref{lbe}) and defining $\rho$ for negative
$\al$ by $\rho(\al)=\rho(-\al)$, we obtain
\be
 p'(\al)+{1\ov 2L}p_{bdry}^{\p}(\al)=\int_{-\pi}^{\pi}\phi'(\al-\ba)\rho(\ba)d
\ba
+2\pi(\rho(\al)+\rho_h(\al)),  \label{inteq}
\ee
with
\be
p_{bdry}^{\p}(\al)=-i{B'(\al,H_1)\ov B(\al,H_1)}-
i{B'(\al,H_2)\ov B(\al,H_2)}+2\phi'(2\al)-
2\pi\delta(\al)-2\pi\delta(\al-\pi). \label{pbdry}
\ee
The presence of delta-functions  in (\ref{pbdry}) is due to the fact that
$\al_j=0$ and $\al_j=\pi$ are always solutions to (\ref{be}),
which should be excluded, since they make the wave-function
(\ref{new:wf}) vanish identically.

\section{Solution for the bulk part}

In eq. (\ref{inteq}) the ``boundary terms''
 are down by the factor  $1/2L$. Neglecting
$p_{bdry}'$ and setting
$\rho_h=0$, we obtain the equation for the ground state density
 of the periodic $XXZ$ chain \cite{XXZ:YY}. Solving it by
the Fourier expansion
\be
f(\al)=\sum_{l=-\infty}^{\infty}\hat{f}(l)e^{il\al}, \qquad
\hat{f}(l)={1\ov 2\pi}\int_{-\pi}^{\pi}f(\al)e^{-il\al}d\al, 
\label{XXZ:fourier}
\ee
 and using (\ref{new:ener}), we recover the result for the
ground state energy of the periodic chain \cite{XXZ:YY}:
\be
2\pi\hat{\rho}_{per}(n)={ \hat{p}'(n) \ov 1+\hat{\phi}'(n)},\qquad
\hat{\phi}'(n)=e^{-2\g|n|}, \qquad \hat{p}'(n)=(-1)^ne^{-\g|n|}
\label{useful}
\ee
\be
E_{gr}^0={L\Delta\ov 2}
-2L\sinh\g\int_{-\pi}^{\pi}\rho_{per}(\al)p'(\al)d\al={L\Delta\ov 2}
-L\sinh\g\sum_{n=-\infty}^{\infty}
{e^{-\g|n|}\ov\cosh\g n}. \label{grener}
\ee
The spin of the ground state is
$S_z=L/2-L\int_{-\pi}^{\pi}\rho_{per} d\al=0,$
which is the well-known result \cite{XXZ:YY}.

An elementary ``bulk'' excitation above the vacuum
in the model (\ref{ham}) is a hole in the distribution
of $I_j$, but only a pair of holes
can occur for the periodic chain, as argued in \cite{FadT}. Thus physical
 excitations contain an even number of holes.
The energy of the hole with rapidity $\theta$ can be easily computed:
\be
\ve_h(\th)=
\sinh\g\sum_{n=-\infty}^{\infty}{(-1)^ne^{in\th}\ov \cosh\g n}>0,\label{hole}
\ee
and the spin with respect to the vacuum is $S_z=
1/2$.  (Our result, eq. (\ref{hole}), differs
from the conventional one by the shift $\th\to\th+\pi$, but the dispersion
relation is unchanged by rapidity reparametrization.)

Analogous arguments can be applied to analyze ``bulk''
string solutions with complex values of $\alpha$. Although there exists
an infinite hierarchy of complex strings of arbitrary length,
and quartets, their energy vanishes with respect to the vacuum
  \cite{BdVV}.

\section{Boundary excitations}

So far we discussed the bulk excitations, which are essentially
the same as in the periodic chain.
Let us now turn to the new solutions of eq. (\ref{be}), boundary strings.
The analysis is close to that of section 4.4. Boundary excitations
have their wave-function (\ref{new:wf}) localized at the left or right
ends of the chain,
and in the limit $L\to\infty$ the two boundaries may be considered separately.
Let us study first the left boundary, $h_1>0$.
The fundamental boundary 1-string
consists of one root located at $\al_0=-i\g H_1$ for $0<h_1<h_{cr}^{(1)}$,
and at $\al_0=\pi-i\g H_1$ for $h_{cr}^{(2)}<h_1<\infty$
(in both cases $-1<H_1<0$).
It is a solution of BE due to the mutual cancellation of the decreasing
 modulus of the first term in (\ref{be}) and the increasing modulus of the
second term $B(\al, H_1)$ as $L\to\infty$ and $\al\to\al_0$.
When $h_{cr}^{(1)}<h_1<h_{cr}^{(2)}$, no such solution exists.
Introduction of such a string into the vacuum with the density of roots 
  $\rho(\al)$ defined from
\be
 p'(\al)+{1\ov 2L}p_{bdry}^{\p}(\al)=\int_{-\pi}^{\pi}\phi'(\al-\ba)\rho(\ba)d
\ba
+2\pi\rho(\al),  \label{inteqII}
\ee
 leads to the redistribution
of roots by $\de\rho\equiv 2L(\tilde{\rho}-\rho)$ satisfying
the equation
\be
0=\int_{-\pi}^{\pi}\phi'(\al-\ba)
\de\rho(\ba)d\ba +\phi'(\al-\al_0)+\phi'(\al+\al_0)
+2\pi\de\rho. \label{densI}
\ee
From the latter we find
\be
2\pi\de\hat{\rho}(n)=-{2\cos n\al_0 e^{-2\g|n|}\ov 1+e^{-2\g|n|}}.
\label{answ}
\ee
For the energy and spin of the boundary 1-string with respect to this vacuum
we obtain:
\be
\tilde{\ve}_b=
-2\sinh(\g) p'(\al_0)-\sinh\g\int_{-\pi}^{\pi}\de\rho p'(\al)d\al=
-\sinh\g\sum_{n=-\infty}^{\infty}{(-1)^ne^{in\al_0}\ov\cosh\g n}, 
 \label{ansb}
\ee
$$ S_z=-\half.$$
We see that the energy (\ref{ansb}) is negative, so the
correct ground state wave-function (\ref{new:wf})
should contain the boundary 1-string root $\al_0$ when the value of
$h_1$ is not in the gap $h^{(1)}_{cr}<h_1<h^{(2)}_{cr}$.
The ground state density $\tilde{\rho}$ in this case satisfies the equation
\be
 p'(\al)+{1\ov 2L}(p_{bdry}^{\p}(\al)-\phi'(\al-\al_0)-\phi'(\al+\al_0))
=\int_{-\pi}^{\pi}\phi'(\al-\ba)
\tilde{\rho}(\ba)d\ba
+2\pi\tilde{\rho}(\al).  \label{inteqI}
\ee
 The boundary excitation
is obtained by removing from vacuum (\ref{inteqI}) 
the root $\al_0$, which means
that it has the energy $-\tilde{\ve}_b>0$
and spin $1/2$, equal to the spin
of the bulk hole. Substituting the value of $\al_0$ into
 (\ref{ansb}), we
get the boundary excitation energy, which precisely agrees with
the one obtained in \cite{miwa}:
\be
\ve_b(h_1)=\sinh\g\sum_{n=-\infty}^{\infty}{(-1)^{\kappa n}e^{\g H_1n}
\ov\cosh\g n},
\qquad -1<H_1<0,
\label{ansen}
\ee
with $\kappa=1$ if $h_1<h_{cr}^{(1)}$ and $\kappa=2$ if $h_1>h_{cr}^{(2)}$.

 From (\ref{hole}) and (\ref{ansen}) we see that for $h_1<h_{cr}^{(1)}$ the
 energy of the boundary
excitation is smaller than the bottom of the energy band of bulk excitations,
 and becomes equal to it at $h_1=h_{cr}^{(1)}$ (see Figure 6.1).
 So in this regime we may interpret
 the boundary excitation as the bound state of the kink, which gets unbound at
 $h_1=h_{cr}^{(1)}$. For $h_1>h_{cr}^{(2)}$ the energy of the boundary
bound state is bigger than the top of the energy band.
Therefore it is stable, in spite of its huge energy.

Besides the fundamental boundary 1-string, there exists an infinite set of
``long'' boundary strings, consisting of roots $\al_0-2ik\g,
 \al_0-2i(k-1)\g,..., \al_0+2ni\g$ with $n,k\geq 0$. We will call such solution
an (n,k) boundary
 string (thus the fundamental boundary string considered above is the (0,0)
 string). One can use the same arguments as given
in section 4.4 to show that the (n,k) string is a solution of BE when its
 ``center of mass'' has positive imaginary part
and the lowest root $\al_0-2ik\g$ lies below the real axis.
Thus, sufficiently long boundary
string solutions exist even in the region $h^{(1)}_{cr}<h_1<h^{(2)}_{cr}$.
However, a direct calculation shows that their energy vanishes with respect to
the vacuum,
 so they represent charged vacua. \footnote{As an example, consider
the boundary (1,0)-string consisting of the roots $\al_0+2i\g, \al_0$. It
exists
if $-1<H_1<1$, although the (0,0)-string exists only if $-1<H_1<0$.
The (1,0)-string has  charge $S_z=-1$ and vanishing energy with respect to
the vacuum.}
(Analogous phenomenon occurs for the
 ``long''
strings in the bulk \cite{BdVV}: if the imaginary part of $\al$
lies outside the strip $-2\g<{\rm Im}\al<2\g$, the root
$\al$  gives no contribution
 to the energy.) For $0<h_1<h_{cr}^{(1)}$ and
$h_{cr}^{(2)}<h_1<\infty$ the (n,0) strings also represent charged vacua, while
(n,k) strings with $k\geq 1$ have the same energy
(\ref{ansen}) as the boundary bound state
found above, and
hence represent charged boundary excitations. \footnote{{\it E.g.},
(1,1) string with roots $\al_0+2i\g, \al_0, \al_0-2i\g$ has $S_z=-3/2$
and energy given by (\ref{ansen}) with respect to the physical vacuum.}

Consider now the right boundary, $h_2<0$ ($H_2<-1$). Now the fundamental
boundary 1-string solution $\al_0=-i\g H_2$ exists for any value of $h_2$ in
the
interval $-h_{lim}<h_2<0$
(resp. $\al_0=\pi-i\g H_2$ for $h_2<-h_{lim}$).
Explicit calculation shows that  it has non-vanishing
energy only if $-2<H_2<-1$, which corresponds to $-h_{cr}^{(1)}<h_2<0$ (resp.
$h_2<-h_{cr}^{(2)}$). For such values of $h_2$ the energy  of the 1-string
with respect to the vacuum (\ref{inteqII}) is positive and equal to
$\ve_b(-h_2)$ (see eq. (\ref{ansen})), and its spin is $S_z=-1/2$.
In some sense the pictures are dual for the positive and negative $h$ cases: 
there exist two states when $|h|$ is not between $|h_{cr}^{(1)}|$ and 
$|h_{cr}^{(2)}|$,
 one with boundary 1-string and one without. One of them
is the ground state and another is the excited state at the boundary, and these
states exchange their roles when the sign of $h$ changes.
 The analysis of long boundary strings is
very similar to that at the left boundary, and therefore  will be
omitted. The net result is again that long boundary strings represent
charged vacua or charged boundary excitations.

In all examples shown above, the charge of boundary excitations turned out to
be half-integer. One can easily check that this is true for all boundary
strings
representing charged excitations.
Since the charge of physical excitations is obviously restricted to be an
 integer (see (\ref{ene})),
we conclude that a boundary excitation can appear only paired with the bulk
excitation of half-integer charge
(i.e. containing an odd number of holes), 
or with a boundary excitation at the other end of the chain.
 We give a qualitative interpretation of this fact below.

\section{The surface energy}

To compute the vacuum surface energy, eq. (\ref{surfen}), of the
model (\ref{ham}),
 one should use eq. (\ref{new:ener}) in the limit $L=\infty$ with the root density
determined from eqs. (\ref{inteqII}) or (\ref{inteqI}) and
the boundary terms (\ref{defBI}) or (\ref{defBII}), depending on
the value of $h$.  Define for convenience
\be
g(\Delta)={\Delta\ov 2}+2\sinh\g\sum_{n=1}^{\infty}{e^{-2n\g}-1\ov\cosh
2n\g}.
\label{fung}
\ee
 We  consider separately the following intervals for positive $h_1$ and 
negative $h_2$:

\begin{description}
\item[1)] $|h_{1,2}|<h_{cr}^{(1)}$. The ground state contains 
one boundary 1-string,
corresponding to $h_1$.
 The spin of the ground state can be found to be $S_z=0$.
Using  eqs. (\ref{defBII}), (\ref{pbdry}) and (\ref{inteqI}),
and subtracting the bulk contribution (\ref{grener})
we get
\be
E_{surf}={1\ov 2}(h_1+h_2)-g(\Delta)
-\sinh\g\sum_{n=1}^{\infty} (-1)^n{e^{-\g H_1n}-e^{\g H_2n}
\ov \cosh\g n}. \label{eqI}
\ee

\item[2)] $|h_{1,2}|>h_{cr}^{(2)}$. The ground state contains
 one boundary 1-string
and has $S_z=0$. From  eqs. (\ref{defBI}) and (\ref{inteqI}) it follows:
\be
E_{surf}={1\ov 2}(h_1+h_2)-g(\Delta)
-\sinh\g\sum_{n=1}^{\infty} {e^{-\g H_1n}-e^{\g H_2n}
\ov \cosh\g n}. \label{eqII}
\ee

\item[3)]  $h_{cr}^{(1)}<|h_{1,2}|<h_{lim}$. The ground state has no boundary strings
and its spin is zero.
{}From (\ref{inteqII}) and (\ref{defBII}) one obtains
the same expression as in case 1.

\item[4)] $h_{lim}<|h_{1,2}|<h_{cr}^{(2)}$. From (\ref{inteqII})
 and (\ref{defBI}) one
 obtains the same expression as in case 2. The ground state has the same
 structure as in case 3.
\end{description}

A qualitative plot of the surface energy as a function of $h$  ($h=h_1=-h_2$)
is given in Figure 6.2.  The apparent difference between (\ref{eqI}) and
(\ref{eqII})
 is an artefact of our parametrization of $h$ in terms of
$H$. In fact, $E_{surf}$ is an analytic function of $h$ in the domain
 $h\in(0,\infty)$, which can be seen after substituting $H$ as a function of
$h$  according to (\ref{maphI})-(\ref{maphII}).
In this sense the fields $h_{cr}^{(1,2)}$ are not actually ``critical.''
We find for $h_1=h_2=0$ the value
\be
E_{surf}=-{\Delta \ov 2}+4\sinh\g\left( {1 \ov 4}+\sum_{n=1}^{\infty}{\frac
{e^{2n\g}-1}{1+e^{4n\g}}}+
\sum_{n=1}^{\infty}{\frac{(-1)^n}{1+e^{2n\g}}}\right).
\label{Esurf}
\ee
Note that one can obtain the boundary magnetization $\langle
\sigma_1^z\rangle$ \cite{miwa}
immediately from the formula for the surface energy (\ref{eqI})-(\ref{eqII})
by differentiating it in $h_1$.

\section{The Ising $\Delta=\infty$ and rational $\Delta=1$ limits}
In the extreme anisotropic limit $\Delta\to\infty$, $h\sim\Delta$ of the $XXZ$
chain (\ref{ham}) one gets the one-dimensional Ising model:
\be
{\cal H}= {1\over 2 }
\left\{\sum_{i=1}^{L-1}
\Delta\sigma^z_i
\sigma^z_{i+1} +h_1\sigma_1^z+h_2\sigma_L^z \right\}, \label{hamising}
\ee
In this limit from (\ref{maphI})-(\ref{maphII}) one has
\be
h\approx \Delta \pm e^{-\g H}, \label{limith}
\ee
and the gap between $h_{cr}^{(1)}$ and  $h_{cr}^{(2)}$ dissapears, so
for any $h$ there
exists a boundary bound state. The energy of the ``bulk'' hole (\ref{hole})
 becomes
$\theta$-independent   and equal to $\Delta$,
since only $n=0$ term contributes to the sum
when $\g\to\infty$. The energy of the boundary bound
state (\ref{ansen})
becomes $\ve_b=\Delta \pm e^{-\g H_1}=
h_1$. This suggests the following interpretation
in terms of the Ising chain. In the Ising ground state the $i$-th spin
has the value $(-1)^i$. Local bulk excitation of the smallest energy $2\Delta$
can be obtained by flipping one spin
(the first and last spins excepted). The arising two surfaces (domain walls)
 separating
 the flipped spin  from its right and left neighbours
are called kinks and carry the energy $\Delta$ each. Kink corresponds
to a hole in the Bethe ansatz picture, and  kinks
 obviously appear only in pairs,
which demonstrates that holes can exist only in pairs, too.  The charge of
the
one-spin-flipped state is equal to one, in agreement with the charge of
two holes in BA. In addition to charge 1 excitation, one has charge 0
excitation of the same energy obtained by flipping any even number of spins
in a row. In the BA this corresponds to the ``2 holes and 2-string'' state.
In the Ising model the left (right) boundary bound state is obtained by 
flipping the first (last)
spin. Such a
state has the energy $h_1+\Delta$ above the vacuum energy,
where $h_1$ is the contribution of the boundary term in  (\ref{hamising})
and $\Delta$ is the energy of the kink created due to the boundary-bulk
interaction. Thus flipping
the boundary spin actually gives
a combination of the boundary excitation and the bulk kink.
Still another possibility is to flip all spins, creating two boundary
bound states, one at each boundary.
 This explains why, in the BA picture, a boundary excitation
can exist only if paired with a hole in the Dirac sea or with another boundary
excitation.
 The vacuum surface energy (\ref{surfen})
of the Ising chain in the thermodynamic limit is $(\Delta-h_1+h_2)/2$.
The $\Delta/2$ contribution here is the bulk interaction energy that we
lost when we
disconnected the periodic chain, and $\pm h_{1,2}/2$ is the contribution of
each
of the boundary terms. Taking the limit $\g\to\infty$ in eqs. (\ref{eqI})-
(\ref{eqII}), we obtain the expected result
 $E_{surf}\to(\Delta-h_1+h_2)/2$.

In the isotropic (rational)
limit $\Delta\to 1$ ({\it i.e.}, $\g\to 0$) one gets
the $XXX$ chain in a boundary magnetic field,
which was discussed in the BA framework in \cite{GMN} for $0<h_{1,2}<2$. From
(\ref{maphI})-(\ref{maphII}) one has in this limit
\be
h={2\ov 1+H}. \label{Hnew}
\ee
There is only one critical field $h_{cr}=2$, which is the limit
of $h_{cr}^{(2)}$. Passing from summation to integration
in eq. (\ref{eqII}), we obtain for $0<h_1<\infty$, $0<-h_2<\infty$:
 \begin{eqnarray}
E_{surf}&=&{1\ov 2}(h_1+h_2)-{1\ov 2} +{\pi\ov 2} -\int_{0}^{\infty}dx
{e^{-({2\ov h_1}-1)x} - e^{({2\ov h_2}-1)x}+ e^{-x}
\ov\cosh x}= \nonumber \\
&=&{1\ov 2}(h_1-h_2)-{1\ov 2} +{\pi\ov 2} -\int_{0}^{\infty}dx
{e^{-({2\ov h_1}-1)x} + e^{({2\ov h_2}+1)x}+ e^{-x}
\ov\cosh x}, \label{xxxlimI}
\end{eqnarray}
where the second line was obtained from the first one after
 a simple manipulation. This agrees with the results of \cite{GMN}.
 For $h_1=h_2=0$
one has from (\ref{xxxlimI})
$E_{surf}=(\pi-1)/2 - \ln 2$.

\section{The case of parallel magnetic fields}

Another aspect is the structure of the ground state
in the regime $h_{1,2}>0$. Assuming that, for example, for
$h_{1,2}>h_{cr}^{(2)}
$ the ground state contains both left and right boundary 1-strings
to minimize the energy, we
end up after a short calculation with a half-integer spin of the vacuum,
which signals that such a state cannot,
in fact, be the vacuum. Hence, the ground state must
have a more intricate structure.
Appealing to the Ising limit $\g\to\infty$, one
sees that for $h_{1,2}>\Delta$ the ground state
 must have both boundary spins  directed opposite to the magnetic field,
 and therefore contain a kink in the bulk (recall that L is even).
On the other hand, for $h_{1,2}<\Delta$ the lowest energy configuration
is such that the boundary spins are antiparallel, which means that the
 physical vacuum contains what was
called a boundary excitation at one of the ends. This suggests that for finite
$\Delta$ the correct ground state
 wave-function of the Hamiltonian (\ref{ham}) should contain a bulk hole
with the minimal possible energy (i.e. the kink with zero rapidity $\theta=0$)
and both boundary 1-strings 
when $h_{1,2}>h_{cr}^{(2)}$.  Such a state has spin zero.
Changing the rapidity of this stationary kink away from zero, one obtains 
in such a way an excited state whose energy can be arbitrarily close to 
the vacuum one, which means that there is
a new gapless branch in the spectrum.
\footnote{In the Ising limit $\g\to\infty$ the energy of the kink is
independent
of rapidity, and therefore this branch degenerates to the vacuum.}
 Similarly, when $h_{cr}^{(1)}<h_{1,2}<h_{cr}^{(2)}$, for the
ground state
to have the integer charge it should also contain a kink in the bulk. When
$h_{1,2}<h_{cr}^{(1)}$ the physical vacuum contains only
one of the two boundary 1-strings 
and no stationary kink in the bulk (when $h_1=h_2$ there are two possibilities
to have either left or right boundary 1-string in the vacuum, corresponding
to the obvious two-fold degeneracy of the Ising ground state in this case).
  Such a state has a smaller energy for $h_{1,2}<h_{cr}^{(1)}$ than the one
with
 a hole in the bulk and two boundary strings,
while for $h_{1,2}>h_{cr}^{(2)}$ the state with the bulk hole is energetically
 preferable, since in this case $\ve_b>\ve_h$ (see Fig.12 and \cite{miwa}).
This situation is in
some sense analogous to the case of the periodic 
antiferromagnetic $XXZ$ chain with odd L, where the ground state contains
a kink.
According to the above
 discussion the surface energy in the case $h_{1,2}>h_{lim}$
is:
\be
E_{surf}={1\ov 2}(h_1+h_2)-g(\Delta)+\varepsilon_h(0)
-\sinh\g\left(1+\sum_{n=1}^{\infty} {e^{-\g H_1n}+e^{-\g H_2n}
\ov \cosh\g n}\right). \label{eqM}
\ee
In the rational ($\g\to 0$) limit  $\varepsilon_h(0)$ vanishes and eq.
(\ref{eqM}) becomes
\be
E_{surf}={1\ov 2}(h_1+h_2)-{1\ov 2} +{\pi\ov 2} -\int_{0}^{\infty}dx
{e^{-({2\ov h_1}-1)x} + e^{-({2\ov h_2}-1)x}+ e^{-x}
\ov\cosh x}. \label{xxxlimIP}
\ee
This expression agrees with the one obtained in \cite{GMN}. Note that
the authors of
\cite{GMN} obtained eq. (\ref{xxxlimIP}) under the assumption that
$0<h_{1,2}<h_{cr}$, while our derivation shows that this result is valid for
$0<h_{1,2}<\infty$. In the Ising limit eq. (\ref{eqM}) gives the correct
result $E_{surf}=(3\Delta-h_1-h_2)/2$.
Observe that for the $XXX$ chain the following equality holds
(see (\ref{xxxlimI}) and (\ref{xxxlimIP})): $E_{surf}(h_1,h_2)=
E_{surf}(h_1,-h_2)$. This is the consequence
of the decomposition $E_{surf}=f(h_1)+f(h_2)+const$,
which  takes place
 in the limit $L=\infty$ when two boundaries
are independent, and the obvious property of the semi-infinite 
chain $f(-h)=f(h)$. The same statements are true for the 
 surface
energy of $XXZ$ chain apart from the $\varepsilon_h(0)$ contribution
(see (\ref{eqM})).  

\section{Remarks}

  We would like to mention also that within the BA technique
it is possible to calculate also the boundary S-matrix for the
scattering
of kinks (represented by holes in the Dirac sea) in the ground state of the
Hamiltonian (\ref{ham}) or in
the excited boundary state.
Such a calculation has been performed
 in \cite{GMN} for the boundary $XXX$ chain
and in chapter 4 above for the boundary sine-Gordon model.
For the boundary $XXZ$ chain these
S-matrices have been obtained by Jimbo et al. \cite{miwa} by the algebraic
approach.

\newpage
\begin{figure}
\epsfxsize=5.92truein
\centerline{\epsfbox{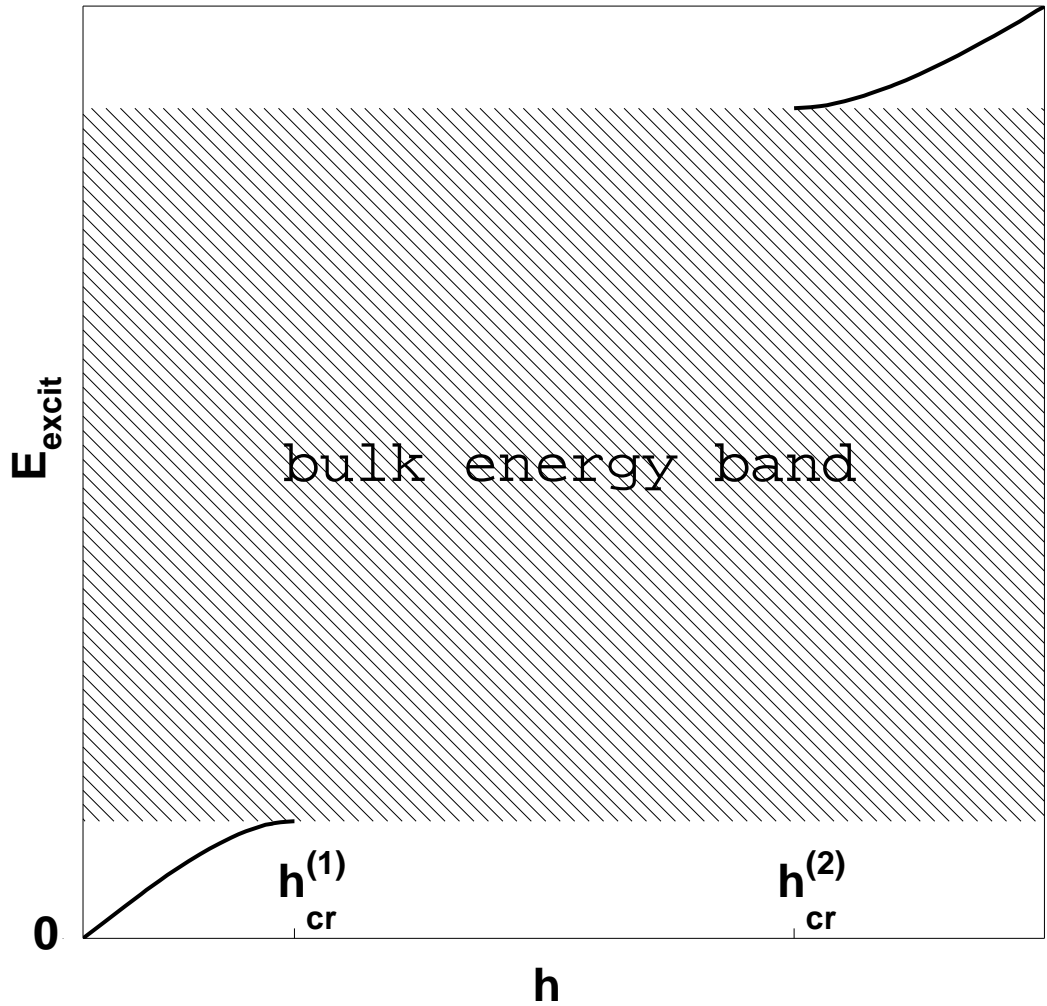}}
\caption{
 Solid line: a schematic plot of the energy of the boundary excitation, $\ve_b(h)$, as a function of
 the boundary magnetic field $h$. Shaded area: the energy band of the bulk
excitations.}
\end{figure}

\newpage
\begin{figure}
\epsfxsize=5.92truein
\centerline{\epsfbox{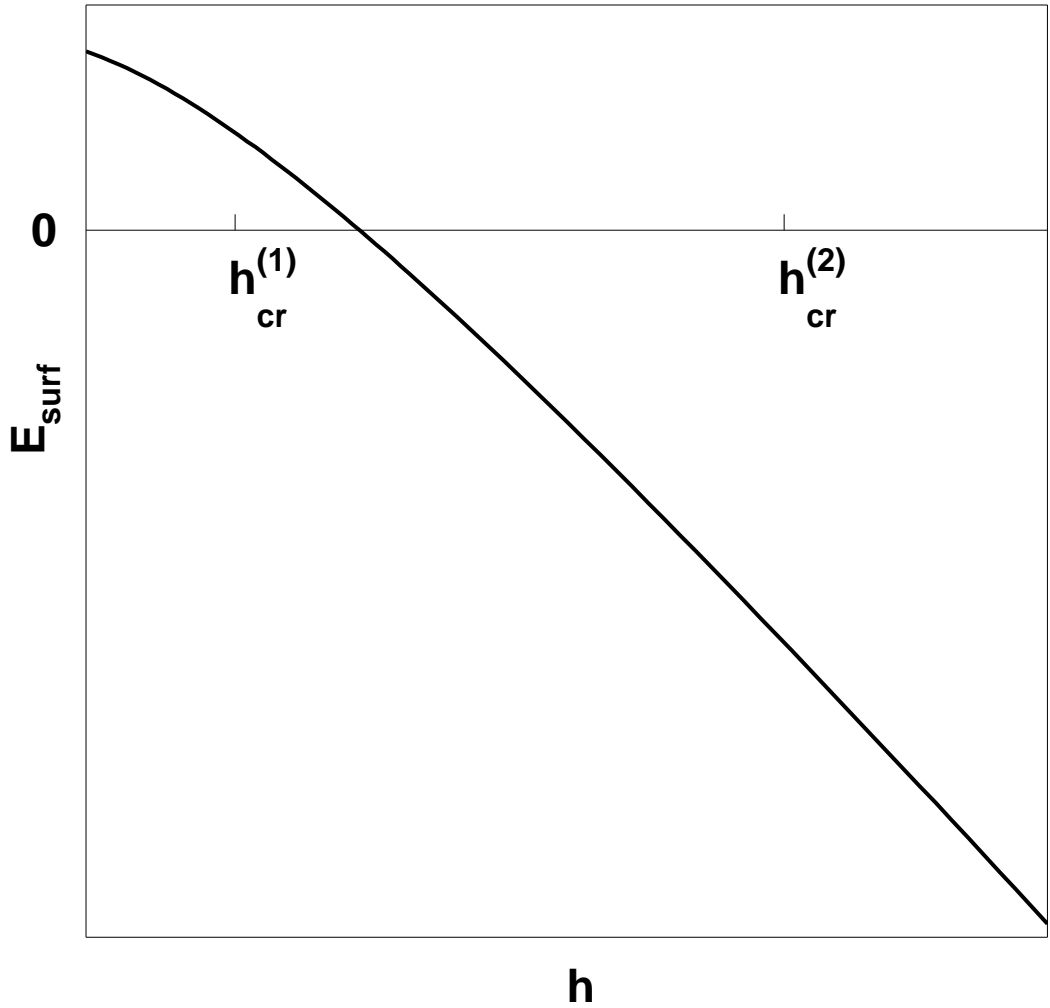}}
\caption{
 A schematic plot of the vacuum surface energy as a function of
 the boundary magnetic field $h=h_1=-h_2$.}
\end{figure}


\chapter{Calculation of correlation functions for the problems with
impurities}

We show how to compute analytically time and space dependent correlations in
one-dimensional quantum integrable systems with an impurity. Our approach is
based on a description of these systems in terms of massless scattering 
of quasiparticles \cite{FS:less}. 
Correlators follow then from matrix elements of local
operators between multiparticle states -- the {\it massless form-factors}.
Although, in general an infinite sum of these form-factors has to be considered,
we find that for the current, spin and energy operators only a few 
(two or three) are necessary to obtain an accuracy of more than 1\%. 
Our results are valid for {\tt arbitrary impurity strength}, in contrary
to the perturbative expansions in the coupling constants. As an example,
we compute the frequency dependent condunctance, 
at zero temperature, in a Luttinger liquid
with an impurity, and also discuss the succeptibility in the Kondo model
and the time-dependent properties of the two-state problem with dissipation.
This chapter is based on \cite{LSS}.

\section{Introduction}

In this chapter we present the technique to calculate current-current
correlation functions in the quantum field theories of the form
\be
H={1\over 2} \int^{\infty}_0 dx \
[8\pi g \Pi^2+{1\over 8\pi g}(\partial_x\phi)^2] + {\cal B},
\ee
where ${\cal B}$ is a problem dependent boundary interaction and the
fields are defined
on the positive half-line. 
The method we adopt  is based on the form-factor
formalism \cite{Muss:Rev} and provides us with the series
expansions of correlators. 
One has to insert intermediate n-particle
states in the correlator $\langle 0|j_\mu j_\nu|0\rangle=\sum_n
\langle 0|j_\mu |n\rangle\langle n|j_\nu |0\rangle$; then
the well-known technique for integrable models  \cite{Smir} gives the
exact values for the above scalar products between the vacuum and n-particle
states (the form-factors). 
The integral representation for each term of the series
is available. Moreover, the observed {\tt rapid convergence at  any scale
from small to large distances}
allows us to  truncate the series after a few terms (typically two
or three) to obtain a 1\% accuracy. 
There are few novelties that 
deserve our attention: first, we have to deal with the {\it massless}
form-factors, following the pioneering work \cite{Muss_massless}.
Second, we work in a half-plane geometry. An instructive example, although
somewhat simple and abstract, is an Ising model in a boundary magnetic field
\cite{KLM:freaks}.

The role, played by integrability in our approach is two-folded.
First, it gives us the {\tt exact} expressions for the form-factors
and other necessary quantities. Second, it provides us with a basis
of particles (intermediate states) which allow us to truncate
the series expansion of correlators due to the rapid convergence. Thus,
the integrability insures the  relatively small
contribution  of the multi-particle terms.

To pave our path through the calculations, we consider
some simple model of the free massless 1-dimensional scalar field
with the boundary interaction of the form ${\cal B}=M_B\cosh\phi$. 
This model
can be viewed as the massless limit of the integrable sinh-Gordon model
with boundary interaction:
\be
S=\int
dxdt\left[{1\over 2}(\partial_\mu\phi)^2-{m^2\over g^2}\cosh(g\phi)\right]
-M_B\int dt \cosh\left({g\phi\over 2}\right)|_{x=0}
\label{mShG}
\ee
The advantage of this approach, as opposed to starting with the free massless
boson quantized in the plane-wave basis, is the following:
 in the  massless limit, obtained by
scaling the energy and momentum of particles along with the boundary
mass $M_B$ to infinity, we get the convenient basis of massless particle states
which are particular combinations of plane waves that scatter diagonally
off the boundary. Corresponding classical solutions are presented in
\cite{FSW}. Working with these massless particles at first sight 
adds some complexity, but it is paid off by the final simple and manageable
results, while in the plane wave picture one has to do infinite summations
and the final result is difficult to extract.

We apply our technique to find the correlation functions in three
models of condensed matter physics: 
the Kondo model, the spin-boson model (two-level system with dissipation), and
the Luttinger liquid with impurity (realized as
the quantum Hall liquid with constriction).
A common
feature to all these models is that they can be reduced to a model
described by massless excitations in the bulk interacting
with an impurity at the boundary.
 The boundary interaction $\cal B$
in the first and second models is ${\cal B}=\lambda \ ( S_+ e^{i\phi(0)/2}+S_- 
e^{-i\phi(0)/2})$, and the third model has ${\cal B}=M\cos\phi(0)$.
The absence of a mass gap leads to a power law behaviour for the
current correlators in both the ultra-violet and the
infra-red regime.  The cross-over between these two regimes is
non-trivial
because of the renormalisation group flow induced by the impurity.
For each model that we study,
 correlation functions can be 
related to the measurable quantities in the model dependent way.
For example, for the Luttinger liquid, the Kubo formula \cite{ref:Kubo}
gives the {\it AC conductivity}:
\be
{
\sigma_{\mu\nu}(\omega)=\int_0^\infty dx\int_0^\infty
\langle 0|j_\nu(0)j_\mu(t+ix)|0\rangle e^{-i\omega t}dt
}\label{last:kubo}
\ee
 The {\it perturbative}
analysis of \cite{CMP} shows that the conductivity scaling function, depending
on the impurity strength, flows from the insulating at $T=0$ to 
the perfect conductance at $T=\infty$ in the repulsive regime. 
The alternative {\it exact} methods, which do not employ the knowledge 
of correlation functions, have been developed in \cite{FLUDS} to find
the static DC conductivity at non-zero temperature and voltage. These methods
do not allow, however, the determination of the AC conductivity.
The experimental data \cite{Millik}, as well as the numerical Monte-Carlo
simulations \cite{moon,Mak}
for the Luttinger liquids with impurity are available.

\section{Integrable models in condensed matter physics}

In this section we introduce and review three prominent
models of condensed matter physics:
the Kondo model, the spin-boson model (two-level system with dissipation), and
the Luttinger liquid with impurity. The results of
application of our technique for computing
correlators to these models are partially shown here and in part further.
 When bosonized, all three models look alike
in the bulk (free massless boson), but have different boundary interactions
$\cal B$:
$ \lambda \ ( S_+ e^{i\phi(0)/2}+S_- e^{-i\phi(0)/2})$ for the first and
second, and $M\cos\phi(0)$ for the third one.

\subsection{Kondo model}

The Kondo model describes the free bulk electrons interacting with an
impurity via the spin:
\be
H_K=\sum_{k,\sigma}\varepsilon(k)c^+_{k\sigma}c_{k\sigma} +J\vec{s}(0)
\vec{S}_{imp}
\label{last:kondo}
\ee
Here, $\vec{S}_{imp}=\half\vec{\sigma}$ 
is the impurity spin, $\vec{\sigma}$ are Pauli matrices,
and $\vec{s}(0)$ is the spin 
 induced by the electrons at the point of impurity,
\be
 \vec{s}(0)=\half\sum_{\sigma\sigma'}\psi^+_\sigma(0)
\vec{\sigma}_{\sigma\sigma'}
\psi_{\sigma'}(0), \qquad \psi_\sigma(0)=L^{-1/2}\sum_k c_{k\sigma}.
\ee
A useful generalization of (\ref{last:kondo}) is an anisotropic Kondo model
\cite{AYU}:
\be
H_K=\sum_{k,\sigma}\varepsilon(k)c^+_{k\sigma}c_{k\sigma} + 
J_{\pm}[\sigma_+\psi^+_\downarrow\psi_\uparrow(0)+
\sigma_-\psi^+_\uparrow\psi_\downarrow(0)] + J_z\sigma_z
\sum_{k,k',\sigma=\pm}\sigma c^+_{k\sigma}c_{k'\sigma},
\label{last:kondo_anis}
\ee
which reduces to (\ref{last:kondo}) if $J_z=J_\pm$.

We need also the bosonized version of (\ref{last:kondo_anis}). For it,
two bosonic fields are necessary: one associated with charge and one with spin.
The charge-density field decouples completely and only the spin-density
 has interaction
at the boundary. The Hamiltonian for the spin-density field is of the form
\be
H_K={1\over 2} \int^{\infty}_0 dx  [8\pi g \Pi^2+{1\over 8\pi
g}(\partial_x\phi)^2]
+ \lambda \ ( S_+ e^{i\phi(0)/2}+S_- e^{-i\phi(0)/2}).
\label{last:bosonizedKondo}
\ee
The coupling constant $\lambda$ is related to $J_{\pm}$.
 The $S_z$ term in
the Hamiltonian has been eliminated by a unitary rotation, but 
dependence on $J_z$ is implicit  in $g$. The case $g=1$ corresponds to the isotropic Kondo model.

The quantities of interest in the Kondo model are the spin-spin
correlation functions for the induced electron spin in the bulk 
(screening cloud
problem \cite{Affl:Kondo}) and for the impurity spin on the boundary,
$\langle S^z_{imp}(t)S^z_{imp}(0)\rangle$. We consider here mostly
the latter ones, since they are relevant for the dissipative two-level
problem. But the screening cloud problem can be approached by the 
form-factors technique, too, and for partial results we refer to \cite{LSS}.
The {\it dynamic susceptibility}, or the {\it response function} is given by
\be
\chi''(\omega)\equiv {1\over 2}\int dt e^{i\omega t}\langle
[S^z(t),S^z(0)]\rangle.
\label{last:chi}
\ee
 In \cite{LSS} it was
shown how to express the spin correlators in terms of the bosonic field
correlators of (\ref{last:bosonizedKondo}). In particular, for
the response function we have\footnote{In the most of this chapter
it is tacitly assumed that $\hbar=1$.}
\be
\chi''(\omega)={1\over (2g\pi)^2}{1\over\omega^2}
{\rm Im} \left[{\cal G}
(-i\omega,\beta_B)-
{\cal G}(-i\omega,-\infty)\right],
\label{last:resprel}
\ee
where
\be
\langle\partial_{\bar{z}}\phi(x,y')
\partial_{z}\phi(x,y'')\rangle_{\lambda}
=\int_0^\infty dE {\cal G}(E,\beta_B)\exp\left[-2Ex-iE(y'-y'')\right],
\ee
and the boundary temperature $\beta_B$ is related to $\lambda$.
The foregoing technique (see section 7.4) allows us to compute
the quantity ${\cal G}(E,\beta_B)$ in the form of series expansion.
 Every term of the series is known analytically.
In practice, it is enough to keep only few first terms of these series
to obtain a very good accuracy of 1\% or more. E.g., for the Toulouse
limit, corresponding to $g=\half$, the series get truncated 
and we obtain the exact result
\bea
\chi''(\omega)&=&{2\over \pi^2}{T_B\over\omega}
{\rm Im}
 \left(
\int_0^\omega dx{1\over  (x+iT_B)(\omega-x+iT_B)}\right)
\nonumber \\
&=&{1\over \pi^2}
{4T_B^2\over  \omega^2 +4T_B^2}\left[{1\over\omega}
\ln\left({T_B^2+\omega^2\over T_B^2}\right)+{1\over T_B}
\tan^{-1}{\omega\over T_B}\right]. \nonumber
\eea
For the less simple $g=1/3$ case the leading contribution
to the response function reads
\be
\delta\chi''(\omega)^{(1)}=-{9 \mu^2  d^2\over 8\pi  \omega }{\cal
R}e
\left[ {\tanh\left(\half\log({\omega\over \sqrt{2}T_B})
-{i\pi\over 8}\right)
\over
\tanh\left(\half\log({\omega\over \sqrt{2}T_B})+{i\pi\over
8}\right)}
-1\right]
\ee
where $\mu\approx 3.14$, $d\approx 0.1414$.
Similar computations give rise to the results in
figure 7.1 where we plotted  $\chi''(\omega)/\omega$  for the values
$g=3/5,1/2, 1/3, 1/4$. Notice the emergence of the peak for $g<1/3$.
It is remarkable that this peak appears at $g=1/3$ and not at $g=1/2$ as was
expected from other means of calculations. 

\begin{figure}
\epsfxsize=100truemm
\centerline{\epsfbox{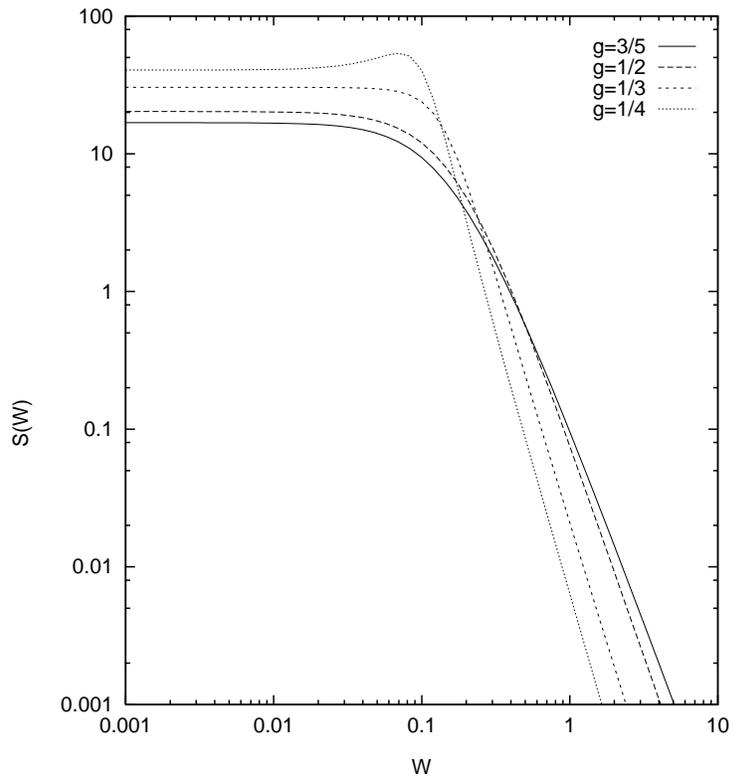}}
\caption{Plot of $\chi''(\omega)/\omega$  for the values $g=3/5,1/2, 1/3, 1/4$}
\end{figure}

Using eq. (\ref{last:resprel}) and the general form of ${\cal G}(E,\beta_B)$
given below, we can find the exact 
static succeptibility $\chi_0$ which is given
by the Shiba's relation \cite{Shiba}:
\be
\lim_{\omega\rightarrow 0} {\chi''(\omega)\over \omega}
=2 \pi g\chi_0^2.
\ee
Omitting the details, we list the result:
\be
\chi_0={1\over \pi^2 g T_B}.
\ee
(but see \cite{LSS} for the complete account of this derivation).

\subsection{Quantum systems with dissipation}

The effect of dissipation on quantum tunnelling has been addressed in the
pioneering work of Caldeira and Leggett \cite{CaldLeg}. Suppose first one has
a system with friction, or dissipation, described by 
the equation of motion
\be
M\ddot{q}+\eta\dot{q}+{\partial V\over \partial q} = F_{ext}(t),
\label{last:classic}
\ee
with the potential $V$ admitting a quasi-stationary state that can tunnel
into the continuum.
Then, speaking loosely, 
the question one asks is whether there is a qualitative change 
in the behavior of such a system for $\eta=0$ and $\eta\neq 0$?

It is an interesting problem in general how to introduce friction
in the formalism of quantum mechanics. In \cite{CaldLeg} the following
phenomenological model is suggested, in analogy with (\ref{last:classic}):
\be
C\ddot{\phi}+R_n^{-1}\dot{\phi}+{\partial U\over\partial\phi}=0
\ee
It is usually refered as RSJ (resistively shunted junction) model. The
nature of the resistance $R_n$ is not clear, however. It is introduced
in the model phenomenologically. 

As the analysis of \cite{CaldLeg} shows,
the net effect of the friction is that the WKB exponent for the tunnelling rate
is increased by dissipation and tunnelling rate decreases:
\be
P\sim \exp\left( {-cV_0\over h\omega_{eff}}\right)
\ee
 As a
consequence of interactions, the characteristic frequency of classical
system $\omega_0$ (defined by the product $LC$) is renormalized to 
$\omega_{eff}$.

As for experimental applications, the theory of \cite{CaldLeg} describes
a trapped flux in SQUID (superconducting interference device). The
phase of the flux, $\phi$, is a quantum degree of freedom governed
by the potential
\be
U(\phi)={(\phi-\phi_x)^2\over 2L} - I_c 
\Phi_0\cos\left({2\pi\phi\over\phi_0}\right),
\ee
 where $\phi_x$ is an external flux, $L$ is a self-inductance of the device,
$I_c$ is a critical current and $\Phi_0=h/2e$ is a flux quantum. 
The kinetic energy is given by $K=\half C\dot{\phi}^2$ with $C$ a capacitance.
Another experimental realization can be found in the single Josephson
junction biased by a fixed external current $I_0$. The quantum degree
of freedom now is the phase difference of Cooper pair across the junction,
$\varphi$, and the potential is
\be
U=-{I_0\Phi_0\over 2\pi}\varphi - {I_c\Phi_0 \over 2\pi} \cos\varphi.
\ee

Another fundamental system has been studied in \cite{dissip_qm}. Here,
instead of tunnelling into the continuum, a double well problem is
being considered with the tunnelling between two separated wells.
Since the particle can be localized in either of two wells, the system
is refered to as a two-level system. More generally, it could be
not necessarily a continuum degree of freedom with two spacially
separated states, but also an isospin degree of freedom with two states,
e.g. a strangeness quantum number in particle physics.
Examples and applications include the strangeness oscillations of a neutral 
K-meson beam, or an inversion resonance of the ${\rm NH}_3$ molecules.

The isolated two-level system can be modeled by a Hamiltonian
\be
H_0=-\half \Delta_0\sigma_x +\half\varepsilon\sigma_z
\label{last:twostatefree}
\ee
where $\Delta_0$ is a matrix element for tunneling, and $\varepsilon$
is a ``detuning'' parameter called {\it bias} (e.g. in the double-well
problem it is the difference of the ground state energies of the wells).
It is obvious that the model (\ref{last:twostatefree}) is the effective
Hamiltonian in the quasiclassical description of the double-well problem.

The dynamics of the two-level system with the Hamiltonian 
(\ref{last:twostatefree})
can be easily
solved. It is well-known that in the absense of bias, $\varepsilon=0$,
system shows coherent behavior with the probability distribution 
\be
P(t)\equiv P_R-P_L=\cos(\Delta_0t),
\label{last:oscil}
\ee
where $P_L$ ($P_R$)
 is the probabilty of finding particle in the left (right) well, and we assumed
that initially the particle was localized in the right well. In the
presence of non-zero bias the oscillatory behavior (\ref{last:oscil})
is destroyed. 

Another factor that can potentially destroy the coherence is an interaction
with the environment with sufficiently strong coupling. 
Environment couples through the $\sigma_z$ term in
(\ref{last:twostatefree}) and can be modeled the by following {\it spin-boson}
Hamiltonian \cite{dissip_qm}:
\be
H_{sb}= H_0+\half q_0\sigma_z\sum_i C_i x_i
+ \sum_i \left(\half m_i\omega_i x_i^2 + {p_i^2\over
2m_i} \right)
\label{last:sboson}
\ee
The environment is modeled by the set of $N$ harmonic oscillators
(phonons)  described by
the last term in (\ref{last:sboson}), and, of course, we are interested in
the thermodynamic limit $N\to\infty$. With the non-zero coupling to environment,
$q_0$, one state of the
two-state system becomes more preferable than the other, depending
on  the ``mood'' of the environment. At $T=0$, classically, all the oscillators
are at $x_i=0$ and have no effect on the two-state system. However,
quantum-mechanically the problem becomes non-trivial even at $T=0$ due
to the quantum fluctuations. The question of theoretical interest
 can be formulated as follows: Can there exist oscillatory behavior 
(\ref{last:oscil})
in a macroscopic system (\ref{last:sboson}), or the coherence will be 
destroyed by the interaction with environment?

The 
time-dependent quantities of interest which are useful in the analysis
of this problem are conveniently encoded in the following:
\be
P(t)=\langle\sigma_z\rangle,
\ee
\be
C(t)=\half \langle [\sigma_z(t),\sigma_z(0)]\rangle.
\ee
Here, $P(t)$ measures how the average of spin varies with time, provided
that at $t=0$ it was in a certain state, while $C(t)$ describes the
probability to be in a state $\sigma_z(t)$ given that the system was
in a state $\sigma_z(0)$ at $t=0$. The operators are understood in 
the Heisenberg representation $\sigma_z(t)=e^{-iH_{sb}t}\sigma_ze^{iH_{sb}t}$.

It turns out that the effect of phonon bath on the two-level  system 
is very non-trivial and depends on the form of the {\it spectral
function of the environment},
\be
J(\omega)={\pi\over 2}\sum_i {C_i^2\over m_i\omega_i}\delta(\omega-\omega_i)
\label{last:spectral}
\ee
It is usually assumed that $J(\omega)=\eta\omega^s\exp(-\omega/\omega_c)$
where $\omega_c$ is a cut-off frequency. According to the analysis of
\cite{dissip_qm}, we have, at $T=0$,
the weakly damped coherent oscillations for $s>1$,
and the complete localization for $s<1$. For $s=1$ the analysis becomes
more complicated and the result depends on the 
value of the dimensionless coupling constant $\alpha$,
\be
\alpha={\eta q_0^2\over 2\pi}
\ee
 The $s=1$ case is usually refered to as the {\it ohmic regime}. 
For $\alpha\geq 1$ and $s=1$
it is believed that the system is completely localized
\cite{dissip_qm}, while for $\alpha<1$ the situation is not clear yet.
It is known that there exists a critical point at the value of
$\alpha$ equal or about $\alpha_c=\half$
where a phase transition in the ground state occurs. 
For $\alpha$ less than this critical
value the damped oscillations are observed, while above it there is an
 incoherent relaxation. 

The value $\alpha=\half$, called the Toulouse limit, is an exactly solvable
point. At this point the model can be mapped onto the following
model:
\be
H_{T}=\sum_k kc^+_kc_k + V\sum_k (d^+c_k + c_k^+d)
\ee
where $d^+, d$ create and annihilate a localized state
(corresponding to the spin degree of freedom in the spin-boson model), while
$c_k^+, c_k$ are the creation and annihilation operators for the
fermions in continuum (corresponding to the bath).\footnote{
The bias $\varepsilon$ corresponds to the energy of localized state.}
 The Toulouse model is a particular case, $U=0$, of
a more general exactly solvable {\it resonance-level model} \cite{Fin},
which can be mapped onto the spin-boson model~:
\be
H_{RL}=H_{T}+U\sum_{k, k'}(c^+_kc_{k'}-\half)(d^+d-\half)
\ee
The correspondence is given by the simple relations
\be
\sigma_+=d^+, \qquad \sigma_-=d, \qquad \sigma_z=d^+d-\half
\ee
and $V$ is directly proportional to the hopping $\Delta_0$, while $U$ is related to
$\alpha$ as follows \cite{bosoniz}~:
\be
\alpha\sim\left(1-{U\over \pi}\right)^2
\ee
 The exact solution of the Toulouse model gives the following behavior
at $\alpha=\half$:
\be
P(t)=\exp(-t/\tau).
\label{last:Toul_exact}
\ee

A particularly useful fact for us  is that the long-time behavior,
$t\gg\omega_c^{-1}$, 
of the spin-boson model in the ohmic regime with $\half<\alpha<1$
is the same as the long-time behavior of the 
anisotropic Kondo model
\footnote{Only the low-lying excitations are relevant in this limit.}
 (\ref{last:kondo_anis}). This result
can be established rigorously by means of the bosonization \cite{bosoniz}.
The harmonic oscillators in the spin-boson model play the role of
the spin-density excitations
in the Kondo case, and the impurity spin ($S_{imp}=\half$)
corresponds to the discrete degree of freedom in (\ref{last:sboson}). 
We have~:
\bea
\Delta_0&\sim& J_\pm \nonumber \\
\alpha&\sim&(1-\half\rho J_z)^2.
\eea
The point $\alpha=1$
maps into $J_z=0$, in such a way separating the ferro and antiferromagnetic
regimes in the Kondo model. Note that the mapping works only for $J_{\pm}, J_z$
small enough, i.e. $\alpha$ close to 1. The regions where $\alpha$ is far from
1 can be approached by the renormalization group analysis \cite{Anderson}.
 By continuity,
the results should hold at least up to the strong coupling fixed point,
$\alpha=\half$, where the exact solution (\ref{last:Toul_exact}) holds.  
Note that the analysis of the previous section, in particular figure 7.X
suggests that the incoherent relaxation takes place in fact up to $\alpha=1/3$.
  This
is supported by a recent RG numerical study \cite{costi}.

\subsection{Quantum Hall liquid with constriction}

Transport in one-dimensional interacting electron systems in the presence
of impurities is an instructive problem with a broad rangle of experimental
applications. Important ingredients, defining a proper physical model, are
the Coulomb correlations in the vicinity of a tunneling point, as well as
the electron interactions in the ``feeding leads''. Namely, when the leads
are one-dimensional, they ought to be described by a Luttinger liquid, rather
than the Fermi liquid since the latter is de-stabilized by interactions.
In real 1D wires the impurities away from the point contact  will complicate
matters, tending to localize electrons. This localization makes it difficult
to realize 1D Luttinger liquids in experiment. 

However, systems are available \cite{moon,Millik} which are free of undesirable
localization -- where the leads feeding a point contact are 2D fractional
quantum Hall (QH) fluids. In the appropriate regime of the QH liquid the
incoming current will be carried by {\it edge states}. Wen has demonstrated
\cite{wen} that the gapless edge excitations of a QH system are {\it chiral}
Luttinger liquids. Let us recall briefly the logic behind Wen's theory.
The long-length-scale physics of the bulk 2D QH state is described
by the massive 2+1 dimensional Chern-Simons theory
\be
S_{bulk}={i\over 4\pi\nu}\int a_\mu\partial_\nu a_\lambda \varepsilon_{
\mu\nu\lambda}d^2xd\tau.
\ee
The coefficient $\nu$ is uniquely fixed by the quantized Hall conductivity.
\footnote{For simplicity, we choose $\nu^{-1}$ to be an odd integer to
have only one branch of edge state.}
In the presence of a boundary, say at $y=0$, one can integrate out the bulk
degrees of freedom. The resulting 1D action for the edge field reads
\be
S_{edge}^R=-{1\over 4\pi\nu}\int dxd\tau (\partial_x\varphi_R)(i\partial_\tau+v
\partial_x)\varphi_R,
\ee
where $v$ is the velocity of edge excitation (we set $v=1$ hereafter).
The boson field $\varphi$ was introduced as $a_j=\partial_j\varphi$ to solve an 
 incompressibility constraint on the electron density,
$\varepsilon_{ij}\partial_ia_j=0$. Similarly, one writes for 
another edge
\be
S_{edge}^L=-{1\over 4\pi\nu}\int dxd\tau (\partial_x\varphi_L)(-i\partial_\tau+v
\partial_x)\varphi_L.
\ee
The charge density along the edge is given by $\rho(x)=\partial_x\varphi/2\pi$,
and the momentum operator conjugate to $\varphi$ is $\Pi=\rho/\nu$. 
Adding an extra electron to the edge is equivalent to creating a soliton
in $\varphi$ with electron creation operator being
\be
\Psi(x)\sim \exp[2\pi i\int^x\Pi(x')dx']=e^{i\varphi(x)/\nu}.
\ee
A quasiparticle of fractional charge $\nu e$ is created by $e^{i\varphi(x)}$
(speaking loosely, an electron is composed of  $\nu^{-1}$ quasiparticles).

Due to the chirality, backscattering is only possible when
opposite edges of the sample are close together, i.e. at the point contact.
Thus, localization in such  leads is absent. The analog of impurity
that causes backscattering is a narrow constriction which brings left
and right edges close enough for Laughlin quasiparticles to tunnel, as
illustrated in figure 7.2. This is achieved by applying a gate voltage $V_G$
across the narrow region in the Hall bar.

\begin{figure}
\epsfxsize=100truemm
\centerline{\epsfbox{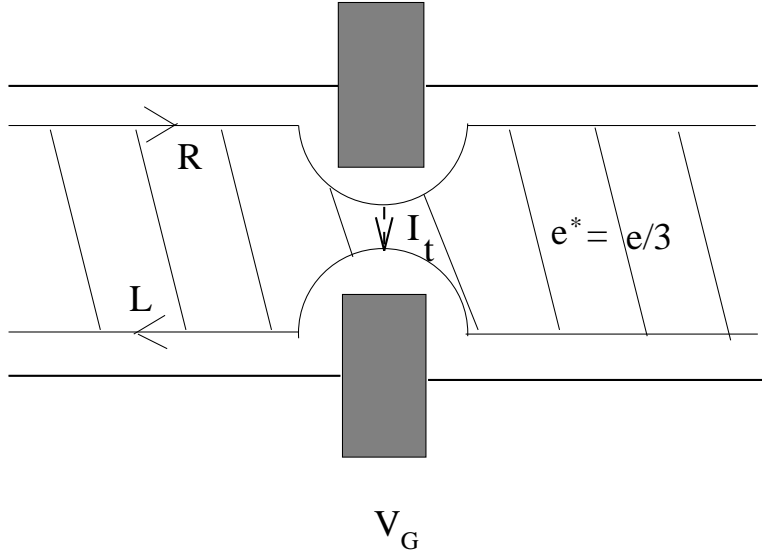}}
\caption{Quantum Hall bar with tunneling between the edge states.}
\end{figure}

Suppose that tunneling takes place at $x=0$. The edges are no longer
independent, but rather coupled via the tunneling term in the total action:
\be
S=S^L_{edge}(\varphi_L)+S^R_{edge}(\varphi_R) + 
\int d\tau V_{imp}(\varphi_L,\varphi_R)_{x=0}.
\ee
The most general form of the tunneling term is
\be
V=\Psi^+_L\Psi_R(0)+ {\rm h.c.}=
\sum_{m=1}^\infty v_m e^{im[\varphi_L(0)-\varphi_R(0)]} + {\rm c.c.}
\label{last:genpotential}
\ee
where $v_m$ are tunneling amplitues. The term with $m=1$ corresponds to 
quasiparticle tunneling, while $m=\nu^{-1}$ term -- to the electron tunneling.
In what follows we choose $\nu=1/3$. 
As was argued in \cite{CMP}, for this value of $\nu$ only the term
$m=1$ in (\ref{last:genpotential}) is relevant, corresponding to
the transfer of the $Q=e/3$ Laughlin quasiparticle. For generic filling
fraction, many types of quasiparticles contribute as relevant charge transfer.
So, the model of interest has the following Hamiltonian
(in a slightly different notations for further convinience):
\be
H=\half\int_{-\infty}^\infty dx[8\pi\nu\Pi^2+{1\over 8\pi\nu}
(\partial_x\varphi)^2]+\lambda\delta(x)\cos(\varphi_L-\varphi_R),
\label{last:QHham}
\ee
where the L and R components depend on $x,t$ as $\varphi_L(x+t), 
\varphi_R(x-t)$. One can tune the tunneling amplitude
$\lambda$ by varying the gate voltage $V_G$. The Hall  voltage $V$ between
the two edges of the QH liquid can be introduced in the model by adding
the term $-e(Q_L-Q_R)V/2\hbar$ to the Hamiltonian. This 
substitution has an effect of placing the charge carriers injected into
the left and right edges at different chemical potentials. As can be
easily seen, by shifting the fields $\varphi_{L,R}$ we can bring the voltage
dependence in the boundary term: $\cos(\varphi_L-\varphi_R+\omega t)$.
We will deal with the $V=0$ and $T=0$ case only.

As discussed in \cite{FLUDS}, in order to map (\ref{last:QHham}) to a boundary
problem on a half-line, it is convenient to proceed
in two steps. First introduce~:
\bea
\phi^e(x+t)&=&{1\over\sqrt{2}}\left[\varphi_L(x,t)+
\varphi_R(-x,t)\right]\nonumber \\
\phi^o(x+t)&=&{1\over\sqrt{2}}
\left[\varphi_L(x,t)-\varphi_R(-x,t)\right] \label{last:remii}
\eea
which are both left moving. The even and odd charges are related to the
charges of the original left- and right-moving edges by
\be
\Delta Q=Q_L-Q_R=\sqrt{2}Q^o, \qquad Q_L+Q_R=\sqrt{2}Q^e.
\ee
The backscattering current is related to $Q^o$, whereas the total
charge on both edges, $Q^e$, is conserved even in the presence of impurity.
It is clear that the interaction term
does not affect the even field, which therefore remains free.
As for the odd term, it can be mapped onto a boundary problem
as follows. Define~:
\bea
\phi^o_L(x,t)&=&\sqrt{2}\phi^o(x+t), \quad x>0, \nonumber \\
\phi^o_R(x,t)&=&\sqrt{2}\phi^o(-x+t), \quad x>0.\label{last:remiii}
\eea
The odd hamiltonian then reads~:
\be
H={1\over 2}\int^{\infty}_0
[8\pi g (\Pi^o)^2+{1\over 8\pi g}(\partial_x\phi^o)^2]
+\lambda \delta(x) \cos{1\over 2}\phi^o,
\label{remiv}
\ee
and in the following we will write  $\phi\equiv \phi^o$ and $g$
instead of
$\nu$.  Thus, for
this problem, ${\cal B}=\lambda  \cos{1\over 2}\phi(x=0,t)$.

The quantity of interest in this case is
the AC conductance at vanishing temperature.
A standard way of representing it is through the Kubo formula 
(\ref{last:kubo}). It is usually easiest to calculate (\ref{last:kubo})
using Matsubara formalism \cite{Mahan}.
First, one defines current-current correlator in the Matrubara
formalism~:
\bea
\sigma_{\alpha\beta}(\omega)&=&{i\over \omega}
\Pi_{\alpha\beta}(\omega)\nonumber \\
\Pi_{\alpha\beta}(x,\tau)&=&-\langle T_\tau j_\alpha(x,\tau)j_\beta(x,0)\rangle
\label{last:manydefs} \\
\Pi_{\alpha\beta}(x,i\omega_M)&=&\int_0^\beta d\tau e^{i\omega_M\tau}
\Pi_{\alpha\beta}(x,\tau) \nonumber
\eea
Then, one substitutes $i\omega_M\to\omega+i\delta$ into $\Pi_{\alpha\beta}$
and sends $\delta\to 0$. Eventually, we get for the real part of conductivity
tensor~:
\be
{\rm Re}\sigma_{\alpha\beta}=-{1\over\omega}{\rm Im}\Pi_{\alpha\beta}(\omega).
\ee
The imaginary part of the spectral function in the latter formula
can be expanded as follows:
\be
-{\rm Im}\Pi_{\alpha\beta}(\omega)=\pi(1-e^{-\beta\omega})e^{\beta\Omega}
\sum_{n,m}e^{-\beta E_n}\langle n|j_\alpha|m\rangle\langle m|j_\beta|n\rangle
\delta(\omega+E_n-E_m).
\ee
In the limit of zero temperature, $\beta\to\infty$, the sum over $n$
contains only one term -- the ground state, so one has~:
\be
-{\rm Im}\Pi_{\alpha\beta}(\omega)=\pi
\sum_{m}\langle 0|j_\alpha|m\rangle\langle m|j_\beta|0\rangle
\delta(\omega-E_m).
\ee
In the above formulas $j_{\alpha,\beta}\equiv j$ 
 is the physical current in the unfolded system,
$j=\partial_t(\varphi_L-\varphi_R)$.  Without impurity, the AC
conductance of the Luttinger liquid is frequency independent, $G=g$.
When adding the impurity, it becomes $G={g\over 2}+\Delta G$. After
some
simple manipulations using the folding, one finds~:
\bea
\Delta G(\omega_M)&=&{1\over 8\pi \omega_M
L^2}\int^{L}_0
dxdx'\int
_{-\infty}^\infty dy e^{i\omega_My} \nonumber \\ 
& &\left[\langle\partial_z\phi(x,y)\partial_{\bar{z}'
}\phi(x',0)\rangle
+\langle\partial_{\bar{z}}\phi(x,y)\partial_{z'}\phi(x',0)\rangle\right],
\label{kubozeri}
\eea
where $z=x+iy$.

\section{Boundary sinh-Gordon model}

We discuss in this section one other boundary integrable model, the boundary
sinh-Gordon model with ${\cal B}=M\cosh\phi(0)$. 
It is related to the boundary sine-Gordon model
by the analytic continuation in the coupling constant $g$, but we do not
make use of this  fact here. Instead, we use it as a toy model for 
studying the correlation functions. It is fairly straighforward then how
to tackle more complicated models. The advantage of the sinh-Gordon
model is its technical simplicity: the particle spectrum of this model
consists of only one scalar particle -- a sinh-Gordon boson.  
The bulk sinh-Gordon model is well-studied in the literature.

\subsection{ The boundary reflection coefficient}

We need to know the boundary reflection coefficient for sinh-Gordon model:
\be
S_{ShG}=\int
dxdt\left[{1\over 2}(\partial_\mu\phi)^2-{m^2\over g^2}\cosh(g\phi)\right]
-{M_B\over g^2}\int dt\cosh\left({g\phi(x=0)\over 2}\right)
\label{ShG}
\ee
For this we use the fact that the action (\ref{ShG}) is related to that of
sine-Gordon model 
\be
S_{SG}=\int
dxdt\left[{1\over 2}(\partial_\mu\phi)^2+{m^2\over\beta^2}\cos(\beta\phi)\right]
+{M_B\over \beta^2}\int dt\cos\left({\beta\phi(x=0)\over 2}\right)
\label{SnG}
\ee
by the analytic continuation in coupling constant $g=i\beta$. Recall that
the sinh-Gordon model has only one particle with neutral charge.  
A useful
observation is that the bulk scattering matrix for this particle,
\be
S(\theta, B)={\tanh{1\over 2}(\theta-i{\pi B\over 2})\over
\tanh{1\over 2}(\theta+i{\pi B\over 2})}, \qquad B={2g^2\over 8\pi+g^2}
\label{sm}
\ee
can be obtained from the scattering matrix of the lightest breather
of the sine-Gordon model \cite{ZZ} by the  mentioned above analytic
continuation in coupling constant. We assume that the same is true for the
boundary reflection matrices. Partial confirmation of this can be found
in \cite{Corr}, where this statement was proven in the semi-classical limit.
So, we will use the result obtained in \cite{Ghosh} for the reflection
coefficient of the lightest breather in the boundary sine-Gordon model.
The general solution of Ghoshal has two free boundary parameters \cite{Ghosh}:
\be
R_{mass}(\theta)=U(\eta, \theta)U(i\vartheta, \theta)R_0(\theta)
\label{Gh}
\ee
One can argue that for the boundary term of the form $M\cos(\beta\phi/2)$
one should set $\eta=0$, the other parameter $\vartheta$ being related
to $M_B$. So, the reflection coefficient for the boundary sinh-Gordon
model with boundary interaction $M_B\cosh(g\phi/2)$ obtained from (\ref{Gh})
by the analytic continuation reads:
\be
R_{mass}(\theta)={\tanh{1\over 2}(\theta+{\vartheta B\over 2}-
{i\pi\over 2})\over
\tanh{1\over 2}(\theta-{\vartheta B\over 2}+{i\pi\over 2})}
{\cosh{1\over 2}(\theta+{i\pi\over 2})\cosh{1\over 2}(\theta-{i\pi B\over 4})
\cosh{1\over 2}(\theta+{i\pi B\over 4}-{i\pi\over 2}) \over
\cosh{1\over 2}(\theta-{i\pi\over 2})\cosh{1\over 2}(\theta+{i\pi B\over 4})
\cosh{1\over 2}(\theta-{i\pi B\over 4}+{i\pi\over 2}) }
\label{Rm}
\ee
The massless limit of (\ref{Rm}) is obtained by sending $\theta$ and $\vartheta$
to infinity. We have in the massles limit
\be
R(\theta)=-\tanh{1\over 2}(\theta-\theta_B-{i\pi\over 2})
\label{mlR}
\ee
where $\theta$ is now massless rapidity related to the momentum as 
in (\ref{eq:enmom}).
This is as well the massless limit of the sine-Gordon reflection matrix.
The boundary parameter $\theta_B$ is related to the boundary mass $M_B$
in the way unknown to us.  Note that in terms of momentum expression (\ref{mlR})
looks like
$$ R(p)={p-ip_B\over p+ip_B} $$
The definition of $K(\theta)$ in the massive case \cite{GZ}
$$K_{mass}(\theta)=R_{mass}\left({i\pi\over 2}-\theta\right)$$
becomes in the massless limit
\be
K(\theta)=R\left({i\pi\over 2}+\theta\right)
\label{eq:defK}
\ee
This follows from
\bea
K(\theta)&=&
\lim_{\theta_0\to\infty}K_{mass}(\theta_0+\theta)=\lim_{\theta_0\to\infty}
R_{mass}({i\pi\over 2}-\theta_0-\theta)= \nonumber \\
&=&\lim_{\theta_0\to\infty}
S(2\theta)R_{mass}({i\pi\over 2}+\theta_0+\theta)
=R({i\pi\over 2}+\theta) \nonumber
\eea
where we used massive crossing-unitarity relation \cite{GZ} above.

\subsection{ Sinh-Gordon form-factors}

The form-facotrs for the massive Sinh-Gordon model were found 
in \cite{Mussardo}. Let us list here some of them that we need.
For the field $\phi$ itself, the form-factors between the ground state
and n-particle states are:
\bea
F_{2n+1}(\beta_1,...,\beta_{2n+1})&=&\langle 0|\phi(0)|\beta_1,...,\beta_{2n+1}
\rangle= \nonumber \\
&=&H_{2n+1}Q_{2n+1}(e^{\beta_1},...,e^{\beta_{2n+1}})
\prod_{i<j}{F_{min}(\beta_i-\beta_j)
\over e^{\beta_i}+e^{\beta_j}}
\label{ff}
\eea
where $H_n$ are normalization constants 
$$H_{2n+1}={1\over\sqrt{2}}\left[{4\sin{\pi B\over 2}\over F_{min}(i\pi,B)}
\right]^{n} $$
$Q_{2n+1}$ are symmetric polynomials in the variables $p_i=e^{\beta_i}$,
$$Q_1=1, \qquad Q_3=\exp(\beta_1+\beta_2+\beta_3),$$
$$Q_5=\sigma_5^{(5)}(\sigma_2^{(5)}\sigma_3^{(5)}-c_1^2\sigma_5^{(5)}),
\qquad c_1=2\cos{\pi B\over 2} $$
By parity, only form-factors with 
odd number of particles are allowed. We used the standard basis of elementary
symmetric functions $\sigma_i^{(n)}$ in variables $p_j$ in the formulas above,
and
$$ F_{min}=D\exp
\left[8\int_0^\infty {dx\over x}{\sinh{Bx\over 4}\sinh(1-{B\over 2}
){x\over 2}\sinh{x\over 2}\over \sinh^2x}\sin^2{x(i\pi-\beta)\over 2\pi}
\right]$$
$$ D=\exp\left[-4\int_0^\infty{dx\over x}{\sinh{Bx\over 4}\sinh(1-{B\over 2}
){x\over 2}\sinh{x\over 2}\over \sinh^2x}\right]$$
The plot of $F_{min}$ is given in Figure 7.3.

\begin{figure}
\epsfxsize=100truemm
\centerline{\epsfbox{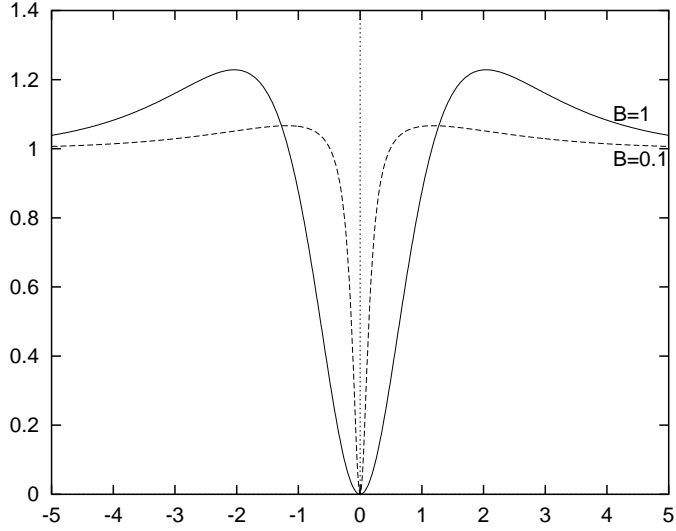}}
\caption{Plot of $|F_{min}(\beta,B)|^2$ for two different couplings.}
\end{figure}

Since the sinh-Gordon form-factors (\ref{ff}) for the field operator $\phi$
are scalars with respect to the Lorentz transformations, they are invariant
under the simultaneous
shift of all the rapidities $\beta_i\to\beta_i+\beta_0$, where $\beta_0$
is some constant.
Taking the massless limit implies making the substitution $\beta_i=\th_i+
\beta_0$ with $\beta_0
\to\infty$ for the right-moving particles, and making the substitution
$\beta_i=-\th_i-\beta_0$ for the left-moving particles. Here
$\th_i$ are the rapidities of massless particles. So, the massless
form-factors derived by this procedure are:
\bea
\langle 0|\phi_R(0)|\th_1,...,\th_{2n+1}\rangle_R&=&
N_0^{-1/2}H_{2n+1}Q_{2n+1}(e^{\th_1},...,e^{\th_{2n+1}})
\prod_{i<j}{F_{min}(\alpha_i-\alpha_j)
\over e^{\th_i}+e^{\th_j}}
\nonumber \\
\langle 0|\phi_L(0)|\th_1,...,\th_{2n+1}\rangle_L&=&
N_0^{-1/2}H_{2n+1}Q_{2n+1}(e^{\th_1},...,e^{\th_{2n+1}})
\prod_{i<j}{\overline{F}_{min}(\th_i-\th_j)
\over e^{\alpha_i}+e^{\alpha_j}}
\nonumber
\eea
where we used the fact that $F_{min}(-\th)=\overline{F}_{min}(\th)$.
The normalization factor $N_0$ can be fixed by comparison with the
massless free boson two-point correlation function
(\ref{last:massless_cf}). Inserting into the vacuum two-point correlation
function of free fields the full set of normalized intermediate states,
and using the form-factors above,
we obtain the contribution from each n-particle state in the form
$$ {1\over 2N_0}\left({a_n\over z^2} + {a_n\over \bar{z}^2}\right) $$
with some constant 
coefficients $a_n$,
$$
a_{n}= \int {d\th_1\ldots d\th_{2n+1}\over
(2\pi)^{n}n!}
 \left(e^{\th_1}+\ldots+e^{\th_{n}}\right)^2
e^{-(e^{\th_1}+\ldots+e^{\th_{n}})}
|F_n(\th_1,\ldots,\th_{n})|^2.
$$
 The normalization factor $N_0$  can then be
obtained by summing the series $\sum a_n$ and requiring the result to be 1.
One can compute $a_n$ numerically and
observe that they decrease very fast with $n$, and so
can be truncated after a few terms. In particular,
using one, three
and five particle form-factors
we found the approximate value of $N_0$:
$$ N_0=1.005\qquad B=1.0$$
$$ N_0=1.0002\qquad B=0.1$$
Note that the rate of convergence of the series $\sum a_n$ gives us a hint
about the convergence of analogous series in the case with the interaction
at the boundary (see below). The rapid convergence holds in the massive
case as well, and has a physical explanation \cite{CardMuss}.

\section{Calculation of correlation functions}

In the massless limit the Green's function of current-current type
on the half-line
can be factorized  as follows:
\bea
G(x_i,t_i,g,M_B)&=&\langle 0|\partial_x\phi(x_1,t_1)\partial_x\phi(x_2,t_2)|0\rangle =\nonumber \\
&=&G_0(x_1-x_2, t_1-t_2, g)+G_1(x_1+x_2, t_1-t_2, g,M_B) \label{factr}
\eea
(this can be easily seen from the form-factor approach below).
The first term $G_0$ does not depend on boundary coupling and is
in fact equal to the current-current correlation function in the 
translationally invariant system on the line. It is the second term
$G_1$ where the breakdown of translational invariance manifests itself
explicitly and which carries all the dependence on the boundary coupling.

In general $M_B$ is the only dimensional coupling that enters the Green's
function.
There are two values of $M_B$ where the theory is scale-invariant. One of
them, $M_B=0$, corresponds to the free boundary condition
 $\partial_x\phi|_{x=0}=0$,
while another one, $M_B=\infty$, corresponds to the fixed boundary condition
$\phi|_{x=0}=0$. For these values of $M_B$
 the Green's functions are known exactly:
$$G|_{M_B=0}=G_0(x_1-x_2)+G_0(x_1+x_2)$$
$$G|_{M_B=\infty}=G_0(x_1-x_2)-G_0(x_1+x_2)$$
where 
\be
G_0(x,t)={1\over 2\pi}{x^2-t^2\over (x^2+t^2)^2}
\label{last:massless_cf}
\ee
Thus, the boundary mass $M_B$ induces the boundary RG flow from short
to large distances, which vary $G_1$ between $-G_0$ and $+G_0$.

We will work on the half-plane which geometry is shown in Figure 7.4.
Based on the euclidean duality,
there are two alternative ways to introduce the Hamiltonian picture \cite{GZ}. 
First, one can take $x$ to be euclidean time. In this case the equal time 
section is an infinite line $x=$const, $y\in(-\infty,\infty)$. Hence the
associated space of states is the same as in the bulk theory. The boundary
at $x=0$ appears as the ``time boundary,'' or initial condition at $x=0$
which is described by the {\it boundary state} $|B\rangle$ (a particular
state from the bulk Hilbert space). The correlators are expressed as
\be
\langle O_1(x_1,y_1)...O_N(x_N,y_N)\rangle = 
{\langle B|{\cal T}_x[O_1(x_1,y_1)...O_N(x_N,y_N)]|0\rangle
\over \langle 0|B\rangle}, \label{last:viewI}
\ee
where $O_i(x,y)$ are the Heisenberg local field operators
\be
O_i(x,y)=e^{-xH}O_i(0,y)e^{xH},
\ee
and ${\cal T}_x$ means x-ordering.

Alternatively, one can take the direction along the boundary to be the time.
In this case boundary appears as a boundary in space, and the Hilbert space
of states is associated with the semi-infinite line $y=$const, $x\in[0,\infty)$.
The correlation functions of any local fields $O_i(x,y)$ can be
computed in this picture as the matrix elements
\be
\langle O_1(x_1,y_1)...O_N(x_N,y_N)\rangle = 
{\langle 0||{\cal T}_y[O_1(x_1,y_1)...O_N(x_N,y_N)]||0\rangle
\over \langle 0||0\rangle}, \label{last:viewII}
\ee
where $||0\rangle$ is the ground state of the boundary Hamiltonian,
$O_i(x,y)$ are understood as the corresponding Heisenberg operators
\be
O_i(x,y)=e^{-yH_B}O_i(x,0)e^{yH_B},
\ee
and ${\cal T}_y$ means y-ordering.

The equality of expressions (\ref{last:viewI}) and (\ref{last:viewII})
can be understood as a definition of the boundary state, which is chosen
such as to provide the equivalence of correlators.

\begin{figure}
\epsfxsize=75truemm
\epsfysize=75truemm
\centerline{\epsfbox{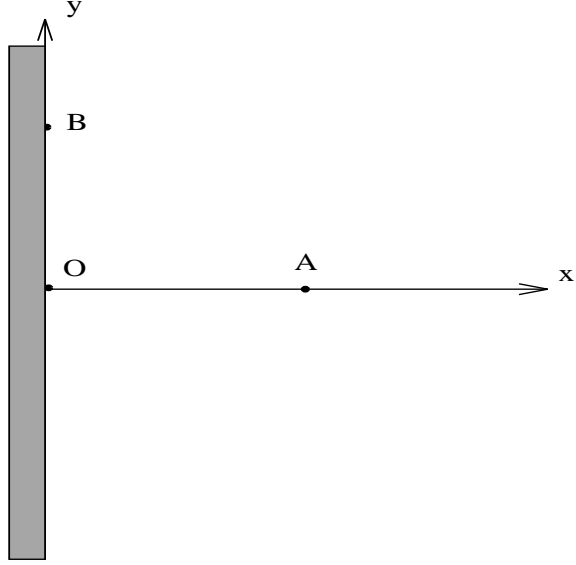}}
\caption{The geometry of space-time.}
\end{figure}

\subsection{Boundary-in-time representation}

First suppose that $x>0$ is imaginary time coordinate and $y$ is
space coordinate, $z=x+iy$. It means that time translation is performed by
the operator $T=\exp(xH)$, with $H$ being the bulk Hamiltonian.

We wish to compute the following matrix element:
\bea
 \langle B|\partial_x\phi(x_1,y_1)\partial_x\phi(x_2,y_2)|0\rangle
&=&
\langle B|\partial_z\phi_R(z_1)\partial_{\bar z}\phi_L(\bar{z_2})|0\rangle
\nonumber \\
&+&
\langle B|\partial_{\bar z}\phi_L(\bar{z_1})\partial_z\phi_R(z_2)|0\rangle
\label{cor}
\eea
where $\phi$ is a massless free field $\phi=\phi_L(\bar{z})+\phi_R(z)$
and $|B\rangle$ denotes the massless boundary state 
of the sinh-Gordon model \cite{GZ}:
\be
{
|B\rangle=|0\rangle+
\sum_{N=1}^\infty\int_{-\infty<\theta_1<...<\theta_N<\infty}
\left[\prod_{i=1}^N{d\theta_i\over 2\pi}K(\theta_i)\right]
A_L(\theta_N)...A_L(\theta_1)A_R(\theta_1)...A_R(\theta_N)|0\rangle
} \label{bs}
\ee
Because $|B\rangle$ has chirality zero,  products of the fields
of the same chirality do not contribute to the right hand side of eq.
(\ref{cor}).
Parameter $\theta$ is a massless rapidity in terms of which the energy
and momentum of particles read:
\footnote{Here $\mu$ is an arbitrary mass scale.}
\be
E_L=-P_L=\mu e^\theta, \qquad E_R=P_R=\mu e^\theta,
\label{eq:enmom}
\ee
and $A_{L,R}(\theta)$ are the left and right moving 
particle creation operators, the particles in our case being the massless
sinh-Gordon bosons. The fact that the sinh-Gordon particle bounces
elastically off the boundary allows us to have a much simpler expression
for the boundary state (\ref{bs}) than that for the plane waves.
Substituting the boundary state (\ref{bs}) into (\ref{cor})
and using the fact that left (right) moving field acts only
on left (right) moving particles, we obtain the following expansion
in terms of the form-factors:
\bea
&&\langle B|\partial_x\phi(x_1,y_1)\partial_x\phi(x_2,y_2)|0\rangle=
\langle 0|\partial_x\phi(x_1,y_1)\partial_x\phi(x_2,y_2)|0\rangle+
\nonumber \\
 &+&
\sum_{N=1}^\infty \mu^2\int_{-\infty<\theta_1<...<\theta_N<\infty}
\left[\prod_{i=1}^N{d\theta_i\over 2\pi}K(\theta_i)\right]
\langle 0|\phi_L(0)|\theta_1,...,\theta_N\rangle_L
\langle 0|\phi_R(0)|\theta_1,...,\theta_N\rangle_R \nonumber \\
&\times&\left(\sum_{i=1}^N e^{\theta_i}\right)^2
\left(e^{-(x_1+x_2+iy_1-iy_2)\mu \left(\sum_{i=1}^Ne^\theta_i\right)} + 
{\rm c.c.}\right) \nonumber \\
&\equiv& G_0(z_1-z_2)-\sum_{n=0}^\infty I_{2n+1} \label{corI}
\eea
The product of two massless form-factors for the  left-moving and the
right-moving field above is in fact equal to $|F_N(\theta_1,...,\theta_N)|^2$,
the modulo square of the bulk massive form-factor (with the only
difference that $\th$ is now the {\it massless} rapidity).

The first term in (\ref{corI}) is  $G_0$, the Green's function of the free massless fields
on the plane:
$$G_0(z,\bar{z})={1\over 4\pi}\left({1\over z^2}+{1\over \bar{z}^2}\right)$$
 The explicit expressions for the first few terms $I_{2n+1}$ are:
\be
I_1={\mu^2\over 2N_0}
\int_{-\infty}^{+\infty}{d\theta\over 2\pi}e^{2\theta}
\left(e^{-(x_1+x_2-iy_1+iy_2)\mu e^\theta} + {\rm c.c.}\right)
\tanh{\theta-\theta_B\over 2} \label{contribI}
\ee
Changing variables, $p=\exp(\theta)$, it can be rewritten as
\be
I_1={\mu^2\over 4\pi N_0}\int_0^\infty pdp {p-T_B\over p+T_B}
\left(e^{-(z_2+\bar{z}_1)\mu p} + c.c\right), \qquad T_B\equiv e^{\theta_B}
\label{last:prepr}
\ee
and
\bea
I_3={\mu^2H_3^2\over 3! N_0}&&\int_{-\infty}^\infty
\prod_{i=1}^3{d\theta_i\over 2\pi}\tanh{\theta_i-
\theta_B\over 2}
\label{contribIII} \\
&\times&\prod_{i<j}^3{|F_{min}(\theta_i-\theta_j)|^2 \over
2(1+\cosh(\theta_i-\theta_j))}\left(\sum_{i=1}^3e^{\theta_i}\right)^2
\left(e^{-(z_2+\bar{z}_1)\mu\left(\sum_{i=1}^3e^{\theta_i}\right)} + 
c.c\right) \nonumber
\eea
Plot of these two contributions to the two-point
correlation function for the points OA ($x_1=y_1=y_2=0$) is 
shown in figures 7.5 and 7.6, and few
values of $I_5$ are given in Table 7.1. It turns out that the series $\sum I_n$
converge fast because each term $I_n$ is by the factor of hundred
smaller than $I_{n-2}$. We checked it up to $n=5$. Each integral
$I_n$ converges for any finite value of $(x_1+x_2)$, but is divergent
for $x_1=x_2=0$. Therefore, we need to do an additional work to find
the correlation function between two points on the boundary, OB (Fig. 7.4).

\begin{figure}
\epsfxsize=100truemm
\centerline{\epsfbox{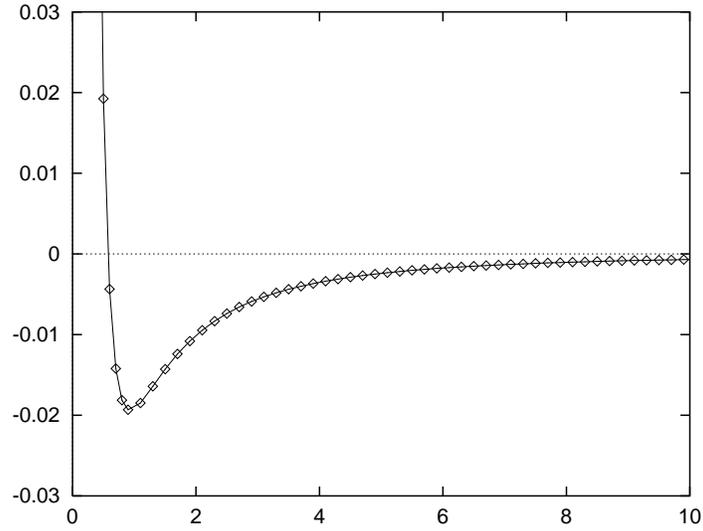}}
\caption{Plot of the one-particle contribution to correlator between
               two points $(0,0)$ and $(x,0)$ as a function of $x$.}
\end{figure}
\begin{figure}
\epsfxsize=100truemm
\centerline{\epsfbox{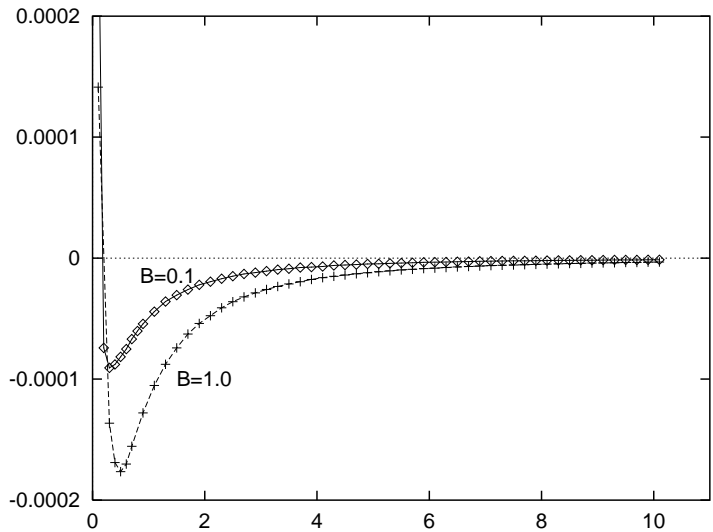}}
\caption{Plot of the three-particle contribution to correlator
          for two different couplings; the curve for B=0.1 is shown scaled
          by the overall factor 10 compared to the true curve.}
\end{figure}
\begin{table}[htb]
\begin{center}
\begin{tabular}{|c|c|} \hline
 $x$   & $I_5$ \\ \hline
$1.1$  & $-4.65436256\cdot 10^{-6}$  \\
$2.1$  & $-2.02534138\cdot 10^{-6}$  \\
$3.1$  & $-1.11571005\cdot 10^{-6}$  \\
$4.1$  & $-7.03902032\cdot 10^{-7}$  \\
$5.1$  & $-4.81526684\cdot 10^{-7}$ \\ \hline
\end{tabular}
\end{center}
\caption{Five-particle contribution for $B=1$.}
\end{table}

Let us find equivalent expressions for the integrals $I_n$
which would be finite for $x_1=x_2=0$. For this, we note
that the contour of integration can be rotated to go from
$0$ to $i\infty$ in the complex $p$-plane, eq. (\ref{last:prepr}).
 This rotation is
equivalent to the shift of countour of integration in $\theta$-plane
up by $i\pi/2$, with having the contour to pass below the poles of $K(\theta)$
(figure 7.7).
 For the complex conjugated term one has to
rotate countour clockwise in p-plane (shift by $-i\pi/2$ in $\theta$-plane).
So, we obtain 
\be
I_n= -\mu^2\int_{-\infty}^\infty \left[\prod_{i=1}^n
 {d\theta_i\over 2\pi}R(\theta_i)\right]
\left(\sum^n_{i=1}e^{\theta_i}\right)^2 |F_n(\theta_1,...,\theta_n)|^2
e^{-(y_1-y_2-ix_1-ix_2)\mu\left(\sum_{i=1}^ne^{\theta_i}\right)} + c.c
\label{analcon}
\ee
where we used eq. (\ref{eq:defK})
and $y_1-y_2>0$ is assumed.

\begin{figure}
\epsfxsize=100truemm
\centerline{\epsfbox{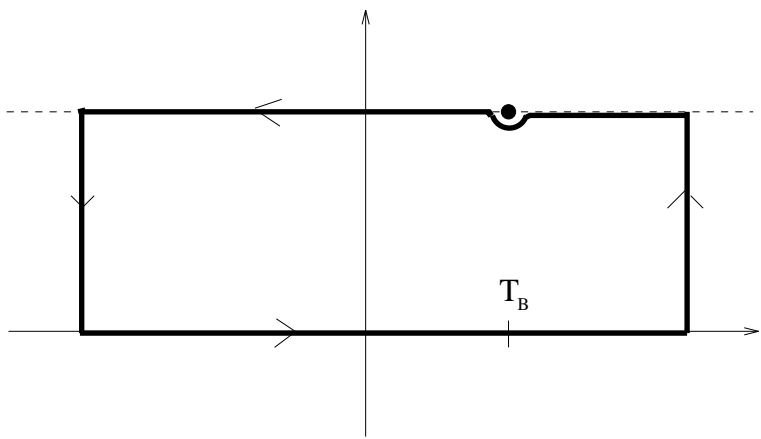}}
\caption{The contour of integration in the complex $\th$ plane.
The vertical intervals of the contour are assumed to be sent to infinity.}
\end{figure}

Notice that this expression could be also obtained if we had
started from the dual picture where time goes along the boundary
and space is a half-line and inserted a full basis of asymptotic
states of the form $|\theta_1,...,\theta_n\rangle_L + R(\theta_1)...R(\theta_n)
|\theta_n,...,\theta_1\rangle_R$ between the fields in the correlator.
This dual picture is to be discussed in detail in the next section. 

For the points OB in Fig. 7.4 ($x_1=x_2=0, y_2=0$)
 expression (\ref{analcon}) reads:
\be
I_n= -\mu^2\int_{-\infty}^\infty {d\theta_1...d\theta_n\over (2\pi)^n}
\left[\prod_{i=1}^n R(\theta_i)+\prod_{i=1}^n\bar{R}(\theta_i)\right] 
\cdot\left(\sum^n_{i=1}e^{\theta_i}\right)^2 |F_n(\theta_1,...,\theta_n)|^2
e^{-y_1\mu\left(\sum_{i=1}^ne^{\theta_i}\right)} 
\label{new:analcon}
\ee
E.g. for $n=1$ we have
$$ I_1=-{\mu^2\over 2\pi N_0}\int_{-\infty}^\infty d\theta 
e^{2\theta-y_1\mu e^{\theta}}\tanh(\theta-\theta_B).$$

\subsection{Boundary-in-space representation}

One can in principle compute the same correlation correlation functions
adopting a different point of view on the space-time geometry
(figure 7.4). In euclidean
field theory the role of space and time can be interchanged. Now, 
in the alternative representation, $x>0$
is a space coordinate, and $y$ is an imaginary time coordinate. The theory
is invariant under the global time shift, which is performed by the
operator $T=\exp(yH)$.

However, the theory, being defined on a semi-infinite line, has rather
intricate structure of the ground state, and the form-factors are not known to
us {\it a priori}. In particular, the ground state is not invariant under
the space translations.
One can speculate \cite{GZ} that the ground state
for a theory on a half-line $||0\rangle$
can be obtained from the bulk vacuum $|0\rangle$ by
the action of a ``boundary creation operator'' $\hat{B}$: $||0\rangle=\hat{B}
|0\rangle$. Speaking loosely, boundary is an infinitely heavy 
inpenetrable particle sitting at the origin. 
The asymptotic particle states are not pure L or R moving, but rather
the superpositions of those. E.g., one-particle states are
\be
||\th\rangle=|\th\rangle_L + R(\th)|\th\rangle_R, \qquad \th>0
\label{last:superp}
\ee
To find the correlation function $\langle 0||\partial_x\phi(x_1,y_1)
\partial_x\phi(x_2,y_2)||0\rangle$ we conjecture that the following equality
holds:
\be
\langle 0||\partial_x\phi(x_1,y_1)\partial_x\phi(x_2,y_2)||0\rangle=
\sum_n {\langle 0|\partial_x\phi(x_1,y_1)||n\rangle\langle n||
\partial_x\phi(x_2,y_2)|0\rangle \over \langle n||n\rangle},
\label{last:conj}
\ee
with the particular ansatz for the intermediate states $||n\rangle$:
\be
||\th_1,\dots,\th_n\rangle = |\th_1,\dots,\th_n\rangle_L + 
R(\th_1)\cdots R(\th_n)|\th_n,\dots,\th_1\rangle_R
\label{last:ansatz}
\ee
(in the {\it bra} states, $\langle n||$, the factors $R(\th)$ become complex
conjugated).
Substituting (\ref{last:ansatz}) into (\ref{last:conj}), combining the complex
conjugated terms in the series and using the {\tt bulk} form-factors, 
we indeed get the correct expression for the correlator
(\ref{analcon}). In the course of this calculation one has to use
the unitarity of the reflection matrix, $\overline{R}(\th)R(\th)=1$,
as well as the property that the vacuum average of the quantum field $\phi$
vanishes, $\langle\phi\rangle=0$. 

Note that in principle it is possible to formulate the set of 
general equations for the form-factors of integrable models on a half-line,
$\langle 0||\phi(0)|n\rangle$,
in analogy with the equations of \cite{Smir}. E.g., it is obvious that
the following should hold:
$$\langle 0||\phi(0)|\th\rangle=R(\th)\langle 0||\phi(0)|-\th\rangle .$$

\subsection{The renormalization group analysis}

Let us study the behavior of integrals $I_n$ under the change
of scale $z\to e^\lambda z$. Such a rescaling can be compensated
by the change $\theta_B\to\theta_B+\lambda$ and by the overall
normalization factor $Z(\lambda)=e^{2\lambda}$ to have the integrals
(and hence the correlator) unchanged. Repeating this RG transformation,
we will flow to the UV or IR fixed points $\theta_B=\pm\infty$
(depending whether $\lambda>0$ or $\lambda<0$).
For
such values of $\theta_B$ the hyperbolic tangent factors in the
integrands are equal to $\pm 1$, and the integrals are proportional
 to $\pm 1/z^2$.
On the plots Figure 7.5 and Figure 7.6 one can see two regimes:
$\mu z\ll 1$ and $\mu z \gg 1$ when the functions behave as $\pm 1/z^2$.
The non-trivial behaviour at the intermediate scales is due to the presence
of boundary, which introduces a scale $\mu e^{\theta_B}$
corresponding approximately to the position of the deep. Shifting
$\theta_B$ corresponds to the motion of the deep to the right or
left on Figures 7.5 and 7.6, untill it will go away completely and one of the
regimes will dominate over all scales.

\subsection{The use of Kubo's formula}

To use the Kubo formula (\ref{last:kubo}), we adopt the first
point of view
where the boundary is taken into account through the
introduction of the boundary state $|B\rangle$. Write~:
\be
\langle\partial_{\bar{z}}\phi(x,y)\partial_{z'}\phi(x',0)\rangle=
\int_0^\infty dE {\cal G}(E) \exp\left[-E(x+x')-iEy\right].
\label{last:only}
\ee
One obtains ${\cal G}(E)$ simply by fixing
the energy to
a particular value in (\ref{corI}). When this is done, the remaining
integrations
occur on a finite domain for each of the individual particle energies
since $\sum_{i=1}^{2n+1} e^{\th_i}=E$, and there is no problem of
convergence
anymore. One then gets~:
\bea
{\cal G}(E)
&= &\sum_{n=0}^\infty \int_{-\infty}^{\ln E}
{d\th_1\ldots
d\th_{2n}\over(2\pi)^{2n+1}(2n+1)!}{E^2\over E-e^{\th_1}-\ldots
-e^{\th_{2n}}}\nonumber \\ 
&&K(\th_1)\ldots K(\th_{2n})
K\left(\ln\left(E-e^{\th_1}-
\ldots-e^{\th_{2n}}\right)\right) \nonumber \\
&&\left|F_{2n+1}\left(\th_1\ldots\th_{2n},\ln\left(E-e^{\th_1}-\ldots
-e^{\th_{2n}}
\right)\right)\right|^2,
\label{last:I}
\eea
with the constraint $\sum_{i=1}^{2n} e^{\th_i}\leq E$.
The denominator might suggest some possible divergences; it is
important
however to realize that it vanishes if and only if
the particle with rapidty $\th_{2n+1}$ has vanishing energy,
in which case the form factor vanishes too. By using the dual picture, one 
writes~:
\be
\langle\partial_{\bar{z}}
\phi(x,y)\partial_{z'}\phi(x',y')>=
\int_0^\infty dE {\cal F}(E) \exp\left[-iE(x+x')-E(y-y')\right],
\ee
The two expressions are in correspondence by the simple analytic
continuation~:
\be
{\cal G}(E)= i{\cal F}(iE).
\ee

Eq. (\ref{last:only}) is the only
correlation contributing to $\Delta G$ for positive Matsubara
frequencies, and
\be
\Delta G(\omega_M)= {{\cal G}(\omega_M)\over 4\
\omega_M}.
\ee
Here we have used the fact that $\omega_M L\ll 1$, i.e. the system,
although
large, is much
smaller than the wavelength associated with the (modulus of the) AC
frequency.
To go  to real frequencies, we can simply substitute
$\omega_M\to -i\omega$ in the $K$ matrices in the integrals
(\ref{corI}):
\be
\Delta G(\omega)={1\over 4\omega}{\rm Im}{\cal
G}(-i\omega)=
{1\over 4\omega}{\rm Re} {\cal F}(\omega).
\ee

\subsection{ The numerical work}

We computed  the integrals for 3 and 5 particle contributions
to correlation functions numerically using Monte-Carlo simulations.
The domain of integration was the hypercube with the length
of the side equal 40. Because of
exponential decay of the integrand, the contribution of the region
outside the box is small withing our accuracy. Since the integrals
diverge as the imaginary time approaches zero, we found it technically
more difficult to find the reliable results for $x<0.1$. 
We took $5\cdot 10^7$ points
to evaluate the integrand inside the box. We checked that the 
Monte-Carlo result is stable by increasing the number of points
and estimated the relative error in 3 and 5 dimensional integrals to
be about $1\%$, while the 1-dimensional integrals were evaluated with
$0.001\%$ accuracy. The higher-particle contributions can be 
computed in the same manner, but require considerable amount of
computer time. Because we expect them to be very small, we found their
evaluation unnecessary for the purpose of the present research.

\section{Boundary sine-Gordon correlators}

We wish to apply now the formalism developed in the previous section
to the massless boundary sine-Gordon model,
with ${\cal B}=M\cos\phi(0)$. The basic procedures and formulas
of the previous section apply directly to the present section, the 
difference appearing in a few technicalities. 

\subsection{Conductance at $g=1/3$}

We are interested
in the value of coupling constant $\beta^2=8\pi/3$
in (\ref{SnG}), which corresponds
to the quantum Hall liquid regime $\nu=1/3$. The particle spectrum
in the sine-Gordon model depends on the coupling constant. At 
$g=1/3$, we have three particles in the spectrum:
a soliton, antisoliton and their bound state -- breather.
The breather is completely analogous to the scalar particle of the sinh-Gordon
model, while the other two particles introduce technical differences. 

Let us describe what these differences are. In the boundary state, (\ref{bs}),
one has to introduce soliton and antisoliton creation operators $A^{(+)}$ and
$A^{(-)}$, and their boundary reflection matrices: 
\bea
|B\rangle=|0\rangle+\sum_{N=1}^{\infty}&&\int_{-\infty<\th_1<\dots<\th_N<\infty}
K^{a_1b_1}(\th_1)\dots K^{a_Nb_N}(\th_N) \nonumber \\
&&A_L^{a_N}(\th_N)\dots A_L^{a_1}
(\th_1)A_R^{b_1}(\th_1)\dots A_R^{b_N}(\th_N)|0\rangle
\label{last:bsI}
\eea  (the summation over the particle indices $a,b$ is assumed)
$$ K^{ab}(\th)=R^a_{\bar{b}}\left({i\pi\over 2}+\th\right) $$
The breather reflection
coefficient is still given by formula (\ref{mlR}), while for the soliton
and antisoliton massless reflection  matrices are 
\bea
R^{\pm}_{\mp}(\th)&=&e^{\sqrt{8\pi/3}}R(\th), \nonumber \\
R^{\pm}_{\pm}(\th)&=&e^{-\sqrt{8\pi/3}}R(\th), \label{eq:Rpm}
\eea
\be
R(\th)={1\over 2\cosh(\th-{i\pi\over 4})} 
{\Gamma({3\over 8}-{i\th\over 2\pi})\Gamma({5\over 8}+{i\th\over 2\pi})
\over\Gamma({5\over 8}-{i\th\over 2\pi})\Gamma({3\over 8}+{i\th\over 2\pi})}
\label{last:R}
\ee
(for the arbitrary $\beta$ expressions we refer to \cite{FSW}).
Correspondingly, there are more terms in the series expansion of the 
correlators since there are more intermediate particles. The leading
contribution, $I_1$, is still the one-particle contribution of the breather.
The next-after-leading contribution comes from two particles in the
intermediate state, soliton and antisoliton. Its magnitude  
is approximately 20\% of the value of $I_1$. Note that soliton and antisoliton 
can appear only in pairs since the total topological charge of the
intermediate states should vanish in order for it to contribute to
the correlator. The next terms come from the three-particle states 
of the breather with soliton and antisoliton, and the 
three breathers. Their magnitude is 1\% of the value
of the leading contribution. The form-factors for different particles
in the sine-Gordon model have been originally obtained by Smirnov
in the massive case \cite{Smir}, and their massless limit is given
in \cite{LSS}. These expressions are much more cumbersome than
that of sinh-Gordon model.

The leading contribution to the conductance computed along the lines
of the discussion above is given by
\be
\Delta G(\omega)^{(1)}=-\mu^2{\pi d^2\over 8}{\rm Re}\tanh\left[
\half\log\left({\omega\over\sqrt{2}T_B}\right)-{i\pi\over 4}\right]
\label{last:deltaGI}
\ee
where $\mu\approx 3.14 $ and $d\approx 0.1414$.
The contribution from the soliton-antisoliton state can be found in
\cite{LSS}. We plot the full function $G(\omega)$ in figure 7.8.

\begin{figure}
\epsfxsize=75truemm
\centerline{\epsfbox{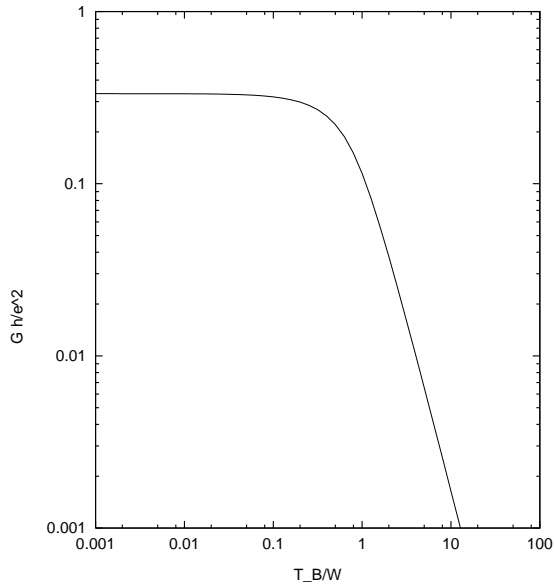}}
\caption{Frequency dependent conductance at $T=0$.}
\end{figure}

\subsection{The free point, $g=1/2$}

In the free case $\beta^2=4\pi$ one has simply:
\bea
R^{\pm}_{\mp}(\th)&=&P(\th)={e^\th\over e^\th + i}, \nonumber \\
R^{\pm}_{\pm}(\th)&=&Q(\th)={i\over e^\th + i}. \label{last:free}
\eea
As for the form-factors, only the soliton-antisoliton form-factor is
non-zero,
$$f(\th_1,\th_2)=2\pi e^{(\th_1+\th_2)/2}. $$
Thus, we find for the conductance
\be
G(\omega)=\half\left[ 1- {T_B\over\omega}\tan^{-1}(\omega/T_B)\right]
\label{last:freecond}
\ee
This is in agreement with the solution of \cite{CMP} and also \cite{Konik}.

\chapter{Conclusion}

This review is concerned with the study of 2D integrable models
with boundary. In chapters 1 through 6 the formalism is developed
for dealing with such models and particular examples are discussed.
It involves a construction of solutions to the classical equations
of motion for a model on a half-line, solution of the quantum XXZ
chain in a boundary magnetic field, investigation of the boundary
bound states (a phenomenon caused by the presence of boundaries),
generalization of the Destri-deVega technique, which allows to find the exact
ground state scaling energy, to the theories with boundaries.
All this work was carried out mostly for the quantum sine-Gordon/Thirring
model, which provides a good basis for theoretical investigations
withing the framework of quantum integrable models.

There is a room for the further development of the above formalism.
For example, we often have chosen the boundary conditions
to be a specific, Dirichlet boundary condition. Consideration
of more general boundary conditions will most certainly be associated
with the  increased complexity, while it is not clear to us whether
the outcome will prize us with a new interesting physics, 
or at least whether the most general boundary condition has any
vital applications. 
An interesting direction of research is associated with the technique
of the thermodynamic Bethe ansatz and  the corresponding finite-difference
equations \cite{LukZam,Roaming}. Another interesting direction
has to deal with the non-integrable deformations of integrable models
\cite{GeniusMussardo}.

In chapter 7 we described some of the applications of boundary
integrable models to condensed matter physics. In particular,
we developed a technique to compute exactly time dependent
properties of the Kondo problem and/or two-state problem
of dissipative quantum mechanics, as well as the conductance
of the one-dimesional wires with a tunneling through impurity.
These problems have potential experimental applications and a room
for further theoretical work. The latter includes the computation
of the voltage and temperature dependent behavior of the conductance,
quantum noise, as well as studying of the Coulomb blocade phenomenon.

\end{document}